\documentclass[aps,superscriptaddress,showkeys,showpacs]{revtex4}

\usepackage{amssymb,hyperref}

\newtheorem{theorem}{Theorem}

\newcommand{\scat}{\hfill $\diamond$ \vspace {3ex}}

\begin{document}
\def\wPm{{$\widehat P$-matrix}}
\def\wPms{{$\widehat P$-matrices}}
\def\Bcal{{\cal B}}
\def\Scal{{\cal S}}
\def\real{\mathbb{R}}
\newcommand\e{{\bf e}}
\newcommand\x{{\bf x}}
\newcommand\y{{\bf y}}
\newcommand\Z{{\mathbb Z}}
\newcommand\R{{\mathbb R}}

\title{On the choice of a basis of invariant polynomials of a Finite Reflection Group. Generating Formulas for $\widehat P$-matrices of groups of the infinite series $S_n$, $A_n$, $B_n$ and $D_n$}

\author{Vittorino Talamini}
\email[]{vittorino.talamini@uniud.it}
%
\affiliation{DCFA, Universit\`a di Udine, via delle Scienze 206, 33100 Udine, Italy
\\and\\  INFN, Sezione di Trieste, via Valerio 2, 34127 Trieste, Italy}


\date{\today}

\begin{abstract}
\noindent Let $W$ be a rank $n$ irreducible finite reflection group and let $p_1(x),\ldots,p_n(x)$, $x\in\mathbb{R}^n$, be a basis of algebraically independent $W$-invariant real homogeneous polynomials. The orbit map $\overline  p:\mathbb{R}^n\to\mathbb{R}^n:x\to (p_1(x),\ldots,p_n(x))$ induces a diffeomorphism between the orbit space $\mathbb{R}^n/W$ and the set ${\cal S}=\overline  p(\mathbb{R}^n)\subset\mathbb{R}^n$. The border of ${\cal S}$ is the $\overline  p$ image of the set of reflecting hyperplanes of $W$. With a given basic set of invariant polynomials it is possible to build an $n\times n$ polynomial matrix, $\widehat P(p)$, $p\in\mathbb{R}^n$, sometimes called $\widehat P$-matrix, such that $\widehat P_{ab}(p(x))=\nabla p_a(x)\cdot \nabla p_b(x)$, $\forall\,a,b=1,\ldots,n$. The border of ${\cal S}$ is contained in the algebraic surface $\det(\widehat P(p))=0$, sometimes called discriminant, and the polynomial $\det(\widehat P(p))$ satisfies a system of differential equations that depends on an $n$-dimensional polynomial vector $\lambda(p)$.  Possible applications concern phase transitions and singularities. If the rank $n$ is large, the matrix $\widehat P(p)$ is in general difficult to calculate. In this article I suggest a choice of the basic invariant polynomials for all the reflection groups of type $S_n$, $A_n$, $B_n$, $D_n$,  $\forall\,n\in \mathbb{N}$, for which I give generating formulas for the corresponding $\widehat P$-matrices and $\lambda$-vectors. These $\widehat P$-matrices can be written, almost completely, as sums of block Hankel matrices. Transformation formulas allow to determine easily both the $\widehat P$-matrix and the $\lambda$-vector in any other basis of invariant polynomials. Examples of transformations into flat bases, $a$-bases, and canonical bases, are considered.
\end{abstract}

\pacs{(Mathematics Subject Classification 2010: 13A50, 20F55)
}
\keywords{Basic invariant polynomials; Basic invariants; Integrity bases; Finite reflection groups; Orbit spaces; Discriminants; Hankel matrices.}

\maketitle


\section{Introduction\label{Intro}}

When a rank $n$ irreducible finite reflection group $W$ acts in $\R^n$, there exist a basis of $n$ algebraically independent $W$-invariant real homogeneous polynomial functions $p_1(x),\ldots,p_n(x)$, $x\in\R^n$, such that all $W$-invariant polynomial (or $C^\infty$) functions $q(x)$ can be uniquely written as polynomial (or $C^\infty$) functions of the basic set: $q(x)=\widehat q(p_1(x),\ldots,p_n(x))$,  with $\widehat q(p)$ a polynomial (or $C^\infty$) function of $p=(p_1,\ldots,p_n)\in\R^n$ \cite{Chev1955,schw-top}. By saying that $q(x)$ is a {\em $W$-invariant function}, one means that $q(gx)=q(x)$, $\forall \,g\in W$, $\forall\, x\in\R^n$. The {\em orbit} (of the $W$ action) containing $x\in\R^n$ is the set $Wx=\{gx \mid g\in W\}$. Any two orbits are either coincident or disjoint. The {\em orbit space} of the $W$-action in $\R^n$ is the quotient space $\R^n/W$, whose points are the orbits.\\

The reflecting hyperplanes of $W$ divide the space $R^n$ in open subsets, called {\em chambers}. The reflecting hyperplanes in the border of a given chamber $C$ are called the {\em walls} of $C$ and their subsets belonging to the topological closure $\overline{C}$ of $C$ are called the {\em faces} of $C$. The closure of a chamber is a fundamental domain of the $W$-action in $\R^n$, and $W$ acts transitively on the chambers (\cite{bourbaki456}, Ch. V, \S 3.1, Lemma 2, and \S 3.3). This means that the closure $\overline{C}$ of a chamber $C$ contains exactly one point for each orbit, that there are ${\rm ord}(W)$ chambers, and that, given any two chambers
$C_1$ and $C_2$, there exists a unique element $g \in W$ that maps in a one to one manner $\overline{C_1}$ in $\overline{C_2}$. Given a chamber $C$, if $x\in C$, the orbit $Wx$ has exactly ${\rm ord}(W)$ points, one for each chamber, and is called a {\em regular} or {\em principal orbit}, while, if $x$ is a point in a face of $C$, the orbit $Wx$ contains less than ${\rm ord}(W)$ points, all lying in reflecting hyperplanes, and is called a {\em singular orbit}.\\

The $W$-invariant polynomial functions are constant on the orbits of $\R^n$, so it is well defined the map $\overline  p:\R^n\to\R^n:x\to (p_1(x),\ldots,p_n(x))$, called the {\em orbit map}. The basic invariant polynomials separate the orbits, that is, the points of $\R^n$ $p=\overline  p(x)$ and  $p'=\overline  p(x')$ are different if and only if the orbits $Wx$ and $Wx'$ are different: $p\neq p'\Leftrightarrow Wx\neq Wx'$. The orbit map $\overline  p$ then represents in a one to one manner the orbits as points of $\R^n$ and the orbit space $\R^n/W$ as a subset ${\cal S}=\overline  p(\R^n)$ of $\R^n$. Actually, ${\cal S}=\overline  p(\overline{C})$, where $\overline{C}$ is the closure of any chamber of $W$, and $\overline  p$ is a diffeomorphism between $\overline{C}$ and ${\cal S}$. This implies that ${\cal S}$ has the geometrical shape of a simplicial cone of $\R^n$, as that one of $\overline{C}$, with its vertex at the origin, because of the homogeneity and the positive degrees of the basic invariant polynomials. The interior of ${\cal S}$ contains the images through the orbit map of the principal orbits, while the border of ${\cal S}$ contains the images through the orbit map of the singular orbits. From a different point of view, the set ${\cal S}$ is a simple example of orbifold (\cite{Thurston1978}, Prop. 13.2.1).\\


With a given set of basic invariant polynomials $p_a(x)$, $a=1,\ldots,n$, it is possible to build an $n\times n$ matrix function, $\widehat P(p)$, with matrix elements that are polynomial functions of $p=(p_1,\ldots,p_n)\in\R^n$, that has the property that when the variables $p_a$ are substituted with the basic invariant polynomials $p_a(x)$, $\forall\,a=1,\ldots,n$, its matrix elements $\widehat P_{ab}(\overline  p(x))$ coincide with the invariant polynomials $P_{ab}(x)$, $\forall\,a,b=1,\ldots,n$, obtained by making the scalar products among the gradients of the basic invariant polynomials, that is: $\widehat P_{ab}(\overline  p(x))= P_{ab}(x)=\nabla p_a(x)\cdot \nabla p_b(x)$, $\forall\,a,b=1,\ldots,n$. Following Ref. \cite{as1983}, I call the matrix $\widehat P(p)$ with the name {\em \wPm}, but the matrix $\widehat P(p)$ has also been used with other notations and called ``convolution of invariants'' matrix \cite{arn1979}, ``complete displacement of invariants'' matrix \cite{giv1980} or it was used without giving it a specific name in Refs. \cite{YaSek1979,SYS1980,ps1985}.
The \wPm\ is strongly related with the set ${\cal S}$: ${\cal S}$ coincides with the subset of $\R^n$ in which the \wPm\ $\widehat P(p)$ is positive semi-definite \cite{ps1985,as1983}, the border of ${\cal S}$ is contained in the algebraic surface determined by the equation $\det(\widehat P(p))=0$, called {\em discriminant}, and at the interior of ${\cal S}$ one has $\det(\widehat P(p))>0$. In addition, $\det(\widehat P(p))$ satisfies a system of differential equations \cite{YaSek1979,Sar1989}, Eq. (\ref{boundaryeqndet}) below, called {\em boundary equation}, that depends on the matrix $\widehat P(p)$ and on an $n$-dimensional polynomial vector $\lambda(p)$. The first appearance of the \wPms\ has been in Ref. \cite{arn1976}, just for the
groups $A_n$, and their use for other finite reflection group has been first made in Refs. \cite{arn1979,Saito1993,SYS1980,YaSek1979,YaSek1981}. The $\lambda$-vectors first appeared in Ref. \cite{YaSek1979}.\\

Both the \wPm\ $\widehat P(p)$ and the $\lambda$-vector $\lambda(p)$ depend on the choice of the basic invariant polynomials $p_1(x),\ldots,p_n(x)$.
The explicit determination of the \wPm\ from its definition requires to calculate the matrix elements $P_{ab}(x)=\nabla p_a(x)\cdot \nabla p_b(x)$, and to express them as polynomials of the basic invariant polynomials. When the rank $n$ of the group is large, these calculations require an enormous effort, even with a computer, and are often impossible to take to completion in a reasonable time. The possibility to use generating formulas to determine the matrix elements of the \wPms\ are then particularly welcome. In Ref. \cite{arn1976}, Remark 3.9,  Arnold writes a generating formula for the \wPms\ corresponding to the groups of type $A_n$, saying that that formula was communicated to him by D. B. Fuks, and gives a hint for its proof (with a typographical error, a definition $\sigma_0=0$ that should be $\sigma_0=1$). The formula is correct but the proof of that formula, to my knowledge, has never been published. In Ref. \cite{giv1980}, Givental writes generating formulas for the \wPms\ corresponding to groups of type $B_n$ and $D_n$
. His results are correct but his proofs are only sketched and based on differential geometry and singularity theory techniques that are difficult to follow by a non-specialist. Without taking into account the generating formulas given by Arnold \cite{arn1976} and Givental \cite{giv1980}, I was looking for bases of invariant polynomials in which the \wPms\ look particularly simple and with many similarities when comparing the \wPms\ corresponding to groups of the same type and with rank differing by one unit. This allowed me do discover bases of invariant polynomials for which one can use generating formulas to calculate the matrix elements of the corresponding \wPms. These generating formulas are Eq. (\ref{PjkdiSn}), for the groups of type $S_n$,  Eq. (\ref{PjkdiAn}), for the groups of type $A_n$,  Eq. (\ref{PjkdiBn}), for the groups of type $B_n$, and  Eq. (\ref{PjkdiDn_divisa}), for the groups of type $D_n$. They are valid for all values of the ranks $n\in\mathbb{N}$. Unexpectedly, the bases I chose for the groups of type $B_n$ and $D_n$ are the same chosen by Givental while those I chose for the groups of type $A_n$ differ only for a scalar multiples from those chosen by Arnold. I give proofs of these generating formulas that are based only on elementary calculus, algebraic manipulations and induction principle, so the proofs I give for the groups $A_n$, $B_n$ and $D_n$ are very different from those suggested by Givental in \cite{giv1980}, although somewhat longer. 
These four generating formulas, with their algebraic proofs, are the main result of this article.
In correspondence with the chosen sets of basic invariant polynomials, generating formulas for the $\lambda$-vectors that appear in the boundary equation (\ref{boundaryeqndet}) are also determined, for all the groups of type $S_n$, $A_n$, $B_n$, $D_n$, and for all ranks $n\in \mathbb{N}$. These results are given in Eqs. (\ref{lambdaSnEq}), (\ref{lambdaAnEq}), (\ref{lambdaBnEq1})--(\ref{lambdaBnEq3}) and (\ref{lambdaDnEq}).
The proofs are based on Theorems  \ref{quattro} and \ref{cinque} below, that are new, and use sums over the positive roots of the reflection group $W$. Almost the same $\lambda$-vectors, for the groups of type $A_n$, $B_n$ and $D_n$, were obtained in Ref. \cite{YaSek1981} by Yano and Sekiguchi using the induction principle on the rank $n$ of the reflection group $W$, and the explicit expressions of the fundamental anti-invariant, product of the linear forms vanishing on the reflection hyperplanes of $W$. These different proofs for the same generating functions, both for the \wPms\ and for the $\lambda$-vectors, are another sign of the linking role that the finite reflection groups have among different branches of mathematics and physics.\\

To keep simpler the reading of the article, the generating formulas of the \wPms\ and of the $\lambda$-vectors for the groups of type $S_n$, $A_n$, $B_n$, $D_n$, are reported in Sections \ref{Sn}-\ref{Dn}, respectively, without proofs. The proofs, some of them quite long, are collected in a separate section, Section \ref{proofs}.\\

As a byproduct, the \wPms\ obtained from the generating formulas for the groups of type $A_n$, $B_n$, $D_n$, show a high level of symmetry and can be written, almost completely, as sums of block Hankel matrices. These results are reported in Section \ref{hankel} without proofs and their proofs are reported in Section \ref{proofs}.\\

Several authors determined explicit bases of invariant polynomials for the irreducible finite reflection groups, for example Coxeter~\cite{cox1951}, Ignatenko~\cite{Ignatenko1984}, Mehta~\cite{Mehta1988}, and many others that I do not cite. A few authors tried to select, among the infinitely many possibilities, some distinguished bases of invariant polynomials, by requiring some supplementary conditions to be satisfied. For example,
Saito, Yano and Sekiguchi~\cite{SYS1980} defined the {\em flat bases}, Sartori and Talamini~\cite{Sar-Tal1991} the {\em $a$-bases} and Flatto~\cite{Flatto1970} and Iwasaki \cite{Iwasaki1997} the {\em canonical bases}. Their results will be recalled at the end of next section.
The bases of invariant polynomials 
that are here suggested are mainly justified by the possibility to use generating functions to determine the corresponding \wPms\ and $\lambda$-vectors. It turns out, however, that the matrix elements of the \wPms\ obtained from the generating functions here reported (Eqs. (\ref{PjkdiSn}), (\ref{PjkdiAn}), (\ref{PjkdiBn}), and (\ref{PjkdiDn_divisa})) are polynomials (with only integer coefficients) with a small number of terms, generally much less than those obtained using other bases of invariant polynomials. At the end of next section I report the number of terms of the matrix elements of the \wPms\ of $B_8$ in 4 different bases of invariant polynomials, precisely, the basis in Eq. (\ref{basicpolynBn}), a flat basis, an $a$-basis and a canonical basis.\\

The generating formulas for the \wPms\ and the $\lambda$-vectors here reported for the groups of type $S_n$, $A_n$, $B_n$ and $D_n$, correspond to a particular choice of the basic invariant polynomials. However, one may easily obtain the \wPms\ and the $\lambda$-vectors corresponding to any other basis of invariant polynomials of these groups, using the transformation formulas between bases of invariant polynomials and the transformation formulas for the \wPms\ and the $\lambda$-vectors, given in Eqs. (\ref{Ptransf}) and (\ref{lambdatransf}). Some examples, reported in Section \ref{examples}, show how one can determine in this way the \wPms\ and the $\lambda$-vectors corresponding to a general flat basis, $a$-basis or canonical basis.\\

The following Section \ref{background} contains more or less known results, but also some new ones. The proofs of the statements reported in Section \ref{background} are generally omitted, because they can be found in the literature, except those of the new Theorems \ref{quattro} and \ref{cinque} and of a few other results, collected in Theorems \ref{uno}, \ref{tre} and \ref{trasflambda}, that will be widely used in the sequel. The proofs written in Section \ref{background} are generally short and simple.\\

Section \ref{exceptional} deals very shortly with the exceptional and the non-crystallographic reflection groups.\\

Orbit spaces and  \wPms\ have interesting applications in mathematics and physics, especially in the study of phase transitions and symmetry breaking~\cite{as1983,MicZhi2001,gufetal,sar-val2005}, in the study of singularities~\cite{arn1976,arn1979,giv1980,Saito1993}, and in constructive invariant theory~\cite{tal-sspcm05}. 
In the rest of this introduction, I recall some of these applications of orbit spaces and \wPms. This part is not necessary to follow the rest of the article.

\begin{itemize}
  \item Phase transitions and symmetry breakings.\\
The {\em isotropy subgroup}, or {\it stabilizer}, of a point $x\in\R^n$ is the subgroup $W_x$ of $W$ such that $W_x=\{g\in W\,\mid\, gx=x\}$. Points in a same orbit have conjugated isotropy subgroups. The {\em orbit type} of the orbit $Wx$ is the conjugacy class $[W_x]$ of subgroups of $W$. An {\em orbit type stratum} is a set of points $x\in\R^n$ that have conjugated isotropy subgroups. The number of orbit type strata is finite and coincides with the number of conjugacy classes of isotropy subgroups of $W$. The images of the orbit type strata through the orbit map form the orbit type strata of ${\cal S}$. An orbit type stratum of ${\cal S}$ of dimension $r$, with $r=0,1,\ldots,n$, is the union of a finite positive number of connected components of dimension $r$, called {\em primary strata}. If $r>0$, the boundary of an $r$-dimensional primary stratum is the union of primary strata of dimension lower than $r$. Each primary stratum of dimension $r$ is the image through the orbit map of the set of fixed points of an isotropy subgroup of $W$ of rank $n-r$. The decomposition of obit type strata in primary strata depends on the reflection group $W$ and is well known for all irreducible reflection groups (see for example Refs. \cite{Carter1972,bordemann} for Weyl groups of simple Lie algebras, and \cite{GeckPfeiffer2000}, Ch. 3, for all irreducible finite reflection groups, including the non-crystallographic ones).


If $W$ is a general compact linear group with a basis of $n$ algebraically independent homogeneous polynomials, the union of the primary strata of dimension $r$, $\forall\,r=0,1,\ldots,n$, coincides with the subset of $\R^n$ in which the \wPm\ $\widehat P(p)$ has rank $r$ and is positive semi-definite, if $r<n$, or positive definite, if $r=n$~\cite{as1983,ps1985}. These conditions on the rank and on the (semi-)positivity of the matrix $\widehat P(p)$, can be translated, using the minors of $\widehat P(p)$, in a (generally complicated and overdetermined) system of algebraic equations and inequalities on the variables $p_1,\ldots,p_n$ of $\R^n$. 
(However, in the case of an irreducible finite reflection group $W$, the easiest way to find the algebraic equations and inequalities defining a primary stratum $\sigma$ of ${\cal S}$ of dimension $r=0,\ldots,n-1$, is to consider the linear equations and inequalities defining the intersection of $n-r$ faces of a chamber $C$, that has $\sigma$ for image in ${\cal S}$ through the orbit map $\overline  p$. Using these linear equations and inequalities as linear equations and inequalities among parameters in the orbit map, one obtains a set of $n$ parametric equations and some inequalities, that depend on $r$ free parameters, and define the primary stratum $\sigma$). Moreover, in any point of the discriminant surface, the rows of the \wPm\ are a set of $n$ vectors tangent to the discriminant surface. With some more complications, one can also determine, in a similar way, starting from the \wPm, also the stratification of the orbit space of a general compact linear group acting in $\R^n$, having a basis of $m>n$ algebraically dependent polynomials. In this case one has to take into account the syzygies, that is the polynomials equations in the variables $p_1,\ldots,p_m$, that are identically satisfied when these variables are substituted with the basic invariant polynomials $p_1(x),\ldots,p_m(x)$, $x\in\R^n$. The set ${\cal S}\subset\R^m$, representing the orbit space, is contained, in this case, in the intersection ${\cal Z}\subset\R^m$ of all the independent syzygy surfaces~\cite{ps1985}.

Let $G$ be a rank $n$ simple Lie group with Weyl group $W$. $W$ is a rank $n$ irreducible finite reflection group. The adjoint representation of $G$, indicated with Ad$(G)$, is a representation of $G$ in its Lie algebra $\mathfrak{g}$, that is a real vector space of dimension $\dim(\mathfrak{g})={\rm Card}({\cal R})+n$, where ${\cal R}$ is the set of roots of $W$. In Ref.~\cite{Chev1952} Chevalley proved that  it is always possible to obtain a basis of $W$-invariant polynomials from a basis of Ad$(G)$-invariant polynomials by taking the restrictions of the basic Ad$(G)$-invariant polynomials to the Cartan subalgebra $\mathfrak{h}$ of $\mathfrak{g}$ (see also \cite{bourbaki78}, Ch. VIII, \S 8, no. 3). The bases of homogeneous $W$-invariant polynomials and of homogeneous Ad$(G)$-invariant polynomials contain then the same number $n$ of elements with the same degrees, but they differ for the number of independent variables: the basic $W$-invariant polynomials depend on $n$ variables, while the basic Ad$(G)$-invariant polynomials depend on $\dim(\mathfrak{g})$ variables. One proves that the \wPm\ corresponding to the $W$-invariant basis obtained by taking the restrictions to the Cartan subalgebra of the basic polynomials in an Ad$(G)$-invariant basis is identical to the \wPm\ corresponding to the Ad$(G)$-invariant basis (\cite{sar-tal1998}, Section III). This implies that $W$ and Ad$(G)$ have orbit spaces represented by the same set ${\cal S}$, stratified in orbit type strata and primary strata in exactly the same way. The isotropy subgroups corresponding to the different primary strata are, in the two cases, subgroups of $W$ or of Ad$(G)$, and the ones are the Weyl groups of the others. The \wPms\ here obtained for the irreducible finite reflection groups of type $A_n$, $B_n$ and $D_n$ can then be used also for the adjoint representations of the compact simple Lie groups of type $A_n$, $B_n$, $C_n$ and $D_n$ (the basic invariant polynomials of $B_n$ and $C_n$ are the same, and so also their orbit spaces).


The $W$-invariant $C^\infty$-functions $f(x)$ can be studied as functions $\widehat f(p)$ defined in ${\cal S}$, if $f(x)=\widehat f(\overline  p(x))$, $\forall\,x\in\mbox{Dom}(f)\subseteq\R^n$, however, $\widehat f(p)$ is, in general, defined also outside ${\cal S}$. The study of extrema of invariant functions can also be performed in the set ${\cal S}$ as a study of constrained extrema in the primary strata of the set ${\cal S}$.
If $V(x)$ is a $W$-invariant potential function, and $x_0\in\R^n$ is a point in which $V(x)$ takes its minimum value, then $\widehat V(p)$ is minimum in the point $\overline  p(x_0)$, because $V(x)=\widehat V(\overline  p(x))$. If $[W_0]$ is the orbit type of the orbit through $x_0$, then $W_0$ is the {\em residual symmetry group} of the minimum point $x_0$, and the orbit type $[W_0]$ is the  {\em residual symmetry type} of the orbit type stratum hosting the minimum point of $V(x)$. If $V(x)$ depends on some external physical parameters (like pressure, temperature, external magnetic or electric field intensity, etc., in the case of thermodynamic potentials, or coupling constants, mixing angles, etc., in the case of Higgs potentials), when these external physical parameters change with continuity, the minimum point changes its position with continuity. If the new position $x_0'$ lies in the same primary stratum of $x_0$, the residual symmetry group is at most changed by conjugation, the same that happens if one moves $x_0$ along its orbit, so minimum points lying in the same primary stratum represent the same residual symmetry type of the physical theory. If instead the new position $x_0'$ lies in another primary stratum, the two primary strata containing $x_0$ and $x_0'$, respectively, have different dimensions and must necessarily be bordering. Then the residual symmetry type $[W_0]$ is changed to $[W_0']$, with the groups in $[W_0]$ proper subgroups of groups in $[W_0']$, (or vice versa, with $W_0$ and $W_0'$ exchanged). In this situation one has a {\em phase transition} (if $V$ is a thermodynamic potential), or a {\em (spontaneous) symmetry breaking} (if $V$ is a Higgs potential). (When the minimum point changes with continuity it cannot shift from a primary stratum of type $[W_0]$ to another primary stratum of type $[W_0]$ without passing through an orbit type stratum of different type $[W_0']$, this is the reason to speak only about primary strata and not about orbit type strata in the previous sentences). The set of orbit type strata then represent the set of different residual symmetry types possible for the physical system, and the bordering relations between primary strata represent possible phase transitions allowed for the physical system. It is easy to build graphs in which the nodes represent the possible primary strata (or orbit type strata) and the lines connecting the nodes the possible phase transitions. These graphs represent in a clear way all the possible phase transitions allowed to a given physical system.


  \item Singularity Theory.\\
One of first goals of Singularity Theory was to study the degenerate critical points of smooth functions of one or several variables. It is better and simpler to work in complex spaces with complex holomorphic functions. A {\em critical} (or {\em singular}) {\em point} is a point $z_0\in\mathbb{C}^k$ in which the gradient $\left.\nabla f(z)\right|_{z=z_0}$ of a function $f:\mathbb{C}^k\to\mathbb{C}$ vanishes and a {\em degenerate critical point} is a critical point $z_0$ in which also the Hessian $\det\|\left.\partial^2 f(z)/\partial z_i\partial z_j \right|_{z=z_0}\|$ vanishes. 
A critical point $z_0\in\mathbb{C}^k$ is {\em isolated} if there is a sufficiently small neighbourhood $U\subset\mathbb{C}^k$ such that $z_0$ in the only critical point in $U$. One can suppose that $z_0=O$, and that $f(O)=0$, because one can always perform a translation of the origin of the system of coordinates of $\mathbb{C}^k$, and add a constant to $f(z)$, without changing the behaviour of the given function at the given critical point.
Near the origin $O\in\mathbb{C}^k$, all germs of holomorphic functions, having $O$ as a critical point and such that $f(O)=0$, can be divided in equivalence classes, two germs being in the same class if there is a biholomorphic coordinate change in $\mathbb{C}^k$, sending $O$ in itself, that takes one germ into the other one. For each equivalence class one picks a polynomial representative of the germs in the class, that is called their {\em normal form}. This is always possible, if the critical point is isolated. Let $f(z)$ be a normal form having a critical point in $O$ in which $f(O)=0$. If $O$ is non-degenerate, $f(z)$ is a quadratic form, while, if $0$ is degenerate, the normal form $f(z)$ has some terms with degrees greater than two. Two normal forms, that have $O$ as a degenerate critical point, and differ only for the number of quadratic terms exhibit in $O$ the same type of singularity. 
One can then study and classify the singularities by considering only normal forms without constant terms, with the smallest possible number of quadratic terms, and differing only for the terms of degrees higher than two.

Families of functions depending on parameters necessarily exhibit degenerate critical points. One of such a families is a {\em deformation} (or {\em universal unfolding}) $F(z,\lambda)$ of the normal form $f(z)$, with $\lambda\in\mathbb{C}^j$ a set of parameters controlling the deformation, such that $F(z,0)=f(z)$. 
The parameter space $\mathbb{C}^j$ is called the {\em base space} of the deformation. If $O$ is an isolated degenerate critical point for the normal form $f(z)$, with a generic deformation of $f(z)$, the degeneracy in $O$ disappears and bifurcates into a finite number $\mu(f)$ of non-degenerate critical points. The positive integer $\mu(f)$ is called the {\em multiplicity}, or {\em Milnor number}, of the singularity $O$ of $f(z)$. The Milnor number $\mu(f)$ can be calculated algebraically. Let ${\cal O}_k$ be the ring of holomorphic function germs in $O$, $g:\mathbb{C}^k\to\mathbb{C}$, and ${\cal I}_f\subset {\cal O}_k$ the ideal generated by the $k$ partial derivatives of $f(z)$. The quotient ${\cal A}_f={\cal O}_k/{\cal I}_f$ is called the local algebra of $f$ in $O$. Its dimension, as a complex vector space is equal to the Milnor number of $f(z)$ in $O$: $\mu(f)=\dim_{\,\mathbb{C}} {\cal A}_f$. One can take $n=\mu(f)$ monomials $\phi_1(z),\ldots,\phi_n(z)$, as a vector space basis of ${\cal A}_f$ (one of the basic monomials is always equal to 1), and write the deformation $F(z,\lambda)$, in the following way: $F(z,\lambda)=f(z)+\sum_{a=1}^n \lambda_a \phi_a(z)$. One obtains a germ deformation dependent on the least possible number of parameters, such that all other germ deformations can be reduced to it after a suitable biholomorphic coordinate change. This is a so-called {\em miniversal} deformation of $f(z)$. For a miniversal deformation of $f(z)$ one then has $j=n=\mu(f)$, and $\lambda\in\mathbb{C}^n$. One can suppose $f(z)$ a weighted homogeneous polynomials, this implies that one has to assign a weight $w_i$ to each of the coordinates $z_i$, $i=1,\ldots,k$, of $z\in \mathbb{C}^k$, in such a way that all monomials in $f(z)$ have the same weight $w$. One has to take for $w$ and $w_i$ the smallest positive integers for which $f(z)$ is weighted homogeneous. A miniversal deformation $F(z,\lambda)$ of $f(z)$ must then also be a weighted homogeneous polynomial of weight $w$, and this leads to assign weights $d_a$ also to the deformation parameters $\lambda_a$, $\forall\,a=1,\ldots,n$.

Let $f(z)=F(z,0)$ have a degenerate critical point in $O$ with critical value $f(O)=0$ and Milnor number $n$.
One is interested to determine all different singularities that can be realized with a deformation of the normal form $f(z)$.
The set of parameters $\lambda\in\mathbb{C}^n$ in the base space for which $F(z,\lambda)$ has zero as a critical value, that is, for which there exists a point $z_\lambda\in\mathbb{C}^k$ such that $F(z_\lambda,\lambda)=0$ and $\left.\nabla_z F(z,\lambda)\right|_{z=z_\lambda}=O$,
forms a hypersurface $\Sigma_f\subset\mathbb{C}^n$, called {\em level bifurcation set} of $f(z)$ (or {\em bifurcation set of zeroes} of $f(z)$). This set can be found from the conditions for which the equation $F(z,\lambda)=0$ has multiple $z$ roots, at least one, and this happens if and only if the discriminant $\Delta_f$ of the polynomial $F(z,\lambda)$ is vanishing (one here considers $z_1,\ldots,z_k$ as variables and $\lambda_1,\ldots,\lambda_n$ as coefficients. The discriminants of polynomials in one or several variables are well described, for example, in Chapters 12--13 of \cite{GKZ94}). The polynomial $\Delta_f$ depends only on the coefficients $\lambda_a$, $a=1,\ldots, n$, and is a weighted homogeneous polynomial, if $F(z,\lambda)$ is
. The set of points $\lambda\in\mathbb{C}^n$, satisfying the equation $\Delta_f=0$, forms an algebraic hypersurface of $\mathbb{C}^n$ that coincides with the bifurcation set of zeroes $\Sigma_f$. For this reason $\Sigma_f$ is also called the {\em discriminant set}. We have then that if $\lambda\in\Sigma_f$, the equation in $z$, $F(z,\lambda)=0$ has multiple roots, and if $\lambda\notin\Sigma_f$, the equation in $z$, $F(z,\lambda)=0$ has all different roots.

In \cite{arn1972}, Arnol'd discovers an unexpected link between the isolated singularities and the irreducible finite reflection groups of types $A_n$, $D_n$, $E_6$, $E_7$, and $E_8$ (these $ADE$ groups are all the crystallographic finite reflection groups with only simple edges in their Coxeter-Dynkin diagrams). To understand this link, one has first to consider the action of a finite irreducible reflection group $W$ on $\mathbb{C}^n$. This action can be obtained with the same real orthogonal matrices of the action of $W$ in $\R^n$, and this implies that the real and the imaginary part of a point $z\in\mathbb{C}^n$ transform without mixing, because, $\forall\,g\in W$ and $\forall\, z\in \mathbb{C}^n$, one has $gz=g(x+iy)=gx+igy$, with $x,y\in\R^n$ and $i=\sqrt{-1}$. The complex reflection hyperplanes have the same equations of the real reflection hyperplanes, but the variables involved are now complex instead of real. The basic invariant polynomials of this actions are the same as those of the action of $W$ in $\R^n$, so, there is a unique corresponding \wPm\ $\widehat P(p)$ for both the real and the complex actions of $W$. In particular, the coefficients of the matrix elements of $\widehat P(p)$ are real numbers, and $\det(\widehat P(p))$ is a polynomial with real coefficients. In contrast, the independent variables on which $\widehat P(p)$ depends are now complex numbers, because $p\in\mathbb{C}^n$. 
Some of the orbits in $\mathbb{C}^n$ are formed by only real points, and these have image in ${\cal S}\subset\mathbb{C}^n$, where ${\cal S}$ is now a subset of the set of points of $\mathbb{C}^n$ with vanishing imaginary part. Orbits formed by only complex numbers have images, through the orbit map, both in points with vanishing imaginary part (lying outside ${\cal S}$ in the real subspace of $\mathbb{C}^n$), and in points with not vanishing imaginary part. In any case, the orbit space of the action of $W$ in $\mathbb{C}^n$ is the whole $\mathbb{C}^n$: $\overline  p(\mathbb{C}^n)=\mathbb{C}^n$.
A point $z\in \mathbb{C}^n$ belonging to a complex reflecting hyperplane have image, through the orbit map, in a point $p=\overline  p(z)\in \mathbb{C}^n$ in which $\det(\widehat P(p))=0$. This because the polynomial $\det(\widehat P(p))$ can be written as a product of squares of linear forms whose vanishing determine the reflection hyperplanes (Eq. (\ref{detP(x)}) below, in which $x\in\R^n$ has to be substituted by $z\in \mathbb{C}^n$). Then, when $z$ belongs to a complex reflecting hyperplane $\det(\widehat P(\overline  p(z)))=0$. The set of the complex reflecting hyperplanes is then mapped to an algebraic surface $\Sigma$ of $\mathbb{C}^n$ with equation $\det(\widehat P(p))=0$ (with $p\in\mathbb{C}^n$). This surface is no more a boundary surface of $\mathbb{C}^n$, but just a set of points of $\mathbb{C}^n$ that contains the images through the orbit map of the singular orbits passing through the reflecting hyperplanes.

For all the simplest types of normal forms $f(z)$, for which $O$ is an isolated singularity with finite Milnor number $n$, and $f(O)=0$, Arnold  \cite{arn1972} found that there is one and only one rank $n$ irreducible finite reflection group $W$, of one of the types $A_n$, $D_n$, $E_6$, $E_7$, and $E_8$, such that there exist a biholomorphic map between the orbit space $\mathbb{C}^n$ of the action of $W$ in $\mathbb{C}^n$ and the base space $\mathbb{C}^n$ of the miniversal deformation $F(z,\lambda)$ of $f(z)$, keeping the origin $O\in\mathbb{C}^n$ fixed, and such that the surface $\Sigma\subset\mathbb{C}^n$ (containing the image through the orbit map of the reflection hyperplanes of the $W$-action in $\mathbb{C}^n$) is mapped to the surface $\Sigma_f\subset\mathbb{C}^n$ (forming the discriminant set of the singularity). For this reason the image of the reflection hyperplanes in the complex orbit space is called {\em discriminant}. One often can choose, without much effort, the basic invariant polynomials of $W$ in such a way to obtain $\Sigma\equiv\Sigma_f$. The equation of $\Sigma_f$ can be obtained in this case using the \wPm:  $\Sigma_f=\{p\in\mathbb{C}^n\;|\;\det(\widehat P(p))=0\}$. The weights $d_a$ that were assigned quite arbitrarily to the parameters $\lambda_a$ of the deformation are then equal to the degrees of the basic invariant polynomials of the group $W$ associated to the singularity: $d_a=w(\lambda_a)=\deg(p_a(z))$, $\forall\,a=1,\ldots,n$. As a singularity determines a finite reflection group, a finite reflection group determines a singularity. The simplest isolated singularities can then be classified using the same symbols of the classification of the irreducible finite reflection groups. There exist then singularities of types $A_n$, $D_n$, $E_6$, $E_7$, and $E_8$. 
The existence of a biholomorphic map between the orbit space of the $W$-action in $\mathbb{C}^n$ and the base space of a miniversal deformation of $f(z)$, such that the origin is mapped to the origin, and $\Sigma\subset\mathbb{C}^n$ (the $\overline  p$-image of the complex reflection hyperplanes of $W$) is mapped to the discriminant set of the singularity $\Sigma_f\subset\mathbb{C}^n$, was actually proved in general by Slodowy \cite{slo1980}, using previous results by Brieskorn \cite{bries1971}.

Other singularities can be associated in a similar manner to the crystallographic finite reflection groups with a non-simple edge in their Coxeter-Dynkin diagrams  \cite{arn1978}. The groups of type $B_n$, $C_n$ and $F_4$ correspond to singularities in $O$ of normal forms invariant for the reflection about a hyperplane containing $O$, that is exhibiting a $\mathbb{Z}_2$ symmetry (this reflecting hyperplane is called {\em boundary} and these singularities are thus called {\em boundary singularities}), while the group $G_2$ correspond to a singularity exhibiting a $\mathbb{Z}_3$ symmetry. In all these cases there exists a biholomorphic map between the orbit space of the $W$-action in $\mathbb{C}^n$ and the base space of a miniversal deformation of the normal form of the singularity $f(z)$, such that $\Sigma\subset\mathbb{C}^n$ (the $\overline  p$-image of the complex reflection hyperplanes of $W$) is mapped to the discriminant set $\Sigma_f\subset\mathbb{C}^n$, and the boundary and the origin are mapped to themselves. The non-crystallographic groups $H_2=I_2(5)$, $H_3$ and $H_4$ are related to singularities that are formed by the time evolution of wave fronts avoiding obstacles (see Ref. \cite{shcherbak1988}). In this case there exist a biholomorphic map between the $\overline  p$-image $\Sigma$ of the complex hyperplanes of the non-crystallographic reflection group and a subset (the closure of a primary stratum) of the discriminant set $\Sigma_f$ of the singularity (see Theorem 4 of Ref. \cite{shcherbak1988}).

The relations between singularities and finite reflection groups are much deeper than how is here written. For example, one can prove that the finite reflection group that is associated to an isolated singularity is the {\em monodromy group} of the singularity that has its simple roots forming a basis of the {\em vanishing cycles}. There are many books and articles treating in many depth levels singularity theory, so I stop here my short review (for details see for example \cite{AGV1,AGV2,Zoladek2006,GLS2007}).

In Refs. \cite{arn1976,arn1979} Arnol'd used the \wPms\ to obtain some other results in singularity theory. In particular, in \cite{arn1979}, he used a linearized form of the \wPms, called {\em linearized convolution} or {\em linearized displacement} matrix, in the cotangent space of the orbit space $\mathbb{C}^n/W$ at the origin, to determine certain integral invariants, called {\em indices} of the singularities. This kind of calculations have been also done by Givental \cite{giv1980}, who first determined explicit formulas for the \wPms\ of the groups $B_n$, $D_n$, $F_4$ and $E_6$.
Saito \cite{Saito1993} proved that on the base space of the universal unfolding of an isolated hypersurface singularity
it is possible to define a flat structure, obtained through the period map associated to a primitive form. This flat structure implies the existence on the base space of the universal unfolding 
of a system of coordinates with a flat metric, i.e. with an everywhere vanishing curvature. The coordinates of the base space
are consequently said flat coordinates. The original references are Ref. 27 of Ref.~\cite{Saito1993} (unpublished) and Ref.~\cite{Saito1983} (all reviewed in Ref.~\cite{oda1987}). Saito, Yano and Sekiguchi proved~\cite{SYS1980,Saito1993}, just using standard properties of finite reflection groups, that in the orbit space $\mathbb{C}^n/W\simeq\mathbb{C}^n$ of the effective action of an irreducible finite reflection group $W$ on $\mathbb{C}^n$, it is possible to choose a system of coordinates (that is, a basis of $W$-invariant polynomials that determines a system of coordinates in the orbit space, obtained by identifying the basic polynomials with the coordinates) with a real and constant flat metric. It is said consequently that the orbit space acquires a flat structure and that the corresponding basic invariant polynomials form a flat basis. To obtain a flat basis of invariant polynomials in Ref. \cite{SYS1980}, Saito, Yano and Sekiguchi used the \wPms. Even if they did not know the complete form of the \wPms, they succeeded to determine the flat bases of all irreducible finite reflection groups, except $E_7$ and $E_8$, just studying the terms of the \wPms\ containing the highest degree basic invariant polynomial. The flat bases of the groups $E_7$ and $E_8$, together with their \wPms, were found in Ref. \cite{tal-jmp2010} using the method suggested in Ref. \cite{SYS1980}.
In 1991 Dubrovin, trying to give a coordinate independent
formulation of the Witten--Dijkstra--Verlinde--Verlinde
associativity equations that appeared those years in the setting
up of the two dimensional topological field theory, was lead to
define a geometric object that he called Frobenius manifold. He
also proved that the (complexified) orbit space of an irreducible finite
reflection group, endowed with the flat structure introduced by
Saito, satisfies all the axioms of a Frobenius
manifold~\cite{Dub1996,Dub1999} (in effect this is the most
elementary example of Frobenius manifold).
Recent articles describing the set up of the flat structure and
the Frobenius manifold structure on the (complexified) orbit space
of a finite reflection group and on the base space of the
universal unfolding of an isolated hypersurface singularity are
Refs.~\cite{Saito2004} and \cite{SaitoTak2008}.\\

\item Constructive invariant theory.\\
The \wPms\ can be defined for a general real compact matrix group $G$, not necessarily a finite group. All such groups admit a finite basis of real $G$-invariant homogeneous polynomials. The basic invariant polynomials are in general not algebraic independent and sometimes it may be difficult to determine all the polynomials in a basis. The construction of the \wPm\ can help to overcome this difficulty. The degrees of some of the basic invariant polynomials, as well as the degrees of some of the syzygies among the basic invariant polynomials can be inferred from the Molien function of $G$, and this is the first calculation to do. Then, one has to determine all the basic invariant polynomials of the lowest degrees, for example by calculating the averages in the group $G$ of some proper monomials. Let these invariant polynomials be $p_1(x),\ldots,p_k(x)$, with the labeling increasing according to their degrees. These are supposed to be only some, and not all, of the basic invariant polynomials of $G$. Then one starts to determine the \wPm\ elements with these basic polynomials, from the lowest degree element to the highest degree element, using the definition, that is by expressing the invariant scalar products $\nabla p_a(x)\cdot \nabla p_b(x)$ as polynomials of the known basic invariant polynomials $p_1(x),\ldots,p_k(x)$. If one is not able do do this, then one sets $p_{k+1}(x)=\nabla p_a(x)\cdot \nabla p_b(x)$, and one discovers so a new basic invariant polynomial (and if $\deg(p_k(x))>\deg(p_{k+1}(x))$ one has to relabel the basic invariant polynomials). When one finishes the construction of the \wPm\ and the degrees suggested by the Molien function are in accord with those of the basic polynomials so found, one has determined the full basis of invariant polynomials  of $G$. There are shortcuts and particular cases. The interested reader can find a wider exposition and some examples in Ref. \cite{tal-sspcm05}.\\
\end{itemize}

\section{Background materials and preliminary results\label{background}}

In this review section I shall report mostly known results, but also some new ones, like Theorems \ref{quattro} and \ref{cinque}. The arguments here reviewed concern the theory of finite reflection groups, the theory of linear actions of  compact groups and of the use of invariant theory to describe their orbit spaces. The proofs are generally omitted, except for some theorems that are widely used in the sequel. For a deeper understanding of arguments related to linear actions of a general compact group or to the use of invariant theory in this context one may read Refs.~\cite{as1983,bredon,schw-ihes}, while among the many good books on the theory of finite reflection groups and their relations with Lie groups I mainly used Refs.~\cite{bourbaki456,Humphreys1990}. There one can find the proofs or the references to many of the results that are here just stated.\\

I consider here group actions in $\R^n$, in which it is given a set of $n$ orthonormal canonical basis vectors $e_1,\ldots,e_n$. A {\em reflection} $R$ of $\R^n$ is a linear transformation of $\R^n$ that leaves pointwise fixed a
$(n-1)$-dimensional hyperplane $\pi_R$ of $\R^n$, called the
{\em reflecting hyperplane} of $R$, and sends each $x\in
\R^n$ to its symmetric $x'\in\R^n$ with respect to $\pi_R$. This means that $\frac{x+x'}{2}\in\pi_R$, and that $x-x'\perp \pi_R$. The line passing through the origin of $\R^n$ orthogonal to $\pi_R$ is mapped into itself by the reflection $R$ but is reflected about the origin. It is called the {\em reflecting line} of $R$.

Given any non-null vector $\alpha\in\R^n$, the reflection $R_\alpha$ that turns $\alpha$ into its opposite, can be defined by the following formula:
\begin{equation}
 R_\alpha x=x-\frac{2\,\alpha\cdot x}{\alpha\cdot\alpha}\,\alpha  \qquad \forall
 x\in\R^n\,,\label{reflection0}
\end{equation}
where $\cdot$ indicates the canonical scalar product in $\R^n$. Any non null vector proportional to $\alpha$ would define the same reflection. The reflection hyperplane of $R_\alpha$ has equation $\alpha\cdot x=0$, unique if one excludes possible multiplicative factors. We will indicate with $$l_\alpha(x)=\alpha\cdot x$$
the linear form that vanishes in the reflection hyperplane of $\alpha$. We will identify the reflection $R_\alpha$, with the orthogonal matrix that realizes the reflection $R_\alpha$ in $\R^n$.

By multiplying a set of reflections in all possible ways one obtains a group of orthogonal matrices, called a {\em reflection group}. In general this group is infinite. The irreducible finite reflection groups require particular angles between the reflection lines and were classified by Coxeter~\cite{cox1934} in the following types:
$A_n$, $n\geq 1$, $B_n$, $n\geq 2$, $D_n$, $n\geq 4$, $E_6$, $E_7$, $E_8$, $F_4$, $H_3$, $H_4$, $I_2(m)$, $m\geq 5$.
The number appearing as a subscript is the rank $n$ of the group and is equal to the dimension of the space relative to which the group action has no fixed points besides the origin.  The limitations exclude group isomorphisms (and for this reason the groups of type $C_n$ are not in the list because, as reflection groups, they coincide with the groups $B_n$, $\forall\, n\geq 3$, and the group $G_2$ is not in the list because of the isomorphism $I_2(6)\simeq G_2$). 
A reflection has determinant equal to $-1$, and hence it is an improper rotation. The product of two reflections has determinant equal to $+1$, and hence it is a proper rotation. 
A finite reflection group has an equal number of proper and of improper rotations. Only a few number of the improper rotations are reflections, the others do not fix pointwise an hyperplane.

Let $n$ be a positive number and $W\subset O(n)$ a rank $n$ irreducible finite reflection group. If we choose a vector $r\in\R^n$ parallel to a reflecting line, the group $W$ maps this vector into a finite set $\Phi_r=\{gr,\ \forall\,g\in W\}$ of vectors of $\R^n$, all parallel to reflecting lines, and all of the same length, because $W$ is orthogonal. For all vector $v\in\Phi_r$, also $-v\in\Phi_r$, because $R_v\in W$. The set $\Phi_r$ is stable with respect to $W$ transformations. The vectors in $\Phi_r$ are called {\em roots} and $\Phi_r$ forms a {\em root system} (according to the definition of root system given in \cite{Humphreys1990}, Section 1.2).
With certain reflection groups $W$ ($B_n$, $\forall\,n\geq 2$, $F_4$, $I_2(m)$, $\forall\,m\geq 6$, $m$ even), there are reflecting lines that are not parallel to any root in $\Phi_r$. Then it is necessary to define a second vector $r'$ parallel to one of these reflecting lines left over and repeat the construction of a new root system $\Phi_{r'}$. The root systems $\Phi_r$ and $\Phi_{r'}$ are disjoint, and at most two root systems are necessary to cover in this way all reflecting lines of $W$. The lengths of the roots $r$ and $r'$ at this point are arbitrary. One can determine, not in a unique way, a set of $n$ linearly independent roots, called {\em simple roots}, such that all roots are linear combinations of the simple roots with real coefficients, all of the same sign. Those with positive coefficients are called {\em positive roots}. Let's use the symbols ${\cal R}$ and ${\cal R}_+$ for the set of roots and the set of positive roots of $W$. Clearly, ${\cal R}=\{ \pm r\,,\  \forall\, r \in {\cal R}_+ \}$. There is a one to one correspondence between the set of reflections of $W$ and the set ${\cal R}_+$, of positive roots of $W$. The reflections obtained with Eq. (\ref{reflection0}) from the simple roots are called {\em simple reflections} and are sufficient to generate the whole group $W$. The $\Z$-span of the simple roots is called the {\em root lattice}. A finite reflection group $W$ for which there exist a root lattice that is stable with respect to $W$ transformations is called {\em crystallographic}. A finite reflection group with two different root systems that is crystallographic has necessarily the roots in the two root systems with lengths in the ratio $\sqrt{2}$ or $\sqrt{3}$. When the ratio is $\sqrt{2}$, and $n\geq 3$, there exist two different $W$-stable root lattices, that correspond to two different crystallographic reflection groups: $B_n$ and $C_n$. There is a one to one correspondence between the crystallographic irreducible finite reflection groups ($A_n$, $n\geq 1$, $B_n$, $n\geq 2$, $C_n$, $n\geq 3$, $D_n$, $n\geq 4$, $E_6$, $E_7$, $E_8$, $F_4$, $G_2\simeq I_2(6)$) and the Weyl groups of the simple Lie algebras. The non crystallographic finite reflection groups are $H_3$, $H_4$ and  $I_2(m)$, $m\geq 5$, $m\neq 6$. The relative length of the roots of the two root systems of the groups $I_2(m)$, $m\geq 5$, $m\neq 6$ is not fixed. If $W$ is crystallographic, all roots belong to the root lattice, and this implies that they are linear combinations of the simple roots with integer coefficients.\\

Consider now the algebra of polynomials of the $n$ variables $x_1,\ldots,x_n$. The $W$-invariant polynomials are such that $p(gx)=p(x)$, $\forall\,g\in W$, $\forall\,x\in\R^n$. There exist $n$ $W$-invariant real homogeneous polynomials $p_1(x),\ldots, p_n(x),$ such that for every invariant polynomial $p(x)$, there exists a unique polynomial $\widehat p(p)$ in $n$ indeterminates $p=(p_1,\ldots,p_n)\in\R^n$, such that $p(x)=\widehat p(p_1(x),\ldots, p_n(x)),\;\forall x\in \R^n$. The polynomials $p_1(x),\ldots, p_n(x)$ are called {\em basic invariant polynomials}. There are infinitely many possible choices of a set of basic invariant polynomials. For the algebraic independence of the basic invariant polynomials, an equation like $\widehat p(p_1(x),\ldots, p_n(x))=0$ cannot be identically satisfied $\forall\,x\in\R^n$.

The degrees $d_1,\ldots,d_n$, of the basic invariant polynomials of all the irreducible finite reflection groups, were determined by Coxeter~\cite{cox1951}, and for a given irreducible finite reflection group they are all different, except in the case of the groups $D_n$, with even $n$, in which there are two basic invariant polynomials of degree $n$. Usually, the basic invariant polynomials are labeled according to their degrees, for example by taking
\begin{equation}\label{labeling}
  d_a\leq d_{a+1}\,,\qquad \forall\,a=1,\ldots,n-1\,.
\end{equation}
If the group has no fixed points besides the origin, $d_1=2$, and one may define
\begin{equation}\label{quadraticinv}
  p_1(x)=x\cdot x=\|x\|^2=\sum_{i=1}^n {x_i}^2\,,
\end{equation}
a natural choice for real orthogonal actions.

Coxeter proved \cite{cox1951}, that if the finite reflection group $W$ is irreducible, the sum and the product of the degrees $d_a$, $a=1,\ldots,n$, are related to the number $N$ of reflections of $W$ and to the order, ${\rm ord}(W)$, of $W$, by the following formulas:
\begin{equation}\label{sommagradi}
\sum_{a=1}^n d_a=N+n\,,\qquad N={\rm card({\cal R}_+)}\,,
\end{equation}
\begin{equation}\label{prodottogradi}
    \prod_{a=1}^n d_a={\rm ord}(W)\,.
\end{equation}
Moreover, for $W$ irreducible, given the usual degree
labeling (\ref{labeling}) of the basic invariant polynomials, one has the identities: $d_a+d_{n-a+1}=d_n+2,\ \forall
a=1,\ldots,\left[\frac{n+1}{2}\right]$, and, using Eq. (\ref{sommagradi}), these imply $N=n\, d_n/2$. ($d_n$ is the greatest degree of the basic invariant polynomials).\\

With a basis $\{p_1(x),\ldots,p_n(x)\}$ of invariant polynomials, one may
calculate an $n \times n$ real symmetric matrix $P(x)$ which has
its elements $P_{ab}(x)$ that are the scalar products of the
gradients of the basic invariants~\cite{arn1976,SYS1980}:
\begin{equation}\label{matriceP(x)}
  P_{ab}(x) = \nabla p_a(x) \cdot \nabla p_b(x)=\sum_{i=1}^n
\frac{\partial p_a(x)}{\partial x_i}\;\frac{\partial
p_b(x)}{\partial x_i}\qquad \forall\ a,b=1,\ldots,n\,.
\end{equation}
The matrix elements $P_{ab}(x)$ are homogeneous polynomials of degrees $d_a+d_b-2$. By convention I shall use the Latin indices $a,b,c,\ldots$, to label the basic invariant polynomials, and the Latin indices $i,j,k,\ldots$, to label the coordinates of $\R^n$ in their domain.
From the homogeneity of the basic invariant polynomials and the standard form (\ref{quadraticinv}) of $p_1(x)$, one obtains, using Euler's theorem on homogeneous polynomials,
that the first row and column of $P(x)$ have fixed form:
$P_{1a}(x)=P_{a1}(x)=2d_a p_a(x)$, $\forall\, a=1,\ldots,n$, where no sum over $a$ is understood.

The matrix $P(x)$ can also be written in the following way:
\begin{equation}\label{P(x)=j^Tj}
  P(x)=j(x)\, j^\top(x),
\end{equation}
in which the exponent $\top$
means transposition and $j(x)$ is the jacobian matrix  $$j_{ai}(x)=\frac{\partial p_a(x)}{\partial x_i}\,,\qquad
\forall\, a=1,\ldots,n,\qquad \forall\, i=1,\ldots,n\,.$$

Then, $P(x)$ is a positive semi-definite matrix, in fact $x^\top P(x) x=\|j(x) x\|^2\geq 0$, $\forall\, x\in \R^n$. Consequently, all principal minors of $P(x)$ are non-negative, in any point $x\in\R^n$.

A classic result by Coxeter~\cite{cox1951}, Theorem 6.2,
claims that $\det(j(x))$ is proportional to the (anti-invariant) product of the linear forms
$l_r(x)$ whose vanishing determine the set of the reflecting
hyperplanes, that is:
\begin{equation}\label{detj(x)}
  \det(j(x))=c\, \prod_{r\in {\cal R}_+}\, l_r(x)\, ,
\end{equation}
in which ${\cal R}_+$ is the set of positive roots of $W$, $l_r(x)=r\cdot x$, and $c$ is a non-zero
constant. Then, one has:
\begin{equation}\label{detP(x)}
\det(P(x))=c^2\, \prod_{r\in {\cal R}_+}\, l_r^2(x)\,,
\end{equation}
that implies $\det(P(x))=0$ if $x$
belongs to the set of the reflecting hyperplanes and $\det(P(x))>0$
otherwise. The definition of $l_r(x)$ as it is given above depends on the normalization of $r$, and the constant $c$ then depends on the normalization of both the roots and the basic invariant polynomials.

The matrix elements $P_{ab}(x)$
are $W$-invariant polynomials. This follows from the
orthogonality of $W$ and the covariance of the gradients of $W$-invariant functions.
In fact, if $p(x)$ is an invariant polynomial,  $p(x)=p(x^g)$, $\forall\,g\in W$, $\forall\,x\in\R^n$, where $x^g=gx$ is the $g$-transformed point of $x$, and one has, $\forall\,g\in W$ and $\forall\, x\in\R^n$:
$$\frac{\partial p(x)}{\partial{x_i}}=\frac{\partial p(x^g)}{\partial{x_i}}=
\sum_{j=1}^n
\frac{\partial p(x^g)}{\partial{x^g_j}}\;\frac{\partial {x^g_j}}
{\partial{x_i}}=
\sum_{j=1}^n
\frac{\partial p(x^g)}{\partial{x^g_j}}\,g_{ji}=
\sum_{j=1}^n g^\top_{ij}\,
\frac{\partial p(x^g)}{\partial{x^g_j}},\quad \forall i=1,\ldots,n$$
that implies: $\nabla p(x)=g^T \nabla^g p(x^g)$ or
$ \nabla^g p(x^g)=g \nabla p(x)$, where $\nabla^g=(\frac{\partial}{ \partial x^g_1},\ldots,\frac{\partial}{ \partial x^g_n})$. Then:
$$\nabla^g p_a(x^g)\cdot\nabla^g p_b(x^g)=
g \nabla p_a(x)\cdot g \nabla p_b(x)=
\nabla p_a(x)\cdot \nabla p_b(x)\,,$$
that is, $P_{ab}(gx)=P_{ab}(x)$, that tells us that the matrix elements of $P(x)$, calculated with Eq. (\ref{matriceP(x)}), using the variables $x$, or the transformed variables $x^g=gx$, $\forall\,g\in W$, are the same. Hence, $P_{ab}(x)$ is a $W$-invariant polynomial.

As a consequence, we have that all the matrix
elements $P_{ab}(x)$ can be expressed as polynomials of the basic invariant polynomials:
\begin{equation}\label{invmatP}
  P_{ab}(x)=\widehat P_{ab}(p_1(x),\ldots,p_n(x)),\quad
\forall x \in \R^n\,.
\end{equation}

One can then define a matrix function of $p\in \R^n$, $\widehat P(p)$, having $\widehat P_{ab} (p)=\widehat P_{ab} (p_1,\ldots,p_n)$ for elements, such that Eq. (\ref{invmatP}) is satisfied. The real symmetric matrix $\widehat P (p)$ is called the {\em $\widehat P$-matrix}. Its explicit form depends strongly on the given basis $p_1(x),\ldots,p_n(x)$ of $W$-invariant polynomials.\\

It is useful to define an operator $\widehat P$ that applied to a given basis of homogeneous invariant polynomials $p_a(x)$, $a=1,\ldots,n$, gives the corresponding \wPm\ $\widehat P(p)$, obtained through the application of Eqs. (\ref{matriceP(x)}) and (\ref{invmatP}). This will be understood in the following.\\

The map $\overline {p}: \R^n\to \R^n:x\to
\overline {p}(x)=(p_1(x),\ldots,p_n(x))$ is called the {\em orbit
map} and maps $\R^n$ into an $n$-dimensional semi-algebraic connected proper subset
${\cal S}$ of $\R^n$. There is a one
to one correspondence between the orbits in $\R^n$ and the points in
${\cal S}$, because the basic polynomials separate the orbits, that is,
given any two orbits in $\R^n$, at least one of the basic invariant polynomials $p_a(x),\
a=1,\ldots,n$, takes a different value in the two orbits. The orbit map $\overline  p$ actually induces a diffeomorphism between
the {\em orbit space} $\R^n/W$ of the $W$ action in $\R^n$ and the set ${\cal S}$, so ${\cal S}$ can be identified with the orbit space $\R^n/W$.
Points lying in a same orbit of $\R^n$ have conjugate {\em stabilizers}, or {\em isotropy
subgroups}. The set of points in $\R^n$ with conjugated isotropy
subgroups is called a {\em stratum} of the $W$ action in $\R^n$. Its
image in the  set ${\cal S}$ through the orbit map is called a {\em stratum} of ${\cal S}$. Clearly there is a one to
one correspondence among strata of $\R^n$ and strata of $\Scal$.
The number of strata is finite and equals the number of different
conjugacy classes of isotropy subgroups of $W$.\\


The \wPm\ $\widehat P(p)$ contains all information needed to characterize
geometrically the connected set ${\cal S}\subset \R^n$ representing the orbit space $\R^n/W$. For
example, the interior of the set ${\cal S}$ is the only subset of
$\R^n$ where the \wPm\ is positive definite \cite{as1983,ps1985}.
The equation $\det \widehat P(p)=0$ determines the {\em discriminant}, an
$(n-1)$-dimensional algebraic surface in $\R^n$ that contains the image,
through the orbit map, of the set of the reflection hyperplanes of
$\R^n$, and this image coincides with the boundary of the set
$\Scal$. The $k$-dimensional strata of $\Scal$ are found from the
equations and inequalities expressing the conditions:
$\mbox{rank}(\widehat P(p))=k$ and $\widehat P(p)\geq 0$ (this last condition means semi-positivity). 
$\Scal$ is a semi-algebraic set because all these defining conditions can be obtained through polynomial
equations and inequalities. This method to determine the algebraic equations and inequalities of the strata of the orbit space is however not very easy to perform, and for the finite irreducible reflection groups there are better methods, shortly recalled in the introduction. 
\\

As the basic invariant polynomials
$p_1(x),\ldots, p_n(x)$ are homogeneous, it is natural to consider the coordinates
$p_1,\ldots, p_n$ as a set of $n$ graded, or weighted, coordinates whose degrees, or {\it weights}, coincide with the degrees of the basic invariant polynomials, that is: $w(p_a)=\deg(p_a(x))=d_a,\
\forall a=1,\ldots,n$.
The variables $p$ and $x$ are used to label points in different spaces $\R^n$ the variables $p$ are weighted, while the variables $x$ are not.
As homogeneous polynomials are important in what follows, it is convenient to recall two definitions.
A polynomial $\widehat p(p)$ is said of {\it weight} $d$ if the $W$-invariant polynomial $p(x)=\widehat p(p_1(x),\ldots,p_n(x))$ is of degree $d$ in $x$. If $p(x)$ is homogeneous then $\widehat p(p)$ is said {\it weighted homogeneous}, or {\it $w$-homogeneous}. Each coordinate $p_a\in\R^n$ is itself a $w$-homogeneous polynomial with weight $d_a=w(p_a)$, the matrix elements $\widehat P_{ab}(p)$ of the \wPm\ $\widehat P(p)$ are $w$-homogeneous polynomials of weights $d_a+d_b-2$, and the determinant of the \wPm, $\det({\widehat P}(p))$, is a $w$-homogeneous polynomial, of weight $\sum_{a=1}^n (2d_a-2)=2N=nd_n$, if $W$ is irreducible, where Eq. (\ref{sommagradi}) has been used, $N$ is the number of roots of $W$, and $d_n$ is the highest degree among the basic invariant polynomials.\\

A $w$-homogeneous irreducible factor $a(p)$ of the determinant ${\rm det}(\widehat P(p))$ of the \wPm\ $\widehat P(p)$ satisfies the following system of differential equations  \cite{YaSek1979,Sar1989}, called the boundary equation:
\begin{equation}\label{boundaryeqn}
  \sum_{c=1}^n{\widehat P}_{bc}(p)\,\frac{\partial
a(p)}{\partial p_c}=\lambda_b^{(a)}(p)\,a(p),\qquad \forall\,b=1,\ldots,n\,,
\end{equation}
where $\lambda^{(a)}(p)=\left(\lambda_1^{(a)}(p),\ldots,\lambda_n^{(a)}(p)\right)$ is a polynomial vector function of $p\in\R^n$, dependent on $a(p)$, whose elements $\lambda_b^{(a)}(p)$ are $w$-homogeneous polynomial in the weighted variables $p_1,\ldots,p_n$, of weights $d_b-2$, $\forall\,b=1,\ldots,n$:
$$w(\lambda_b^{(a)}(p))=d_b-2\,,\qquad \forall\,b=1,\ldots,n\,.$$
The relation $w(\lambda_b^{(a)}(p))=d_b-2$ easily follows from Eq. (\ref{boundaryeqn}) and the $w$-homogeneity of $a(p)$ and of ${\widehat P}_{bc}(p)$, and in particular from $w({\widehat P}_{bc}(p))=d_b+d_c-2$. The first element $\lambda_1^{(a)}(p)$ is a constant, because $d_1=2$. From Eq. (\ref{boundaryeqn}), with $b=1$, using ${\widehat P}_{1c}(p)=2d_c p_c$ (no sum over $c$) and Euler's theorem on homogeneous polynomials, that for $w$-homogeneous polynomials, like $a(p)$, has the following form:
$$  \sum_{c=1}^n\,d_c\,p_c\,\frac{\partial
a(p)}{\partial p_c}=w(a)\,a(p)\,,$$
one obtains the following general result, that fixes the form of $\lambda_1^{(a)}(p)$: $$\lambda_1^{(a)}(p)=2\,w(a(p))\,.$$
Eq. (\ref{boundaryeqn}) was first given in \cite{YaSek1979}, and further studied in \cite{YaSek1981}, for the case of an irreducible finite reflection group, and in which for $a(p(x))$ it was considered just the invariant polynomial $\det(P(x))$. Eq. (\ref{boundaryeqn}) was rediscovered in \cite{Sar1989}, and extended to the case of a general effective linear group having a basis of $n$ algebraically independent polynomials, and in which for $a(p)$ it was considered a general irreducible factor of $\det(\widehat P(p))$. In \cite{Sar-Tal1991} there were further examined the general properties of arbitrary $w$-homogeneous polynomials $a(p)$, satisfying Eq. (\ref{boundaryeqn}) together with a proper $n$-dimensional vector $\lambda^{(a)}(p)$, whose components $\lambda_b^{(a)}(p)$ are $w$-homogeneous polynomials of degree $d_b-2$, $b=1,\ldots,n$.
Two definitions are convenient. A $w$-homogeneous polynomial $a(p)$, that satisfies the boundary equation (\ref{boundaryeqn}), together with a proper vector $\lambda^{(a)}(p)$, is called an {\em active polynomial} (of $\widehat P(p)$, or corresponding to $\widehat P(p)$), and the vector $\lambda^{(a)}(p)$, without imagination, is called the {\em $\lambda$-vector} (of $a(p)$, or relative to $a(p)$).
Active polynomials differing only for constant multiplicative factors should be identified because they satisfy the boundary equation with the same $\lambda$-vector. In \cite{Sar-Tal1991}, among other things, it was proved that:
\begin{enumerate}
  \item  all irreducible (in the complexes, as well as in the reals) active polynomials are irreducible factors of $\det({\widehat P}(p))$
 ;
  \item all products $a(p)=\prod_{i=1}^r (a_i(p))^{m_i}$, $m_i\in {\mathbb{N}}$, of active irreducible polynomials $a_i(p)$, $i=1,\ldots,r$, are active polynomials too, for which $\lambda^{(a)}(p)=\sum_{i=1}^r m_i\,\lambda^{(a_i)}(p)$, in which $\lambda^{(a_i)}(p)$, $\forall\,i=1,\ldots,r$, are the $\lambda$-vectors of the active irreducible polynomials $a_i(p)$, $i=1,\ldots,r$.
\end{enumerate}
The determinants of the \wPms\ of the irreducible finite reflection groups are generally irreducible polynomials, except for the groups of type $B_n$, $\forall\,n\geq 2$, $F_4$, $I_2(m)$, $\forall\,m\geq 6$, even $m$, in which cases there are two irreducible active factors of $\det({\widehat P}(p))$. These are exactly the cases in which there are two different root systems.


Given a \wPm\ ${\widehat P}(p)$ of an irreducible finite reflection group, the boundary equation, Eq. (\ref{boundaryeqn}), is satisfied by all the irreducible factors of $\det({\widehat P}(p))$, and in any case by $\det({\widehat P}(p))$ itself:
\begin{equation}\label{boundaryeqndet}
  \sum_{b=1}^n{\widehat P}_{ab}(p)\frac{\partial
\det({\widehat P}(p))}{\partial p_b}=\lambda_a^{(\det({\widehat P}))}(p)\,\det({\widehat P}(p)),\qquad a=1,\ldots,n\,.
\end{equation}

There is a general formula giving the $\lambda$-vector $\lambda^{(\det({\widehat P}))}(p)$ corresponding to the active polynomial $\det(\widehat P(p))$ that appears in the boundary equation (\ref{boundaryeqndet}) of a general irreducible reflection group. It is given by the following theorem.

\noindent\begin{theorem}\label{quattro}
Let $W$ be an irreducible finite reflection group, and ${\cal R}_+$ a set of positive roots of $W$. For all $r\in {\cal R}_+$ let $l_r(x)=r\cdot x$,  be the linear form vanishing on the reflection hyperplane of $r$. The $\lambda$-vector $\lambda^{(\det({\widehat P}))}(p)=(\lambda_1^{(\det({\widehat P}))}(p),\ldots,\lambda_n^{(\det({\widehat P}))}(p))$, that appears in the boundary equation (\ref{boundaryeqndet}), can be calculated with the following formula:
\begin{equation}\label{lambdaD}
  \lambda_a^{(\det({\widehat P}))}(p(x))=2\,\sum_{r\in{\cal R}_+}\frac{\nabla p_a(x)\cdot r}{l_r(x)}
  \,,\qquad \forall\,a=1,\ldots,n\,,
\end{equation}
valid $\forall\,x\in\R^n$
.
\end{theorem}

\noindent\textbf{Proof}. Eqs. (\ref{invmatP}) and (\ref{P(x)=j^Tj}) imply $\det( \widehat P(p(x)))=\det( P(x))=(\det(j(x)))^2$. Using this result in Eq. (\ref{boundaryeqndet}), written in terms of the original variables $x\in\R^n$, one has, at the first member:
$$
\left.\sum_{b=1}^n{\widehat P}_{ab}(p)\frac{\partial
\det({\widehat P}(p))}{\partial p_b}\right|_{p=p(x)}=
\sum_{b=1}^n\,\nabla p_a(x)\cdot \nabla p_b(x)\;\frac{\partial
(\det(j(x))^2}{\partial p_b(x)}=
\nabla p_a(x)\cdot \nabla
(\det(j(x)))^2=
$$
$$=2\;\det(j(x))\;
\nabla p_a(x)\cdot \nabla
\det(j(x))\,.
$$
Using Eq. (\ref{detj(x)}), one obtains:
$$\nabla
\det(j(x))=
c\; \nabla\!\prod_{r\in {\cal R}_+}\, l_r(x)=
c\, \sum_{r\in {\cal R}_+}\left( \nabla l_r(x)\,\prod_{\tiny \begin{array}{c}
                                                     r'\in {\cal R}_+ \\
                                                     r'\neq r
                                                   \end{array}
}\,  l_{r'}(x)\right)=
\det(j(x))\, \sum_{r\in {\cal R}_+}\frac{ \nabla l_r(x)}{l_r(x)}=
\det(j(x))\, \sum_{r\in {\cal R}_+}\frac{ r}{l_r(x)}\,,
$$
because $\nabla l_r(x)=r$.
Using this result in the preceding equation, one obtains:
$$
\left.\sum_{b=1}^n{\widehat P}_{ab}(p)\frac{\partial
\det({\widehat P}(p))}{\partial p_b}\right|_{p=p(x)}=2\,(\det(j(x)))^2\;
\nabla p_a(x)\cdot
 \sum_{r\in {\cal R}_+}\frac{ r}{l_r(x)}=
2\,\det({\widehat P}(p(x)))\,
 \sum_{r\in {\cal R}_+}\frac{
\nabla p_a(x)\cdot r}{l_r(x)}\,.$$
Comparing with the second member of Eq. (\ref{boundaryeqndet}), written in terms of the original variables $x\in\R^n$, $\left.\lambda_a^{(\det({\widehat P}))}(p)\,\det({\widehat P}(p))(p)\right|_{p=p(x)}$, one finds that
$$\lambda_a^{(\det({\widehat P}))}(p(x))=2\,
 \sum_{r\in {\cal R}_+}\frac{
\nabla p_a(x)\cdot r}{l_r(x)}
\,,\qquad a=1,\ldots,n\,.$$
\scat

As the first member of Eq. (\ref{lambdaD}) is an invariant polynomial in $x\in\R^n$, also the second member is an invariant polynomial in $x\in\R^n$, and can be expressed as a polynomial of the basic invariant polynomials.\\

A straightforward consequence is the following.

\noindent\begin{theorem}\label{cinque}
Let $W$ be an irreducible finite reflection group with two different root systems, that is $W$ is one of the groups $B_n$, $n\geq 2$, $F_4$, $I_2(m)$, $m\geq 6$, $m$ even, ($B_n$, $n\geq 2$, $C_n$, $n\geq 3$, $F_4$ and $G_2\simeq I_2(6)$, if one considers Weyl groups of simple Lie algebras). Then $\det(\widehat P(p))$ has two different irreducible active factors, $s(p)$ and $l(p)$, such that $\det(\widehat P(p))=c^2\,s(p)\,l(p)$, with $s(p(x))=\prod_{r\in {\cal R}_{+,s}}\, l_r(x)^2$, and $l(p(x))=\prod_{r\in {\cal R}_{+,l}}\, l_r(x)^2$, respectively, where ${\cal R}_{+,s}$ and ${\cal R}_{+,l}$ are the sets of positive short roots and of positive long roots of $W$ (more in general, the sets of roots of the two root systems, respectively), $c^2$ is the constant appearing in Eq. (\ref{detP(x)}), and for any root $r$, $l_r(x)$ is the linear form $l_r(x)=r\cdot x$. The $\lambda$-vectors $\lambda^{(s)}(p)$ and  $\lambda^{(l)}(p)$, corresponding to the active polynomials $s(p)$ and $l(p)$, respectively, 
that appear in the boundary equation (\ref{boundaryeqn})
, can be obtained from the following formulas:
\begin{equation}\label{lambdasl}
  \lambda_a^{(s)}(p(x))=2\,\sum_{r\in{\cal R}_{+,s}}\frac{\nabla p_a(x)\cdot r}{l_r(x)}\,,\qquad
  \lambda_a^{(l)}(p(x))=2\,\sum_{r\in{\cal R}_{+,l}}\frac{\nabla p_a(x)\cdot r}{l_r(x)}\,,\qquad \forall\,a=1,\ldots,n\,,
\end{equation}
valid $\forall\,x\in\R^n$, 
and one has:
\begin{equation}\label{lambdas+lambdal=lambdaD}
  \lambda_a^{(s)}(p)+\lambda_a^{(l)}(p)=\lambda_a^{(\det({\widehat P}))}(p)\,,\qquad  \forall\,a=1,\ldots,n\,.
\end{equation}
\end{theorem}
\noindent\textbf{Proof}.
A group of orthogonal matrices, as is $W$, acts on a vector space by maintaining the vector lengths. Then the long roots and the short roots are transformed into themselves by $W$ transformations (each root system is stable with respect to $W$-transformations). This implies that the set of points contained in the union of the reflection hyperplanes orthogonal to the long roots is invariant under $W$ transformations, and so does the set of points contained in the union of the reflection hyperplanes orthogonal to the short roots. We can then rewrite Eq. (\ref{detP(x)}), using Eq. (\ref{invmatP}), in the following way:
$$\det(\widehat P(p(x)))=c^2\,\left(\prod_{r\in {\cal R}_{+,l}}\, l_r(x)^2\right)\, \left(\prod_{r\in {\cal R}_{+,s}}\, l_r(x)^2\right)=c^2\,s(p(x))\,l(p(x))\,,
$$
where
$$s(p(x))=\prod_{r\in {\cal R}_{+,s}}\, l_r(x)^2\,,\qquad l(p(x))=\prod_{r\in {\cal R}_{+,l}}\, l_r(x)^2$$ are invariant polynomials, and so they can be written in terms of the basic invariant polynomials $p_1(x),\ldots,p_n(x)$. The proof of Theorem \ref{quattro} can then be repeated with $s(p(x))$ or $l(p(x))$ in place of $\det(\widehat P(p(x)))$, and it gives Eqs. (\ref{lambdasl}). By comparing Eq. (\ref{lambdaD}) and Eqs. (\ref{lambdasl}), one immediately finds Eq. (\ref{lambdas+lambdal=lambdaD}).\scat

In the hypothesis of Theorem \ref{cinque}, the equations $s(p)=0$ and $l(p)=0$ contain different orbit type strata of ${\cal S}$ because points in the set of reflecting hyperplanes with images in $s(p)=0$ have isotropy subgroups that cannot be conjugate to any isotropy subgroup of points in the set of reflecting hyperplanes with images in $l(p)=0$.\\

It is interesting to see how the \wPms\ transform when the basic invariant polynomials are transformed. We will examine two classes of basis transformations. The first one is related to a change of the canonical basis in $\R^n$, that naturally implies, among others, a change of the reflection hyperplanes, of the matrix representatives of the reflection group, and of the basic invariant polynomials. The second one is related to a change of the basic invariant polynomials performed without changing the basis of $\R^n$, that is, obtained through algebraic transformations of the basic invariant polynomials such to maintain their degrees and algebraic independence.\\

A change of the (orthonormal) coordinate system used in $\R^n$: $x\to
x'=R x$, $R\in O(n)$, determines a change of the matrix representatives of any matrix group acting in $\R^n$. Let's call $W$ and $W'$ the matrix group before and after the transformation. It is well known that $W'=\{g'\ |\ g'=R g R^\top,\ \forall\,g\in W\}$. In fact, if $y=gx$, $y'=Ry$, $x'=Rx$, one has $y'=R y= R g x=RgR^\top x'=g'x'$ and by comparison $g'=R g R^\top$. The basic invariant polynomials change accordingly from the expressions $p_a(x)$ to other expressions $p'_a(x')$, so that, $\forall \, x\in\R^n$, one has: $p_a'(x')=p_a(x)$, $p_a(gx)=p_a(x)$, $\forall\,g\in W$, and $p'_a(g'x')=p'_a(x')$, $\forall\,g'\in W'$. This happens in particular to the matrix elements of $P(x)$, as they are invariant polynomials in $x$: $P_{ab}(x)= P'_{ab}(x')$. 
It is convenient to state and prove the following well known theorem, because we will use it very often in the following.

\noindent\begin{theorem}\label{uno}
Let $x\to x'=Rx$, with $R\in O(n)$, be an orthogonal coordinate transformation of $\R^n$, $W$ a reflection group, defined in the old coordinates $x$, and $W'=\{g'\ |\ g'=R g R^\top,\ \forall\,g\in W\}$ the corresponding reflection group in the new coordinates $x'$. One has then:
\begin{enumerate}
  \item \label{th1it2}The $W'$-invariant polynomials $p'(x')$ are related to the $W$-invariant polynomials $p(x)$ by the following equation: $$p'(x')= p(R^\top x')= p(x)\,.$$
  \item \label{th1it3}If $p_1(x),\ldots,p_n(x)$, is a basis of $W$-invariant homogeneous polynomials, then $p'_1(x')=p_1(R^\top x'),\ldots,p'_n(x')=p_n(R^\top x')$, is a basis of $W'$-invariant homogeneous polynomials, and
  if $p(x)=\widehat p(p_1(x),\ldots,p_n(x))$ is a $W$-invariant polynomial, then the $W'$-invariant polynomial $p'(x')=p(R^\top x')$ has the same functional dependence $\widehat p$ on the new basis, as that one of $p(x)$ on the old basis, that is:
      $$p(x)=\widehat p(p_1(x),\ldots,p_n(x))\qquad \Leftarrow\!\Rightarrow\qquad p'(x')=\widehat p(p'_1(x'),\ldots,p'_n(x'))\,.$$
  \item \label{th1it4} The \wPms\ $\widehat P(p)$ and $\widehat P(p')$, constructed with the bases $p_a(x)$ and $p'_a(x')$, $a=1,\ldots,n$, respectively, are identical, except for the substitution of the coordinates $p_a$ with the coordinates $p'_a$, $\forall\,a=1,\ldots,n$, that is:
$$\widehat P(p')=\left.\widehat P(p)\right|_{p=p'}\,.$$\\
\end{enumerate}
\end{theorem}
\noindent\textbf{Proof}.
\noindent
\ref{th1it2}. If $p(x)$ is a $W$-invariant polynomial, $p(gx)=p(x)$, $\forall\,g\in W$ and $\forall\,x\in \R^n$. One defines the polynomial $p'(x')$ in such a way that $p'(x')=p(x)$, that also implies $p'(x')=p(R^\top x')$, because $x=R^\top x'$. It is then just to prove the $W'$-invariance of $p'(x')$.
One has: $p(x)=p(gx)=p((R^\top g' R)(R^\top x'))=p(R^\top g' x')$, but also $p(x)=p(R^\top x')$, so that: $p(R^\top g' x')=p(R^\top x')$, $\forall\,g'\in W'$ and $\forall\,x'\in \R^n$, that is $p'(g' x')=p'(x')$, $\forall\,g'\in W'$ and $\forall\,x'\in \R^n$, and the $W'$ invariance of $p'(x')$ is proved.\\
\ref{th1it3}. With the basis of $W$-invariant polynomials $p_a(x)$, $a=1,\ldots,n$, one can form $n$ $W'$-invariant polynomials  $p'_a(x')=p_a(R^\top x')$, $\forall\,a=1,\ldots,n$. These $n$ polynomials are such that $p_a(x)=p'_a(x')$, $\forall\,a=1,\ldots,n$, and form a basis for the $W'$-invariant polynomials. In fact, if $p'(x')$ is a $W'$-invariant polynomial, $p'(x')=p(x)$, with $p(x)$ a $W$-invariant polynomial. But then $p'(x')=p(x)=\widehat p(p_1(x),\ldots,p_n(x))=\widehat p(p'_1(x'),\ldots,p'_n(x'))$, and this proves that $p'(x')$ can be written as a polynomial of the basic invariant polynomials $p'_a(x')$, $a=1,\ldots,n$, with the same functional dependence $\widehat p$ on the new basis, as that one of $p(x)$ on the old basis.\\
\ref{th1it4}. By substituting in the \wPm\ $\widehat P(p)$ the variables $p_a$, $a=1,\ldots,n$, with the basic $W$-invariant polynomials $p_a(x)$, $a=1,\ldots,n$, one obtains the matrix $\widehat P(p_1(x),\ldots,p_n(x))=P(x)$, whose elements are determined by Eq. (\ref{matriceP(x)}) and are homogeneous $W$-invariant polynomials of $x$. From item 2, using the variables $x'$,  one has: $P'(x')=P(R^\top  x')=\widehat P(p_1(R^\top x'),\ldots,p_n(R^\top x'))=\widehat P(p'_1(x'),\ldots,p'_n(x'))$, whose matrix elements are homogeneous $W'$-invariant polynomials of $x'$, with the same functional dependence on the new basic invariant polynomials as those of $P(x)$ on the old basic invariant polynomials.
It remains to show that the matrix $P'(x')$ coincides with the matrix $P(x')$ obtained from Eq. (\ref{matriceP(x)}), using the basic polynomials $p'_1(x'),\ldots,p'_n(x')$. One has, $\forall\,a,b=1,\ldots,n$:
$$\widehat P_{ab}(p'_1(x'),\ldots,p'_n(x')) = P_{ab}(x')=\nabla p'_a(x') \cdot \nabla p'_b(x')=\sum_{i=1}^n
\frac{\partial p'_a(x')}{\partial x'_i}\;\frac{\partial
p'_b(x')}{\partial x'_i}=\sum_{i=1}^n
\frac{\partial p_a(x)}{\partial x'_i}\;\frac{\partial
p_b(x)}{\partial x'_i}=
$$
$$=\sum_{i,j,k=1}^n
\frac{\partial p_a(x)}{\partial x_j}\;\frac{\partial
R^\top_{jl}x'_l}{\partial x'_i}\;\frac{\partial
p_b(x)}{\partial x_k}\;\frac{\partial
R^\top_{km}x'_m}{\partial x'_i}=\sum_{i,j,k=1}^n
\frac{\partial p_a(x)}{\partial x_j}\;R^\top_{ji}\;\frac{\partial
p_b(x)}{\partial x_k}\;R^\top_{ki}=\sum_{j,k=1}^n
\frac{\partial p_a(x)}{\partial x_j}\;\delta_{jk}\;\frac{\partial
p_b(x)}{\partial x_k}=
$$
$$=\sum_{j=1}^n
\frac{\partial p_a(x)}{\partial x_j}\;\frac{\partial
p_b(x)}{\partial x_j}=\nabla p_a(x) \cdot \nabla p_b(x)=P_{ab}(x)=\widehat P_{ab}(p_1(x),\ldots,p_n(x))\,.$$
The \wPm\ $\widehat P(p')$ is then obtained from the \wPm\ $\widehat P(p)$ just by replacing the variables $p_a$ with the variables $p'_a$, $\forall\,a=1,\ldots,n$.
\scat

The basic invariant polynomials, as well as the matrices representing the group of linear transformations, are dependent on the system of coordinates used in $\R^n$, but the \wPm\ is independent from this system of coordinates.
The \wPm\ depends only on how a given basis of $W$-invariant polynomials generates the ring of $W$-invariant polynomials $[\R^n]^W$
. For this reason the \wPm\ can be considered a basic tool
of constructive invariant theory~\cite{tal-sspcm05}. Theorem \ref{uno} holds for a general compact linear group $G$, because the action of $G$ can always be obtained with orthogonal matrices and $G$ possesses a finite basis of homogeneous invariant polynomials.\\

From Theorem \ref{uno}, item \ref{th1it3}, we know that under an orthogonal coordinate transformation of $\R^n$, $x\to x'$, the basic invariant polynomials change: $p_a(x)\to p'_a(x')$, $\forall a=1,\ldots,n$, so the weighted coordinates of $\R^n$, that are their representatives in the space $\R^n$ hosting the set $\Scal$, also change: $p_a\to p'_a$. However, for Theorem \ref{uno}, item \ref{th1it4}, the \wPm\ changes only for the substitutions $p_a\to p'_a$: $\widehat P(p')=\left.\widehat P(p)\right|_{p=p'}$, and this implies that the set $\Scal$ (with its stratification), that is determined by the \wPm, only, does not change under orthogonal coordinate transformation of $\R^n$.\\


Let's now examine transformations of the basic invariant polynomials obtained without changing the coordinate basis in $\R^n$.
One changes the basic set of invariant
polynomials: $p_a(x)\to p'_a(x)$, $a=1,\ldots,n$, in such a way to maintain the
homogeneity, the degrees $d_a$, $a=1,\ldots,n$, and the algebraic
independence of the $p_a'(x)$, $a=1,\ldots,n$. 
As $\forall\,a=1,\ldots,n$, the $p'_a(x)$ are
$W$-invariant homogeneous polynomials of degrees $d_a$, it is possible to write them as
polynomials of the basic invariant polynomials $p_a(x)$, $\forall\,a=1,\ldots,n$, in such a way that:
$$p'_a(x)={\widehat p\,}_a'(p_1(x),\ldots,p_n(x)),\qquad \forall\,a=1,\ldots,n,\qquad\forall\, x\in
\R^n\,.$$
Removing the dependence on $x$, this basis transformation can be seen as a particular coordinate transformation of $p\in\R^n$, that is:
$$p'_a={\widehat p\,}_a'(p_1,\ldots,p_n)={\widehat p\,}_a'(p)\,,\qquad \forall\,a=1,\ldots,n\,,$$
in which ${\widehat p\,}_a'(p)$ is a $w$-homogeneous polynomial of weight $d_a$ of the weighted variables $p_1,\ldots,p_n$.

The jacobian matrix of this transformation, $J(p)$, has elements:
\begin{equation}\label{jacobian}
  J_{ab}(p)=\frac{\partial {\widehat p\,}_a'(p)}{\partial p_b},\qquad \forall\,a,b=1,\ldots,n\,,
\end{equation}
that are $w$-homogeneous polynomial functions of weights $d_a-d_b$ of
the weighted variables $p_1,\ldots,p_n$, so they are vanishing if
$d_a<d_b$, and constant if $d_a=d_b$.

The requirement that the transformed invariant polynomials be algebraic independent gives some constraints. One can require, for example, that the expression of ${\widehat p\,}_a'(p)$ contains necessarily the variable $p_a$, $\forall\,a=1,\ldots,n$.
For simplicity, let's suppose to adopt the convention (\ref{labeling}) on the labeling the basic invariant polynomials according to their degrees.
If all the degrees $d_a$ are different, that is, in the case of
all irreducible finite reflection group different from $D_n$, with
even $n$, the diagonal elements $J_{aa}(p),\ \forall a=1,\ldots,n$, must be non-zero constants, and the non-diagonal elements $J_{ab}(p)$, with $a<b$, must vanish, and this implies that $J(p)$ is a lower triangular matrix with non-zero elements along its diagonal.
In the case of $D_n$, with even $n$, two degrees are equal and it
is possible to have a constant non-singular non-diagonal $2\times 2$ sub-matrix along the
diagonal of $J(p)$, in correspondence with the rows and columns corresponding to the two coordinates of weight $n$, but for the rest $J(p)$ is a lower triangular matrix, with non zero constant elements along its diagonal. The only jacobian matrices that can be accepted as giving rise to transformations between basis of (algebraically independent) invariant polynomials must meet all the above requirements.

This implies that, in any case then, $\det(J(p))$ is a non-zero real constant. The inverse transformation $p'\to p$ is then everywhere well defined. The inverse transformation $p'\to p$ is a $W$-homogeneous polynomial
map, like the transformation $p\to p'$, that is, $p_a=\widehat
p_a(p'),\ \forall a=1,\ldots,n$, are $w$-homogeneous polynomial
functions of the weighted variables $p'_1,\ldots,p'_n$, whose weights are $w(p_a(p'))=d_a$, and such that $\widehat
p_a(p')$ contains necessarily $p'_a$, $\forall\,a=1,\ldots,n$.

The transformation $p\to p'$ causes a transformation of the \wPm\
and this implies also a transformation of the geometric shape of the set ${\cal S}$ describing the orbit space. Because the maps $p\to p'$ and $p'\to p$ are polynomial,
this transformation of ${\cal S}$ is a diffeomorphism, and this is enough to know that ${\cal S}$ maintains in any case its topological shape and stratification.\\

It is convenient to give the following definition. An {\em allowed basis transformation} is a transformation $p'={\widehat p\,}'(p)$ with Jacobian matrix $J(p)$ that satisfies the following requirements:
\begin{enumerate}
  \item $J_{ab}(p)$ is a $w$-homogeneous polynomial of weight $d_a-d_b$;
  \item $\det(J(p))$ is a non-zero constant.
\end{enumerate}

Let $\widehat P_{ab}(p')$  and $\widehat P_{ab}(p)$,
$\forall\,a,b=1,\ldots,n$, be the matrix elements of the \wPms\
$\widehat P(p')$ and $\widehat P(p)$, corresponding to the bases
$p'_1(x),\ldots,p'_n(x)$, and $p_1(x),\ldots,p_n(x)$, respectively. The relation between the matrices $\widehat P(p')$ and $\widehat P(p)$ is given by the following theorem:

\noindent\begin{theorem}\label{tre}
If $J(p)=\|\partial {\widehat p\,}_a'(p)/\partial p_b\|$ is the jacobian matrix of an allowed basis transformation $p'={\widehat p\,}'(p)$, whose inverse transformation is $p=\widehat p(p')$, the relation between the \wPms\ $\widehat P(p')$ and $\widehat P(p)$ corresponding to the bases $p'_1(x),\ldots,p'_n(x)$, and $p_1(x),\ldots,p_n(x)$,
is given by the following {\em \wPm\ transformation formula}:
\begin{equation}\label{Ptransf}
  \widehat P(p')=\left.J(p)\; \widehat
P(p)\; J^\top(p)\right|_{p= \widehat p(p')}\,.
\end{equation}
\end{theorem}

\noindent\textbf{Proof}.
$$\left.\widehat P_{cd}(p')\right|_{p'= {\widehat p\,}'(p(x))}=\left(\nabla p'_c(x) \cdot \nabla
p'_d(x)\right)\Bigr|_{p'(x)= {\widehat p\,}'(p(x))}=\nabla {\widehat p\,}_c'(p(x)) \cdot \nabla
{\widehat p\,}_d'(p(x))=$$
$$=\sum_{a,b=1}^n \left.\frac{\partial
{\widehat p\,}_c'(p)}{\partial p_a}\right|_{p= p(x)}\nabla p_a(x)
\cdot \left.\frac{\partial {\widehat p\,}_d'(p)}{\partial
p_b}\right|_{p= p(x)}\nabla p_b(x)=\left.\sum_{a,b=1}^n
J_{ca}(p)\widehat P_{ab}(p)J^\top_{bd}(p)\right|_{p= p(x)}$$
that is:
\begin{equation}\label{Ptransf2}
 \left.\widehat P(p')\right|_{p'= {\widehat p\,}'(p)}=J(p)\; \widehat
P(p)\; J^\top(p)
\end{equation}
Using the inverse transformation $p=\widehat p(p')$, this formula can be written in the form of Eq. (\ref{Ptransf}).\scat

It is interesting to see how an active polynomial $a(p)$ (possibly coincident with $\det({\widehat P}(p))$), and its corresponding $\lambda$-vector $\lambda^{(a)}(p)$ transform under transformations of the basis of invariant polynomials.
As we will use these results in the following we will state and prove the following theorem.

\noindent\begin{theorem}\label{trasflambda}
Let $W$ be a rank-$n$ irreducible finite reflection group, $p_1(x),\ldots,p_n(x)$, a basic set of $W$-invariant polynomials, and $\widehat P(p)$ the corresponding \wPm. Let $a(p)$ be an active polynomial of $\widehat P(p)$ and $\lambda^{(a)}(p)$ its $\lambda$-vector.
\begin{enumerate}
  \item \label{thlit1}If one performs an orthogonal coordinate transformation $x\to x'=Rx$, and $p'_a(x')=p_a(R^\top x')$, $a=1,\ldots,n$, are the basic invariant polynomials in the new coordinates $x'$, the active polynomial $a'(p')$, corresponding to $a(p)$ in the new coordinates $x'$, and its $\lambda$-vector $\lambda^{(a')}(p')$, are the following:
$$a'(p')=\left.a(p)\right|_{p= p'}\,,\qquad \lambda^{(a')}(p')=\left.\lambda^{(a)}(p)\right|_{p= p'}\,.$$
  \item \label{thlit2}If one performs an allowed basis transformation $p'={\widehat p\,}'(p)$, with Jacobian matrix $J(p)=\|\partial {\widehat p\,}_a'(p)/\partial p_b\|$, the active polynomial $a'(p')$, corresponding to $a(p)$ in the new basis of invariant polynomials $p'(x)$, and its $\lambda$-vector $\lambda^{(a')}(p')$, are the following:
\begin{equation}\label{lambdatransf}
  a'(p')=\left.a(p)\right|_{p=\widehat{p}(p')}\,,\qquad \lambda_b^{(a')}(p')=\left.\sum_{c=1}^n\,J_{bc}(p)\,\lambda_c^{(a)}(p)\right|_{p=\widehat{p}(p')}\,,\qquad \forall\, b=1,\ldots,n\,.
\end{equation}
\end{enumerate}
\end{theorem}
\noindent\textbf{Proof}.
\ref{thlit1}. Let $a(p)$ be irreducible (in the reals or in the complexes). Then $a(p)$ is a factor of $\det(\widehat P(p))$, and satisfies the boundary equation (\ref{boundaryeqn}). Theorem \ref{uno}, item \ref{th1it4}, establishes that in the new coordinates $x'=Rx$, the \wPm\ is unchanged in form, only one has to replace the old variables $p$ with the new ones $p'$. Then $\det(\widehat P(p'))$ has the irreducible factor $a'(p')=\left.a(p)\right|_{p= p'}$, that satisfies the boundary equation (\ref{boundaryeqn}), in which all variables $p_a$ have been replaced by the variables $p'_a$, $\forall \, a=1,\ldots,n$. Then $a'(p')$ is active and its $\lambda$-vector $\lambda^{(a')}(p')$ is obtained from $\lambda^{(a)}(p)$ just by replacing the variables $p_a$ with the variables $p'_a$, $\forall \, a=1,\ldots,n$, that is $\lambda^{(a')}(p')=\left.\lambda^{(a)}(p)\right|_{p= p'}$. If $a(p)$ is a reducible active polynomial, it is the product of irreducible active factors of $\det(\widehat P(p))$, and its $\lambda$-vector is the sum of the $\lambda$-vectors of its irreducible factors. Then, $a'(p')=\left.a(p)\right|_{p= p'}$ is a reducible polynomial in the variables $p'$, that is the product of irreducible active factors of $\det(\widehat P(p'))$, and so it is an active polynomial, and its $\lambda$-vector is the sum of the $\lambda$-vectors of its irreducible active factors.\\
\ref{thlit2}. 
 Let $a(p)$ be irreducible (in the reals or in the complexes). Theorem \ref{tre} establishes that $\widehat P(p')=\left.J(p)\; \widehat
P(p)\; J^\top(p)\right|_{p= \widehat p(p')}$. Then $\det( \widehat P(p'))=j^2\,\left.\det( \widehat P(p))\right|_{p=\widehat p(p')}$, where $j=\det(J(p))$ is a non-zero real constant, and $a'(p')=\left.a(p)\right|_{p= \widehat p(p')}$ must be an irreducible factor of $\det( \widehat P(p'))$. It results that $a'(p')$ is an active polynomial because it satisfies the boundary equation (\ref{boundaryeqn}). In fact, $\forall \, a=1,\ldots,n$,  we have, at the first member of Eq. (\ref{boundaryeqn}), written in the new coordinates $p'$, and understanding the summation symbols:
$$ {\widehat P}_{bc}(p')\,\frac{\partial
a'(p')}{\partial p'_c}=
\left.\left(J_{bd}(p)\; \widehat
P_{de}(p)\; J_{ec}^\top(p)\right)\right|_{p= \widehat p(p')}\,\left.\left(\frac{\partial
a(p)}{\partial p_f}\right)\right|_{p= \widehat p(p')}\,\frac{\partial
\widehat p_f(p')}{\partial p'_c}=
$$
$$=
\left.\left(J_{bd}(p)\; \widehat
P_{de}(p)\right)\right|_{p= \widehat p(p')}\,\frac{\partial
\widehat p_f(p')}{\partial p'_c}\,\left.\left(\frac{\partial
\widehat p'_c(p)}{\partial p_e}\right)\right|_{p= \widehat p(p')}\,\left.\left(\frac{\partial
a(p)}{\partial p_f}\right)\right|_{p= \widehat p(p')}=
$$
$$=
\left.\left(J_{bd}(p)\; \widehat
P_{de}(p)\right)\right|_{p= \widehat p(p')}\,\delta_{ef}\,\left.\left(\frac{\partial
a(p)}{\partial p_f}\right)\right|_{p= \widehat p(p')}=
\left.J_{bd}(p)\right|_{p= \widehat p(p')}\,\left.\left( \widehat
P_{de}(p)\frac{\partial
a(p)}{\partial p_e}\right)\right|_{p= \widehat p(p')}=
$$
$$=
\left.J_{bd}(p)\right|_{p= \widehat p(p')}\,\left.\left( \lambda_d^{(a)}(p)\,
a(p)\right)\right|_{p= \widehat p(p')}=\left.\left(J_{bd}(p)\, \lambda_d^{(a)}(p)\right)\right|_{p= \widehat p(p')}\,
a'(p')\,.
$$
By comparing with Eq. (\ref{boundaryeqn}), written for the variables $p'$, and $a'(p')$, we see that $a'(p')$ is an active polynomial and that its $\lambda$-vector has the elements $ \lambda_b^{(a')}(p')$ given by the equation:
$\lambda_b^{(a')}(p')=\left.\left(J_{bd}(p)\, \lambda_d^{(a)}(p)\right)\right|_{p= \widehat p(p')}$, $\forall\,b=1,\ldots,n$.
\scat



Many of the definitions and results given in this Section can be extended to a general compact linear group, not necessarily a finite reflection group, because its algebra of invariant polynomials admits a basis made by a finite number of homogeneous invariant polynomials. In particular, the \wPm\ can be defined for all these groups. 
\\

As already recalled in the introduction, the construction of the \wPm\ from the definition, that is using Eqs. (\ref{matriceP(x)}) and (\ref{invmatP}), when the rank $n$ of the group is large, requires an enormous effort, computer memory and computing time, and is often impossible to take to completion. The generating formulas (\ref{PjkdiSn}), (\ref{PjkdiAn}), (\ref{PjkdiBn}) and (\ref{PjkdiDn_divisa}), obtained in Sections \ref{Sn}--\ref{Dn}, for a given choice of the basic invariant polynomials, are then very useful, especially because, using Eq. (\ref{Ptransf}), they allow to determine the explicit form of the \wPms\ in all bases of invariant polynomials of the groups of type $S_n$, $A_n$, $B_n$ and $D_n$.
Examples \hyperlink{ex2}{2}--\hyperlink{ex4}{4}, in Section \ref{examples} show the use of the \wPm-transformation formula, Eq. (\ref{Ptransf}), and the {\em $\lambda$-vector transformation formula}, Eq. (\ref{lambdatransf}).\\

It is now possible to recall the results of Refs. \cite{SYS1980}, \cite{Sar-Tal1991} and \cite{Flatto1970} with some precision. All of them are concerned with the determination of distinguished bases of invariant polynomials satisfying some supplementary conditions.\\

In Ref. \cite{SYS1980} Saito, Yano and Sekiguchi proved that, for all irreducible finite reflection groups, it is always possible to choose the basis of homogeneous invariant polynomials $p_1(x),\ldots,p_n(x)$, in such a way that the derivative of the \wPm\ $\widehat P(p)$, made with respect to the highest weight variable $p_h$, is a constant matrix, that is, it is always possible to have
\begin{equation}\label{flatcondition}
  A_{ab}=\frac{\partial \widehat P_{ab}(p)}{{\partial p_h}}=\mbox{a real number},\qquad \forall\,a,b=1,\ldots,n\,.
\end{equation}
Moreover, given two bases for which this property
holds, their basic invariant polynomials may differ only for scale
constant factors. This means that, if all the basic polynomials and the corresponding variables $p_1,\ldots,p_n$ are ordered according to their degrees, or weights, (so $p_h\equiv p_n$), and because the degrees of the basic invariant polynomials satisfy the relations $d_a+d_{n-a+1}=d_n+2$, $\forall\,a=1,\ldots,n$, it is possible to choose the basic invariant polynomials in such a way that $p_n$ appears in $\widehat P(p)$ only along the auxiliary diagonal connecting the elements $\widehat P_{1n}(p)$ and $\widehat P_{n1}(p)$. In~\cite{SYS1980} these bases were called {\em
flat}, because they establish a flat structure on the orbit space, that is a system of coordinates with a flat metric.  The authors of Ref. \cite{SYS1980} also determined explicitly the flat bases of invariant polynomials of all irreducible finite reflection groups, except $E_7$ and $E_8$ (and they also calculated explicitly the \wPms\ of the groups $E_6$, $F_4$, $G_2$, $H_3$, $H_4$, $I_2(m)$, $m\geq 2$). The flat bases of invariant polynomials for $E_7$ and $E_8$, together with the corresponding \wPms\ were found in \cite{tal-jmp2010}. Ex. \hyperlink{ex2}{2} in Section \ref{examples} shows how one can determine the \wPm\ and the $\lambda$-vector of its determinant in a flat basis of $A_3$, starting from the \wPm\ and the $\lambda$-vector obtained with the generating formulas here given for the groups of type $A_n$, and using Theorems \ref{tre} and \ref{trasflambda}.\\

In Ref. \cite{Sar-Tal1991} Sartori and Talamini proved that, in the case of a compact linear orthogonal group having a basis of algebraically independent invariant polynomials, 
with $p_1(x)=\|x\|^2$ (and this excludes groups of type $S_n$), it is always possible to choose the basis of invariant polynomials in such a way that the only non zero element of the $\lambda$-vector $\lambda^{(a)}(p)$, of any active polynomial $a(p)$, is the first one, that is: \begin{equation}\label{abasiscondition}
  \lambda_1^{(a)}(p)=2\,w(a)\,, \qquad\lambda_b^{(a)}(p)=0\,,\qquad \forall\,b=2,\ldots,n\,.
\end{equation}
Moreover, given two bases for which this property
holds, their basic invariant polynomials are related by an allowed basis transformation not dependent on $p_1$. In Ref.~\cite{Sar-Tal1991} these bases were called {\em $a$-bases}. Ex. \hyperlink{ex3}{3} in Section \ref{examples} shows how one can determine the \wPm\ $\widehat P(p)$ and the $\lambda$-vector of its determinant $\lambda^{(\det(\widehat P))}(p)$ in a $\det(\widehat P)$-basis (that is in an $a$-basis in which the active polynomial $a(p)$ is $\det(\widehat P(p))$) of $A_3$, starting from the \wPm\ and the $\lambda$-vector obtained with the generating formulas here given for the groups of type $A_n$, and using Theorems \ref{tre} and \ref{trasflambda}.\\

In Ref. \cite{Flatto1970} Flatto defined a bilinear map that maps two polynomials $p(x)$ and $q(x)$ into a polynomial $\langle p,q\rangle(x)$, obtained in the following way:
$$\langle p,q\rangle(x)=p(\nabla)\,q(x)\,,$$
where $p(\nabla)$ means that in $p(x)$ one has to replace all occurrences of $x_i^k$ with $\frac{\partial^k}{\partial x_i^k}$, $\forall\,i=1,\ldots,n$, $\forall \,k\in\mathbb{N}$, and  $p(\nabla)\,q(x)$ is the polynomial obtained by the application of the differential operator $p(\nabla)$ to the polynomial $q(x)$. If $p(x)$ and $q(x)$ are homogeneous of degree $d_p$ and $d_q$, respectively, then $\langle p,q\rangle(x)$ is 0 if $d_p>d_q$, a real number if $d_p=d_q$, a homogeneous polynomial of degree $d_q-d_p$ if $d_p<d_q$. This also shows that the map $\langle\cdot,\cdot\rangle$ is non-symmetric in the two arguments. 
In Ref. \cite{Flatto1970} Flatto proved that, for all effective irreducible finite reflection groups $W$, besides multiples, there exist a unique basis of homogeneous invariant polynomials, $p_1(x),\ldots,p_n(x)$, if one requires that $p_1(x)=\|x\|^2$ and that all basic invariant polynomials satisfy the conditions:
\begin{equation}\label{flattocondition1}
  \langle p_a,p_b\rangle(x)=0\,,\qquad \forall\, a,b=1,\ldots,n,\qquad \mbox{such that}\ d_a<d_b\,,
\end{equation}
\begin{equation}\label{flattocondition2}
  \langle p_a,p_b\rangle(x) =\langle p_b,p_a\rangle(x) =0\,,\quad \mbox{if}\ d_a=d_b=n\,,\quad a\neq b\,,
\end{equation}
where the second one concerns the case $ W\equiv D_n,\ \mbox{even}\ n$, only.
In addition to the conditions (\ref{flattocondition1}) and  (\ref{flattocondition2}), Iwasaki \cite{Iwasaki1997} proposed to fix the normalization of the basic polynomials, by requiring $\langle p_a,p_a\rangle (x)=1$, $\forall\,a=1,\ldots,n$, and called {\em canonical} such a basis. However, both the basic invariant polynomials and the \wPms\ of the canonical bases (according to Iwasaki's definition) contain irrational coefficients, and $p_1(x)$ has not the standard form (\ref{quadraticinv}). For this reason, I decided to leave undetermined the normalization of the basic invariant polynomials, and call {\em canonical} any basis of invariant polynomial satisfying Eqs. (\ref{flattocondition1}) and  (\ref{flattocondition2}). Two canonical bases differ then only for scalar multiples of the basic polynomials. Because of the special form of $p_1(x)$, all polynomials of degrees greater than 2 in a canonical basis satisfy the differential equation $\langle p_1,p_a\rangle(x) =\nabla^2 p_a(x)=0$, $\forall\,a=2,\ldots, n$, and are thus harmonic polynomials. The canonical bases of invariant polynomials of the reflection groups of types $A_n$, $B_n$, $D_n$ and $I_2(m)$, were determined in Ref. \cite{Iwasaki1997} and those of the reflection groups $H_3$, $H_4$ and $F_4$ were determined in Ref. \cite{IwasakiKenmaMatsumoto2002}.
Ex. \hyperlink{ex4}{4} in Section \ref{examples} shows how one can determine the \wPms\ and the $\lambda$-vectors in a {canonical} basis of $A_3$, starting from the \wPm\ and the $\lambda$-vector obtained with the generating formulas here given for the groups of type $A_n$, and using Theorems \ref{tre} and \ref{trasflambda}.\\

I calculated the \wPms\ of the group $B_8$ for the following bases: the basis given in Section \ref{Bn}, a flat basis, a $\det(\widehat P)$-basis and a canonical basis. The form and complexity of the \wPm\ depends strongly on the basis. This can be seen from the matrices reported here under, that, for each of these 4 choices of the basis, they have in their matrix elements the number of terms of the $w$-homogeneous polynomials forming the corresponding matrix elements of the \wPm.
All flat bases differ only for scalar multiples of the basic invariant polynomials, and so do all canonical bases. This implies that the number of terms in the matrix elements of the \wPms\ of all flat bases are the same and that the number of terms in the matrix elements of the \wPms\ of all canonical bases are the same. On the contrary, the number of terms in the matrix elements of the \wPms\ of the $a$-bases depend on the $a$-basis chosen. For example, there are $\det(\widehat P)$-bases of $B_8$ in which the element $\widehat P_{88}(p)$ has 146 terms, the maximum number of terms that is possible for a homogeneous invariant polynomial of degree $d_8+d_8-2=30$, when expressed in terms of basic invariant polynomials of $B_8$. The  $\det(\widehat P)$-basis used to obtain the matrix written above, was obtained from the basis defined in Section \ref{Bn}, by avoiding, in the basis transformation, all terms independent from $p_1$.
Anyway, this calculation seems to suggest that one of the simplest form of the \wPm\ is obtained with the bases here proposed.
\begin{center}
Number of terms in the matrix elements of the \wPms\ of $B_8$\\ in correspondence with different bases of invariant polynomials.\\
\vskip 0.3cm
\begin{tabular}{ccc}
  basis of Section \ref{Bn} & & flat basis \\
$ \left(
            \begin{array}{cccccccc}
              1 & 1 & 1 & 1 & 1 & 1 & 1 & 1 \\
              1 & 2 & 2 & 2 & 2 & 2 & 2 & 1 \\
              1 & 2 & 3 & 3 & 3 & 3 & 2 & 1 \\
              1 & 2 & 3 & 4 & 4 & 3 & 2 & 1 \\
              1 & 2 & 3 & 4 & 4 & 3 & 2 & 1 \\
              1 & 2 & 3 & 3 & 3 & 3 & 2 & 1 \\
              1 & 2 & 2 & 2 & 2 & 2 & 2 & 1 \\
              1 & 1 & 1 & 1 & 1 & 1 & 1 & 1 \\
            \end{array}
          \right)
   $
   & \qquad\qquad &$\left(
            \begin{array}{cccccccc}
              1 & 1 & 1 & 1 & 1 & 1 & 1 & 1 \\
              1 & 3 & 4 & 5 & 7 & 8 & 10 & 10 \\
              1 & 4 & 7 & 10 & 13 & 18 & 21 & 25 \\
              1 & 5 & 10 & 15 & 21 & 26 & 33 & 39 \\
              1 & 7 & 13 & 21 & 28 & 36 & 44 & 54 \\
              1 & 8 & 18 & 26 & 36 & 46 & 57 & 66 \\
              1 & 10 & 21 & 33 & 44 & 57 & 68 & 79 \\
              1 & 10 & 25 & 39 & 54 & 66 & 79 & 88 \\
            \end{array}
          \right) $\\
   &  \\
  $\det(\widehat P)$-basis && canonical basis \\
  $\left(
            \begin{array}{cccccccc}
              1 & 1 & 1 & 1 & 1 & 1 & 1 & 1 \\
              1 & 3 & 4 & 5 & 5 & 4 & 5 & 4 \\
              1 & 4 & 7 & 7 & 9 & 9 & 8 & 7 \\
              1 & 5 & 7 & 12 & 14 & 14 & 13 & 11 \\
              1 & 5 & 9 & 14 & 19 & 19 & 18 & 16 \\
              1 & 4 & 9 & 14 & 19 & 26 & 25 & 22 \\
              1 & 5 & 8 & 13 & 18 & 25 & 31 & 27 \\
              1 & 4 & 7 & 11 & 16 & 22 & 27 & 33 \\
            \end{array}
          \right) $& \qquad\qquad& $\left(
            \begin{array}{cccccccc}
              1 & 1 & 1 & 1 & 1 & 1 & 1 & 1 \\
              1 & 3 & 5 & 7 & 11 & 15 & 22 & 29 \\
              1 & 5 & 7 & 11 & 15 & 22 & 29 & 40 \\
              1 & 7 & 11 & 15 & 22 & 29 & 40 & 52 \\
              1 & 11 & 15 & 22 & 29 & 40 & 52 & 70 \\
              1 & 15 & 22 & 29 & 40 & 52 & 70 & 89 \\
              1 & 22 & 29 & 40 & 52 & 70 & 89 & 116 \\
              1 & 29 & 40 & 52 & 70 & 89 & 116 & 146 \\
            \end{array}
          \right)$ \\
\end{tabular}
\end{center}

\section{Generating formulas for the symmetric groups $S_n$ \label{Sn}}
The symmetric groups $S_n$ are not irreducible finite reflection group but they are strictly connected with the irreducible finite reflection groups of type $A_n$, and we will use them to obtain results for the groups $A_n$.\\
Let $e_1,\ldots,e_n$ be the unit vectors of a canonical basis of $\R^n$, and $x_1,\ldots,x_n$ be the corresponding coordinates of a vector $x\in\R^n$, so that: $x=\sum_{i=1}^n x_i e_i$.
$S_n$ acts on $\R^n$ by permuting in all possible ways the $n$ coordinates of the vectors $x\in\R^n$. Its order is $n!$. $S_n$ can be generated by the ${n}\choose{2}$ transpositions $t_{ij}$, $1\leq i<j\leq n,$ that exchange two coordinates $x_i$ and $x_j$ leaving all the others unchanged. 
The transpositions $t_{ij}$ are reflections of the space $\R^n$: $t_{ij}$ corresponds to the reflection about the hyperplane of equation $x_i-x_j=0$. Defining the vector $\alpha=e_i-e_j$ (or any non zero multiple of it), the action of $t_{ij}$ on any vector $x\in\R^n$ is given by Eq. (\ref{reflection0}). Hence, $S_n$ is a finite reflection group. The positive roots are for example the ${n\choose 2}$ vectors $e_i- e_j$, $\forall\, i<j=1,\ldots,n$.\\

A possible choice of the basis of invariant polynomials of $S_n$, is obtained by using the $n$
elementary symmetric polynomials:
\begin{eqnarray}\label{elemsymmpolSn}
\nonumber  p_1(x)=s_1(x_1,\ldots,x_n) &=& x_1+\ldots +x_n, \\
\nonumber  p_2(x)=s_2(x_1,\ldots,x_n)&=& x_1 x_2 + x_1 x_3 +\ldots + x_{n-1} x_n, \\
           p_3(x)=s_3(x_1,\ldots,x_n)&=& x_1 x_2 x_3 + x_1 x_2 x_4 +\ldots+ x_{n-2} x_{n-1} x_n, \\
\nonumber   &\vdots& \\
\nonumber  p_n(x)=s_n(x_1,\ldots,x_n) &=& x_1 x_2\cdots x_n.\\
\nonumber\end{eqnarray}

The degrees of the basic invariant polynomials are $1,2,\ldots,n$. 
It is clear that the elementary symmetric polynomials $s_a(x)$, $a=1,\ldots,n$, are invariant for all permutations of the coordinates $x_1,\ldots,x_n$ of $x$.

\noindent\begin{theorem}\label{lambdaSn}
\begin{enumerate}
  \item The \wPm\ corresponding to the basis of invariant polynomials of $S_n$ given in Eq. (\ref{elemsymmpolSn}), has matrix elements that can be obtained from the following generating formula:
\begin{equation}\label{PjkdiSn}
    {\widehat P}_{ab}(p)= [n+1-\min(a,b)]\;p_{a-1}\;p_{b -1}- \sum_{i=\max(1,a+b-n-1)}^{\min(a,b)} \,(a+ b - 2i)\;p_{i-1}\;p_{a+b - 1-i}\,,\quad \forall\,a,b=1,\ldots,n\,,\quad
\end{equation}
in which one has to consider $p_0=1$.\\
  \item The determinant of the \wPm\ $\widehat P(p)$ given by Eq. (\ref{PjkdiSn}) satisfies the boundary equation (\ref{boundaryeqndet}) with the following $\lambda$-vector:
$$\lambda^{(\det(\widehat P))}(p) = (\lambda^{(\det(\widehat P))}_1(p),\ldots,\lambda^{(\det(\widehat P))}_{n}(p))\,,$$
\begin{equation}\label{lambdaSnEq}
\lambda^{(\det(\widehat P))}_{a}(p)=-(n - a+2)(n - a+1)\,p_{a - 2}\,,\qquad \forall\,a=1,\ldots,n\,,
\end{equation}
in which one has to consider $p_{-1}=0$ and $p_0=1$.
\end{enumerate}
\end{theorem}
\noindent\textbf{Proof}. In Section \ref{proofSn}.\scat

Using Eq. (\ref{detP(x)}), we have a formula to express the discriminant $\Delta(x)$, that is the lowest degree invariant polynomial vanishing in the set of the reflecting hyperplanes of $S_n$, in terms of the elementary symmetric polynomials:
$$\Delta(x)=\prod_{i<j=1}^{n}(x_i-x_j)^2=\det(\widehat P(p(x)))\,,$$
in which $\widehat P(p)$ is the \wPm\ obtained with the generating formula (\ref{PjkdiSn}), and in which the variables $p_a$, $a=1,\ldots,n$, have been substituted with the elementary symmetric polynomials in Eq. (\ref{elemsymmpolSn}). A formula of this kind was, for example, invoked by Artin in Ref. \cite{Artin1991}, p. 548.

\section{Generating formulas for the groups of type $A_n$ \label{An}}
The symmetric group $S_{n+1}$ acts in $\R^{n+1}$ by permuting the $n+1$ variables $x_1,\ldots,x_{n+1}$, in all possible ways. This action is completely reducible. The hyperplane $H$ having equation $x_1+x_2+\ldots+ x_{n+1}=0$ is invariant under this action (and so is the line orthogonal to it). The group $A_n$ is the group acting in the $n$-dimensional subspace $H$ of $\R^{n+1}$ in the same way as $S_{n+1}$. As $S_{n+1}$ is a finite reflection group, $A_n$ is a finite reflection group. The positive roots of $A_n$, as vectors of $\R^{n+1}$, have the same expressions as the positive roots of $S_{n+1}$, and are, for example, the $n+1\choose 2$ vectors $e_i-e_j$, $\forall\,i<j=1,\ldots,n+1$. The order of $A_n$ is  $(n+1)!$, the same as that one of $S_{n+1}$. The basic invariant polynomials of $A_n$ for this action are the basic invariant polynomials of $S_{n+1}$, of degrees $2,3,\ldots,n+1$, for example those obtained from Eq. (\ref{elemsymmpolSn}), with $n$ replaced by $n+1$. One can verify that Eqs. (\ref{sommagradi}) and (\ref{prodottogradi}) are satisfied by these degrees. These invariant polynomials, are however expressed in terms of $n+1$ variables, and one wants to write them in terms of only the $n$ variables of the $n$-dimensional subspace $H\subset\R^{n+1}$ where $A_n$ acts effectively. To this goal one can consider the unit vector ${u}_{n+1}=(1,1,\ldots,1)/\sqrt{n+1}\in\R^{n+1}$, orthogonal to the invariant hyperplane $H$ and a rotation matrix $R_{n+1}\in O(n+1)$ that transforms $u_{n+1}$ into the canonical unit vector $e_{n+1}$. As a consequence of this rotation, the invariant subspace $H$ is mapped to the invariant subspace of $\R^{n+1}$ with equation $x_{n+1}=0$, and the invariant polynomials forming a basis of $S_{n+1}$ (and of $A_n$) change form, according to Theorem \ref{uno}, item \ref{th1it3}.
At this point if one puts $x_{n+1}=0$ in the expressions of the ``rotated'' basic invariant polynomials of $S_{n+1}$, the linear invariant becomes identically zero in $\R^{n+1}$, and the others become the $n$ basic invariant polynomials of $A_n$ expressed in terms of the $n$ variables $x_1,\ldots,x_n$, only.
A renaming of the basic invariant polynomials of $A_n$ obtained in this way is also convenient because one usually wants the index $i$, that labels the basic invariant polynomials of $A_n$, start at $1$ and end at $n$.
As the basic invariant polynomials of $A_n$ are obtained from those of $S_{n+1}$, one expects that also the \wPm\ of these basic invariant polynomials of $A_n$ can be obtained from that one of $S_{n+1}$, but this is not at all straightforward, as it can be seen from the proof of Theorem~\ref{lambdaAn}.\\

To put all this in a more formal way, also convenient for the proofs, let's introduce the following notation. A point of $\R^{n+1}$ will be indicated with $x$, and its projection in $\R^n$, obtained by taking $x_{n+1}=0$, with $\bar x$. The basic invariant polynomials of $S_{n+1}$, obtained from Eq. (\ref{elemsymmpolSn}), with $n$ replaced by $n+1$, are indicated with $s_a(x)$, $a=1,\ldots,n+1$. In the rotated coordinates $x'=R_{n+1}x$, because of Theorem~\ref{uno}, item \ref{th1it3}., the basic polynomials $s_a(x)$, $a=1,\ldots,n+1$, of $S_{n+1}$ become $s'_a(x')=s_a(R^\top_{n+1}x')$, and, in particular, $s'_1(x')\propto x'_{n+1}$. Then, by taking $x'_{n+1}=0$, that corresponds to take $s'_1(x')=0$, one finds $n$ basic invariant polynomials of $A_{n}$: $s'_{a+1}(\bar{x}')$, $\forall\,a=1,\ldots,n$, whose degrees are $d_a=a+1$, $\forall\,a=1,\ldots,n$.
There are however infinitely many ways to define a set of basic invariant polynomials for the group $A_n$. We require two main things. The first one is that we want the standard form of the quadratic invariant, that is, we want that $p_1(\bar x')=\sum_{i=1}^n {x'_i}^2$. The second one is that we want that the \wPm, that is originating from the basic invariant polynomials of $A_n$, has only integer coefficients. This will strongly reduce the possible choices of the basic invariant polynomials of $A_n$. One of these choices is the one presented in the following, in items 1., 2., 3. and 4..\\

Let's first define the matrix $R_{n+1}\in O(n+1)$ that transforms $u_{n+1}$ into $e_{n+1}$, that is used in the proof of Theorem \ref{lambdaAn}. In the definition below, it is written $R$ in place of $R_{n+1}$ to leave free space for the indices.\\
\begin{equation}\label{matRn+1}
R_{n+1}=R\,:\qquad\left\{
\begin{array}{rclll}
\nonumber  R_{k,i} &=& 1/\sqrt{k(k+1)}\,,\qquad & 1\leq k\leq n\,,\quad & 1\leq i\leq k\,, \\
\nonumber  R_{k,k+1} &=& -k/\sqrt{k(k+1)}\,,\qquad & 1\leq k\leq n\,,& \\
\nonumber  R_{k,i} &=& 0\,,\qquad & 1\leq k\leq n\,,\quad & k+1< i\leq n+1\,, \\
  R_{n+1,i} &=& 1/\sqrt{n+1}\,,\qquad & & 1\leq i\leq n+1\,. \\
\end{array}\right.
\end{equation}
To say it in words, the $k$-th row of
$R_{n+1}$ (with $1\leq k\leq n$) is the vector that has the first $k$
elements equal to 1, the next element equal to $-k$ and all other
elements equal to $0$, all multiplied by a normalization factor equal
to $1/\sqrt{k+k^2}$, and the last row of
$R_{n+1}$, the $(n+1)$-th row, has all its $n+1$ elements equal to $1/\sqrt{n+1}$.\\
By construction all row vectors are
mutually orthogonal and normalized to 1, and this is sufficient to say that the matrix $R_{n+1}$
is orthogonal. An easy calculation shows that its
determinant is 1, so it is a proper rotation matrix. To see this substitute column 2 with the sum of column 2 and column 1, substitute column 3 with the sum of column 3 and the new column 2, and so on. The determinant does not change for these substitutions and the resulting matrix is a lower triangular matrix with its diagonal elements equal to $d_k=k/\sqrt{k^2+k}=\sqrt{k/(k+1)}$, $\forall \,k=1,\ldots,n$, and $d_{n+1}=(n+1)/\sqrt{n+1}=\sqrt{n+1}$, so its determinant is equal to $\prod_{k=1}^{n+1}d_k=\sqrt{n+1}\,\prod_{k=1}^{n}\sqrt{k/(k+1)}=1$.\\

We can now define the basic invariant polynomials of $A_n$ we shall use in Theorem \ref{lambdaAn} below.\\

1. Start from the $n+1$ elementary symmetric polynomials $s_1(x),\ldots,s_{n+1}(x)$, obtained from Eq. (\ref{elemsymmpolSn}), with $n$ replaced by $n+1$, that form a basis for the action of the group $S_{n+1}$ in $\R^{n+1}$:
\begin{eqnarray}\label{elemsymmpolSn+1}
\nonumber  s_1(x) &=& x_1+\ldots +x_{n+1}, \\
\nonumber  s_2(x)&=& x_1 x_2 + x_1 x_3 +\ldots + x_{n} x_{n+1}, \\
  s_3(x)&=& x_1 x_2 x_3 + x_1 x_2 x_4 +\ldots+ x_{n-1} x_{n} x_{n+1}, \\
\nonumber   &\vdots& \\
\nonumber  s_{n+1}(x) &=& x_1 x_2\cdots x_{n+1}.\\
\nonumber\end{eqnarray}

2. Rotate the system of coordinates with the matrix $R_{n+1}$, that is $x'=R_{n+1}\,x$. Theorem \ref{uno}, item \ref{th1it3}, states that the basic invariant polynomials of $S_{n+1}$ in the rotated system of reference have the following expressions:
\begin{equation}\label{trasfAn1}
  s'_a(x')=
  s_a(R^\top_{n+1}x')\,,\qquad a=1,\ldots,n+1\,.
\end{equation}
With this transformation the polynomial $s'_1(x')$ becomes proportional to $x'_{n+1}$. Any orthogonal matrix $R_{n+1}$ transforming $u_{n+1}$ into $e_{n+1}$ can be used. In the proofs we shall use the matrix $R_{n+1}$ defined in Eq. (\ref{matRn+1}). All the other orthogonal matrices transforming $u_{n+1}$ into $e_{n+1}$ differ from the given $R_{n+1}$ by orthogonal transformations mixing the first $n$ variables, only. Theorem \ref{uno}, item \ref{th1it4}., ensures that the \wPm\ corresponding to the basic invariant polynomials $s_a'(x')$, $a=1,\ldots,n+1$, has the same form of the \wPm\ (given by Eq. (\ref{PjkdiSn})) corresponding to the basic polynomials $s_a(x)$, $a=1,\ldots,n+1$, and differs only for the substitutions $s_a\to s'_a$, $a=1,\ldots,n+1$.
It is clear from Theorem \ref{uno} that any other orthogonal matrix of order $n+1$ transforming $u_{n+1}$ into $e_{n+1}$ would give different expressions for the basic invariant polynomials but the same form of the \wPm.\\

3. Perform a scale transformation on the $n+1$ basic invariant polynomials $s'_1(x'),\ldots,s'_{n+1}(x')$ in the following way:
\begin{equation}\label{trasfAn2}
  t_a(x')=-2\, \sqrt{(n+1)^{a-2}}\,s'_{a}(x')\,,\qquad a=1,\ldots,n+1\,.
\end{equation}
The transformations (\ref{trasfAn1}) and (\ref{trasfAn2}) can be done in the reverse order without changing the resulting polynomials $t_a(x')$, $\forall\, a=1,\ldots,n$. The scale transformation (\ref{trasfAn2}) is necessary only if one wants the quadratic invariant polynomial of the standard form (\ref{quadraticinv}) and a \wPm\ with only integer coefficients.\\

4. Put $x'_{n+1}=0$ in the arguments of the basic polynomials $t_a(x')$, $\forall\, a=1,\ldots,n+1$, that is, take the restrictions of the basic polynomials $t_a(x')$, $\forall\, a=1,\ldots,n+1$, to the hyperplane $x'_{n+1}=0$. One finds $t_1(\bar x')=\left.t_1(x')\right|_{x'_{n+1}=0}=0$, and the remaining $n$ basic invariant polynomials, $t_2(\bar x'),\ldots,t_{n+1}(\bar x')$, that are expressed in terms of the $n$ variables in $\bar x'$, only, form the $n$ basic invariant polynomials of $A_n$.
Rename and renumber these basic invariant polynomials of $A_n$ in the following way:
\begin{equation}\label{basicpolynAn}
  p_a(\bar x')=t_{a+1}(\bar x')=\left.t_{a+1}(x')\right|_{x'_{n+1}=0}\,,\qquad \forall\,a=1,\ldots,n\,.
\end{equation}
 In Example 1, in Section \ref{examples}, it is shown how to write the explicit form of the matrix $R_4$, and how one can determine the basic invariant polynomials of $A_3$ from those of $S_4$, following the rules written above.

\noindent\begin{theorem}\label{lambdaAn}
\begin{enumerate}
  \item The \wPm\ corresponding to the basis of invariant polynomials of $A_n$ obtained by Eqs. (\ref{elemsymmpolSn+1})--(\ref{basicpolynAn}), has matrix elements that can be obtained from the following generating formula:
 $${\widehat P}_{ab}(p)=[(n+1)\,\min(a,b)-a\,b\,]\;p_{a-1}\;p_{b -1}+2\,(a+b)\,p_{a+b-1}\,+$$
 \begin{equation}\label{PjkdiAn}
 -\,(n+1)\, \sum_{i=\max(2,a+b-n-1)}^{\min(a,b)-1} \,(a+ b - 2i)\;p_{i-1}\;p_{a+b - 1-i}\,,\qquad \forall\,a,b=1,\ldots,n\,,
\end{equation}
where $p_0=0$ and $p_a=0$, $\forall\,a>n$. The matrix ${\widehat P}(p)$ obtained with Eq. (\ref{PjkdiAn}) is symmetric.\\
  \item The determinant of the \wPm\ ${\widehat P}(p)$, given by Eq. (\ref{PjkdiAn}), satisfies the boundary equation (\ref{boundaryeqndet}) with the following $\lambda$-vector:
$$\lambda^{(\det(\widehat P))}(p) = (2\,n\, (n + 1),\,\lambda^{(\det(\widehat P))}_2(p),\,\ldots,\,\lambda^{(\det(\widehat P))}_{n}(p))\,,$$
\begin{equation}\label{lambdaAnEq}
\lambda^{(\det(\widehat P))}_{a}(p)=- (n + 1)(n - a+2)(n - a+1)\,p_{a - 2}\,,\qquad \forall\,a=2,\ldots,n\,,
\end{equation}
in which one has to consider $p_0=0$ (that implies $\lambda^{(\det(\widehat P))}_{2}(p)=0$).
\end{enumerate}
\end{theorem}
\noindent\textbf{Proof}. In Section \ref{proofAn}.\scat

Using Eq. (\ref{detP(x)}), we have a formula to express the discriminant $\Delta(x)$, that is the lowest degree invariant polynomial vanishing in the set of the reflection hyperplanes of $A_n$, in terms of the basic invariant polynomials of $A_n$:
$$\Delta(x)=\det(\widehat P(p(x)))\,,$$
in which $\widehat P(p)$ is the \wPm\ obtained with the generating formula (\ref{PjkdiAn}), and in which the variables $p_a$, $a=1,\ldots,n$, have been substituted with the basic invariant polynomials $p_1(x),\ldots,p_n(x)$, of $A_n$, obtained in Eqs. (\ref{elemsymmpolSn+1})--(\ref{basicpolynAn}). The polynomial $\Delta(x)$, as calculated above, is unique, modulo constant factors.\\

A generating formula like Eq. (\ref{PjkdiAn}) was written without proof in Remark 3.9 of Ref. \cite{arn1976}, for a basis $q_a(x)$, $a=1,\ldots,n$, that differs from the basis $p_a(x)$, $a=1,\ldots,n$, defined above, only for scalar factors, precisely:
$$q_a(x)=(-1)^a\,\left[2(n+1)^{\frac{a-1}{2}}\right]^{-1}\,p_a(x)\,,\qquad \forall\,a=1,\ldots,n\,.$$

An equation that differ very little from Eq. (\ref{lambdaAnEq}), was written in Eq. (2.1.5.2) of Ref. \cite{YaSek1981}, and proved in a different manner.

\section{Generating formulas for the groups of type $B_n$ \label{Bn}}
Let $e_1,\ldots,e_n$ be the unit vectors of a canonical basis of $\R^n$. A set of simple roots for $B_n$ is the following one:
$$\alpha_i=e_i-e_{i+1},\quad \forall\,i=1,\ldots,n-1,\qquad \alpha_n=e_n\,.$$
A set of simple roots for $C_n$ differs from that one for $B_n$ only for the length of $\alpha_n$: instead to be a factor $\sqrt{2}$ shorter than the others, is a factor $\sqrt{2}$ longer than the others. One can then take the last root of $C_n$ equal to $\alpha_n=2\,e_n$.
The $n$ simple roots determine $n$ reflections with the formula (\ref{reflection0}).
The $n$ reflections obtained from the simple roots of $C_n$ are the same as those obtained from the simple roots of $B_n$. By multiplying in all possible ways these $n$ reflections one obtains the reflection group $B_n$. Then, the simple roots of $B_n$ and the simple roots of $C_n$ determine the same finite reflection group $B_n$. The reflections determined by the simple roots $\alpha_i$, $\forall \, i=1,\ldots,n-1$, when applied to a general vector $x\in\R^n$ exchange the coordinates $x_i$ and $x_{i+1}$ of $x$, leaving all the other coordinates unchanged, while the reflection determined by $\alpha_n$ changes the sign of the coordinate $x_n$ of $x$. $B_n$ then acts on $\R^n$ by permuting in all possible ways the $n$ variables $x_1,\ldots,x_n$ and by changing in all possible ways their signs. The order of $B_n$ is then ${\rm ord}(B_n)=n!\,2^n$. One of the possibilities is to consider the vectors $e_i$, $i=1,\ldots,n$, as the positive short roots, and the vectors $e_i\pm e_j$, $1\leq i<j\leq n$ as the positive long roots.\\

A possible choice of the basis of invariant polynomials of $B_n$, already proposed by Coxeter in \cite{cox1951}, is obtained by using the $n$
elementary symmetric polynomials in the variables $x_i^2$, that is:
\begin{eqnarray}{\label{basicpolynBn}}
 \nonumber  p_1(x)=s_1(x_1^2,\ldots,x_n^2) &=& x_1^2+\ldots +x_n^2, \\
 \nonumber  p_2(x)=s_2(x_1^2,\ldots,x_n^2)&=& x_1^2 x_2^2 + x_1^2 x_3^2 +\ldots + x_{n-1}^2 x_n^2, \\
  p_3(x)=s_3(x_1^2,\ldots,x_n^2)&=& x_1^2 x_2^2 x_3^2 + x_1^2 x_2^2 x_4^2 +\ldots+ x_{n-2}^2 x_{n-1}^2 x_n^2, \\
 \nonumber   &\vdots& \\
  \nonumber p_n(x)=s_n(x_1^2,\ldots,x_n^2) &=& x_1^2 x_2^2\cdots x_n^2.
\end{eqnarray}\\
The degrees of the basic invariant polynomials are then $2,4,\ldots,2n$, and one can verify they satisfy Eqs. (\ref{sommagradi}) and (\ref{prodottogradi}). It is also clear that the polynomials $p_a(x)$, $a=1,\ldots,n$, defined in Eq. (\ref{basicpolynBn}), are invariant for permutations and sign changes of the variables.

\noindent\begin{theorem}\label{lambdaBn}
\begin{enumerate}
  \item The \wPm\ corresponding to the basis of invariant polynomials of $B_n$ given in Eq. (\ref{basicpolynBn}), has matrix elements that can be obtained from the following generating formula:
\begin{equation}\label{PjkdiBn}
  {\widehat P}_{ab}(p)=\sum_{i=\max(0,a+b-n-1)}^{\min(a,b)-1}\;4\,(a+b-1-2i)\;p_i\;p_{a+b-1-i}\,,\qquad \forall\,a,b=1,\ldots,n\,,
\end{equation}
in which one has to consider $p_0=1$.  The matrix ${\widehat P}(p)$ obtained with Eq. (\ref{PjkdiBn}) is symmetric.\\
  \item The determinant $\det(\widehat P(p))$ of the \wPm\ $\widehat P(p)$ obtained from Eq. (\ref{PjkdiBn}) has two active factors. One active factor is the polynomial $s(p)\equiv p_n$, representing the highest degree basic invariant polynomial $p_n(x)$, as well as the invariant polynomial square of the product of the $n$ linear forms defining the reflection hyperplanes corresponding to the short roots. The other active factor is the polynomial $l(p)$, representing the invariant polynomial square of the product of the $2 {n\choose 2}$ linear forms defining the reflection hyperplanes corresponding to the long roots.\\
 $s(p)\;(\equiv p_n)$  satisfies the boundary equation (\ref{boundaryeqn}) with the following $\lambda$-vector:
\begin{equation}\label{lambdaBnEq1}
  \lambda^{(s)}(p) = (\lambda^{(s)}_1(p),\ldots,\lambda^{(s)}_n(p)),\qquad \lambda^{(s)}_a(p)=4\,(n - a + 1)\,p_{a - 1}\,,\qquad \forall\,a=1,\ldots,n\,,
\end{equation}
 $l(p)$  satisfies the boundary equation (\ref{boundaryeqn}) with the following $\lambda$-vector:
\begin{equation}\label{lambdaBnEq2}
  \lambda^{(l)}(p) = (\lambda^{(l)}_1(p),\ldots,\lambda^{(l)}_n(p)),\qquad \lambda^{(l)}_a(p)=4\,(n - a + 1)(n-a)\,p_{a - 1}\,,\qquad \forall\,a=1,\ldots,n\,,
\end{equation}
 $\det(\widehat P(p))$  satisfies the boundary equation (\ref{boundaryeqndet}) with the following $\lambda$-vector:
\begin{equation}\label{lambdaBnEq3}
  \lambda^{(\det(\widehat P))}(p) = (\lambda^{(\det(\widehat P))}_1(p),\ldots,\lambda^{(\det(\widehat P))}_n(p)),\qquad \lambda^{(\det(\widehat P))}_a(p)=4\,(n - a + 1)^2\,p_{a - 1}\,,\qquad \forall\,a=1,\ldots,n\,,
\end{equation}
where, in Eqs. (\ref{lambdaBnEq1}), (\ref{lambdaBnEq2}) and (\ref{lambdaBnEq3}), one has to consider $p_0=1$.
\end{enumerate}
\end{theorem}
\noindent\textbf{Proof}. In Section \ref{proofBn}.\scat

Using Eq. (\ref{detP(x)}), we have a formula to express the discriminant $\Delta(x)$, that is the lowest degree invariant polynomial vanishing in the set of the reflection hyperplanes of $B_n$, in terms of the basic invariant polynomials of $B_n$:
$$\Delta(x)=\prod_{i=1}^{n}x_i^2\;\prod_{i<j=1}^{n}(x_i^2-x_j^2)^2=s(p(x))\;l(p(x))=4^{-n}\;\det(\widehat P(p(x)))\,,$$
in which $\widehat P(p)$ is the matrix obtained with the generating formula (\ref{PjkdiBn}), and in which the variables $p_a$, $a=1,\ldots,n$ have been substituted with the basic invariant polynomials in Eq. (\ref{basicpolynBn}). Here, $s(p(x))$ ($\equiv p_n(x)$) and $l(p(x))$ are the two irreducible factors of $\det(\widehat P(p(x)))$.\\

Eq. (\ref{PjkdiBn}) was first written in Eq. (**) of Ref. \cite{giv1980} and proved in different manner, using techniques common in singularity theory.

An equation that differs very little from Eq. (\ref{lambdaBnEq3}) was first written in Eq. (2.2.5.2) of Ref. \cite{YaSek1981} and proved in a different manner.

\section{Generating formulas for the groups of type $D_n$ \label{Dn}}
Let $e_1,\ldots,e_n$ be the unit vectors of a canonical basis of $\R^n$. A set of simple roots for $D_n$ is the following one:
$$\alpha_i=e_i-e_{i+1},\quad \forall\,i=1,\ldots,n-1,\qquad \alpha_n=e_{n-1}+e_n\,.$$
The given set of simple roots of $D_n$ differs from that one of $B_n$ only for the last root $\alpha_n$.
The $n$ simple roots of $D_n$ determine $n$ reflections using Eq. (\ref{reflection0}).
It is clear that the $n-1$ reflections obtained from the simple roots $\alpha_1,\ldots,\alpha_{n-1}$ are the same as those of $B_n$, and only the reflection relative to $\alpha_n$ differs from the corresponding one of $B_n$. The reflection determined by the simple root $\alpha_i$, $\forall \, i=1,\ldots,n-1$, when applied to a general vector $x\in\R^n$ exchanges the coordinates $x_i$ and $x_{i+1}$ of $x$, leaving all other coordinates unchanged, while the reflection determined by $\alpha_n$ exchanges the coordinates $x_{n-1}$ and $x_{n}$ of $x$ and changes their signs, leaving all other coordinates unchanged. By multiplying in all possible ways these $n$ reflections one obtains the reflection group $D_n$. It acts on $\R^n$ by permuting in all possible ways the $n$ variables $x_1,\ldots,x_n$ and by changing in all possible ways an even number of their signs. The order of $D_n$ is then ${\rm ord}(D_n)=n!\,2^{n-1}$.
The $2{n\choose 2}$ positive roots can be taken, for example, to be $e_i\pm e_j$, $\forall\,i,j=1,\ldots,n$, $i<j$.\\

A possible choice of the basis of invariant polynomials of $D_n$, already proposed by Coxeter in \cite{cox1951}, is obtained by using the $n-1$ lowest degrees
elementary symmetric polynomials in the variables $x_i^2$, and the elementary symmetric polynomial of degree $n$ in the variables $x_i$, that is:
\begin{eqnarray}{\label{basicpolynDn}}
 \nonumber p_1(x)=s_1(x_1^2,\ldots,x_n^2) &=& x_1^2+\ldots +x_n^2, \\
 \nonumber  p_2(x)=s_2(x_1^2,\ldots,x_n^2)&=& x_1^2 x_2^2 + x_1^2 x_3^2 +\ldots + x_{n-1}^2 x_n^2, \\
  p_3(x)=s_3(x_1^2,\ldots,x_n^2)&=& x_1^2 x_2^2 x_3^2 + x_1^2 x_2^2 x_4^2 +\ldots + x_{n-2}^2 x_{n-1}^2 x_n^2, \\
 \nonumber   &\vdots& \\
 \nonumber  p_{n-1}(x)=s_{n-1}(x_1^2,\ldots,x_n^2) &=& x_1^2 x_2^2\cdots x_{n-1}^2+\ldots +x_2^2 x_3^2\cdots x_{n}^2.\\
 \nonumber  p_n(x)=
  s_n(x_1,\ldots,x_n) &=& 
  x_1 x_2\cdots x_n.
\end{eqnarray}\\
The degrees of the basic invariant polynomials are then $2,4,\ldots,2(n-1),n$, and one can verify they satisfy Eqs. (\ref{sommagradi}) and (\ref{prodottogradi}).  If $n$ is even there are then 2 different basic polynomials of degree $n$: $p_{n/2}(x)$ and $p_n(x)$, using the labeling in Eq. (\ref{basicpolynDn}). It is clear that the polynomials $p_a(x)$, $a=1,\ldots,n$, are invariant for permutations of the variables and for an even number of sign changes of the variables. The labeling of the basic invariant polynomials given in Eq. (\ref{basicpolynDn}) is not the standard one based on the degrees, but it turns out to be more convenient for what we are going to do.\\ These basic invariant polynomials differ from those of $B_n$, given in Eq. (\ref{basicpolynBn}), only for the last one, but not very much: the square of $p_n(x)$ of $D_n$ is 
the basic invariant $p_n(x)$ of $B_n$, that is: $p_n^{B}(x)=({p_n^{D}(x)})^2$, using the notation, also used in the proof, with the indices $B$ and $D$ telling us the group to which $p_n(x)$ refers.

\noindent\begin{theorem}\label{lambdaDn}
\begin{enumerate}
  \item The \wPm\ corresponding to the basis of invariant polynomials of $D_n$, given in Eq. (\ref{basicpolynDn}), has matrix elements that can be obtained from the following generating formula:
%
%
\begin{equation}\label{PjkdiDn_divisa}
\left\{\begin{array}{rcll}
\nonumber  {\widehat P}_{ab}(p)&=& \left.\sum_{i=\max(0,a+b-n-1)}^{\min(a,b)-1}\;4\,(a+b-1-2i)\;p_i\;p_{a+b-1-i}\,\right|_{p_n\to p_n^2} & \quad \forall\,a,b=1,\ldots,n-1\,,\\
\nonumber   {\widehat P}_{an}(p)={\widehat P}_{na}(p) &=& 2\,(n-a+1)\,p_{a - 1}\,p_n & \quad \forall\,a=1,\ldots,n-1\,,\\
  {\widehat P}_{nn}(p) &=& p_{n-1}\,, &
\end{array}\right.
\end{equation}
in which one
has to consider $p_0=1$. The matrix ${\widehat P}(p)$ obtained with Eq. (\ref{PjkdiDn_divisa}) is symmetric.\\
  \item The determinant of the \wPm\ ${\widehat P}(p)$, given by Eq. (\ref{PjkdiDn_divisa}), satisfies the boundary equation (\ref{boundaryeqndet}) with the following $\lambda$-vector:
$$\lambda^{(\det(\widehat P))}(p) = (\lambda^{(\det(\widehat P))}_1(p),\ldots,\lambda^{(\det(\widehat P))}_{n-1}(p),0)\,,$$
\begin{equation}\label{lambdaDnEq}
  \lambda^{(\det(\widehat P))}_a(p)=4(n-a + 1)(n - a)\,p_{a - 1}\,,\qquad \forall\,a=1,\ldots,n-1\,,
\end{equation}
in which one
has to consider $p_0=1$.\\

\end{enumerate}
\end{theorem}
\noindent\textbf{Proof}. In Section \ref{proofDn}.\scat

Using Eq. (\ref{detP(x)}), we have a formula to express the discriminant $\Delta(x)$, that is the lowest degree invariant polynomial vanishing in the set of the reflection hyperplanes of $D_n$, in terms of the basic invariant polynomials of $D_n$:
$$\Delta(x)=\prod_{i<j=1}^{n}(x_i^2-x_j^2)^2=4^{1-n}\;\det(\widehat P(p(x)))\,,$$
in which $\widehat P(p)$ is the matrix obtained with the generating formula (\ref{PjkdiDn_divisa}), and in which the variables $p_a$, $a=1,\ldots,n$ have been substituted with the basic invariant polynomials in Eq. (\ref{basicpolynDn}).\\

Eq. (\ref{PjkdiDn_divisa}) was first written in Ref. \cite{giv1980}, at the end of p. 88. It was obtained from the analogous expression for the groups of type $B_n$, Eq. (**) of Ref. (\cite{giv1980}), in essentially the same way of how Eq. (\ref{PjkdiDn_divisa}) is here obtained from Eq. (\ref{PjkdiBn}).

An equation that differs very little from Eq. (\ref{lambdaDnEq}) was first written in Eq. (2.3.5.2) of Ref. \cite{YaSek1981} and proved in a different manner. 

\section{Hankel matrix decompositions of the P-matrices of $A_n$, $B_n$, $D_n$\label{hankel}}
The \wPms\ obtained from the generating formulas (\ref{PjkdiAn}), (\ref{PjkdiBn}) and (\ref{PjkdiDn_divisa}), for the groups of type $A_n$, $B_n$ and $D_n$, respectively, have an interesting Hankel matrix decomposition, that will be pointed out in this section.

Let's recall the definition of a Hankel matrix. A Hankel matrix is a matrix that has equal elements along a same upgoing diagonal. For a Hankel matrix $H$ one then has the following equality among matrix elements:
$$H_{i,j}=H_{i-1,j+1}$$
In other words, the elements $H_{i,j}$ of a Hankel matrix $H$ depend only on the sum $i+j$, and the matrix elements with equal sum of the indices are equal. A square Hankel matrix of order $n$ is completely determined by the $2n-1$ elements located in the first column and in the last row of $H$. We will need only Hankel matrices with zero elements above the auxiliary diagonal (the {\em auxiliary diagonal} of a square matrix $H$ is the diagonal going from the element $H_{n,1}$ to the element $H_{1,n}$). Let's use the symbol $H(v)$ to define a square Hankel matrix that has its last row equal to the vector $v$, and is zero above the auxiliary diagonal. If $n$ is the dimension of $v$, then $n$ is the order of $H(v)$.

With the groups $B_n$ and $D_n$ we will need the matrices $H_n(p)$ and $H_a^{(n)}(p)$. $H_n(p)$ is defined as follows:
\begin{equation}\label{hankelmatrixhn}
  H_n(p)=H(v(p))\,,\qquad v(p)=(v_1(p),v_2(p),\ldots,v_n(p))\,,\qquad v_a(p)=(n-a+1)\,p_{a-1}\,,\qquad \forall\,a=1,\ldots,n\,,
\end{equation}
in which one has to consider $p_0=1$, and $H_a^{(n)}(p)$, with $a\leq n$, is the $n\times n$ block Hankel matrix that has the Hankel matrix $H_a(p)$ in its upper left corner, and is zero elsewhere. For example,
$$H_5(p)=\left(
        \begin{array}{ccccc}
          0 & 0 & 0 & 0& 5\\
          0 & 0 & 0 & 5& 4 p_1\\
          0 & 0 & 5 & 4  p_1& 3 p_2\\
          0 & 5 & 4 p_1 & 3 p_2& 2 p_3\\
          5 & 4p_1 & 3 p_2 & 2 p_3 &p_4\\
        \end{array}
      \right),\qquad H_3^{(5)}(p)=\left(
                             \begin{array}{ccccc}
                               0 & 0 & 3 & 0 & 0 \\
                               0 & 3 & 2p_1 & 0 & 0 \\
                               3 & 2p_1 &p_2 & 0 & 0 \\
                               0 & 0 & 0 & 0 & 0 \\
                               0 & 0 & 0 & 0 & 0 \\
                             \end{array}
                           \right)
$$\\

For the groups $B_n$, one proves the following theorem.

\noindent\begin{theorem}\label{hankelBn}
If the basic invariant polynomials of $B_n$ are those given in Eq. (\ref{basicpolynBn}), the corresponding \wPm, that can be determined using Eq. (\ref{PjkdiBn}), can be written in the following way:
\begin{equation}\label{PBnHankel}
  \widehat P(p)=4\,\sum_{a=1}^n\,p_a\,H_a^{(n)}(p)
\end{equation}
\end{theorem}
\noindent\textbf{Proof}. In Section \ref{proofhankelBn}.\scat

For the groups of type $D_n$ there is a similar theorem expressing their \wPms\ in terms of Hankel matrices, but some definitions have to be first given. The \wPm\ $\widehat P(p)$, given in Eq. (\ref{PjkdiDn_divisa}), corresponding to the basic invariant polynomials of $D_n$ given in Eq. (\ref{basicpolynDn}), distinguishes the elements in the last row and column of $\widehat P(p)$ from the others. For this reason, it is convenient to define the symbol $I^{(n)}$ for the $n\times n$ diagonal matrix with the unit matrix $I_{n-1}$, of order $n-1$, in its upper left corner and is zero in all its last row and column, and the symbol $I_0^{(n)}$ for the matrix $I_0^{(n)}=I_n-I^{(n)}$, where $I_n$ is the unit matrix of order $n$. The diagonal of the matrix $I^{(n)}$ has $n-1$ 1's and a 0 in its last element, vice versa, the diagonal of the matrix $I_0^{(n)}$ has $n-1$ 0's and a 1 in its last element. For example:
$$I^{(5)}=\left(
            \begin{array}{ccccc}
              1 & 0 & 0 & 0& 0 \\
              0 & 1 & 0 & 0& 0 \\
              0 & 0 & 1 & 0& 0 \\
              0 & 0 & 0 & 1& 0 \\
              0 & 0 & 0 & 0& 0 \\
            \end{array}
          \right)\,,\qquad
         I_0^{(5)}= \left(
            \begin{array}{ccccc}
              0 & 0 & 0 & 0& 0 \\
              0 & 0 & 0 & 0& 0 \\
              0 & 0 & 0 & 0& 0 \\
              0 & 0 & 0 & 0& 0 \\
              0 & 0 & 0 & 0& 1 \\
            \end{array}
          \right)\,.
          $$
The matrices $I^{(n)}$ and $I_0^{(n)}$ can be used to define the matrices $H_t^{(n)}(p)$ and $H_0^{(n)}(p)$ in the following way:
$$H_t^{(n)}(p)=I^{(n)}\,H_n(p)\,I^{(n)}\,,\qquad\qquad H_0^{(n)}(p)=H_n(p)-H_t^{(n)}(p)-p_n\,I_0^{(n)}\,.$$ The matrix $H_t^{(n)}(p)$ has the same elements of the matrix $H_n(p)$ but with zero's in the last row and column, and $H_0^{(n)}(p)$ contains the non-diagonal elements of the last row and column of $H_n(p)$ and is zero elsewhere. For example:
$$ H_t^{(5)}(p)=\left(
            \begin{array}{ccccc}
              0 &0 & 0 & 0 & 0 \\
              0 &0 & 0 & 5 & 0 \\
              0 &0 & 5 & 4\,p_1 & 0 \\
              0 &5 & 4\,p_1 & 3\,p_2 & 0 \\
              0 &0 & 0 & 0 & 0 \\
            \end{array}
          \right)\,,\qquad
H_0^{(5)}(p)=\left(
            \begin{array}{ccccc}
              0 &0 & 0 & 0 & 5 \\
              0 &0 & 0 & 0 & 4\,p_1 \\
              0 &0 & 0 & 0 & 3\,p_2 \\
              0 &0 & 0 & 0 & 2\,p_3 \\
              5 & 4\,p_1 & 3\,p_2 & 2 p_3&0 \\
            \end{array}
          \right)\,.
$$

For the groups $D_n$, we can now give the following theorem.

\noindent\begin{theorem}\label{hankelDn}
If the basic invariant polynomials of $D_n$ are those given in Eq. (\ref{basicpolynDn}), the corresponding \wPm, that can be calculated from Eq. (\ref{PjkdiDn_divisa}), can be written in the following way:
\begin{equation}\label{PDnHankel}
  \widehat P(p)=4\,\sum_{a=1}^{n-1}\,p_a\,H_a^{(n)}(p)+4\,p_n^2\,H_t^{(n)}(p)+2\,p_n\,H_0^{(n)}+p_{n-1}\,I_0^{(n)}\,.
\end{equation}\\
\end{theorem}
\noindent\textbf{Proof}. In Section \ref{proofhankelDn}.\scat

The \wPm\ of $A_n$, corresponding to the basic invariant polynomials defined by Eqs. (\ref{elemsymmpolSn+1})--(\ref{basicpolynAn}), that can be constructed using
Eq. (\ref{PjkdiAn}), can be written in terms of block Hankel matrices, too, but it is first convenient to give the following definitions:
$$Y_n=\,H(e_1)\,,\qquad e_1=(1,0,\ldots,0)\,,\qquad\mbox{($n-1$ zero's)}\,$$
$$K_n(p)=\,H(u(p))\,,\qquad u(p)=(u_1(p),\ldots,u_{n}(p))\,,\qquad u_a(p)=v_{a-1}(p)=(n-a+2)\,p_{a-2}\,,\qquad \forall\,a=1,\ldots,n\,,$$
where one has, by definition, $p_0=p_{-1}=0$, (so $u_1(p)=u_2(p)=0$). Note that $K_n(p)$ can be obtained from $H_n(p)$ by considering $p_0=0$ instead of $p_0=1$ (and this means a matrix equal to $H_n(p)$ but with 0's along its auxiliary diagonal), and by shifting all columns of the resulting matrix one position to the right or, by shifting all rows of the resulting matrix one position towards the bottom, if one prefers).\\
Both $Y_n$ and $K_n(p)$ are $n\times n$ Hankel matrices, with $Y_n$ that has $1$'s along the auxiliary diagonal and zero elsewhere.\\
Let's define $Y_a^{(n)}$ and $K_a^{(n)}(p)$, with $a\leq n$, be the $n\times n$ block Hankel matrices that have the Hankel matrices $Y_a$ and $K_a(p)$ in their upper left corner, and are zero elsewhere.\\
For example,
$$K_5(p)=\left(
        \begin{array}{ccccc}
          0 & 0 & 0 & 0& 0\\
          0 & 0 & 0 & 0& 0\\
          0 & 0 & 0 & 0& 4 p_1\\
          0 & 0 & 0 & 4 p_1& 3 p_2\\
          0 & 0 & 4 p_1 & 3 p_2 &2 p_3\\
        \end{array}
      \right),\qquad K_4^{(5)}(p)=\left(
                             \begin{array}{ccccc}
                               0 & 0 & 0 & 0 & 0 \\
                               0 & 0 & 0 & 0 & 0 \\
                               0 & 0 &0 &  3p_1 & 0 \\
                               0 & 0 &   3p_1 & 2 p_2 & 0 \\
                               0 & 0 & 0 & 0 & 0 \\
                             \end{array}
                           \right),\qquad Y_4^{(5)}=\left(
                             \begin{array}{ccccc}
                               0 & 0 & 0 & 1 & 0 \\
                               0 & 0 & 1 & 0 & 0 \\
                               0 & 1 &0 &  0 & 0 \\
                               1 & 0 &   0 & 0 & 0 \\
                               0 & 0 & 0 & 0 & 0 \\
                             \end{array}
                           \right)
$$
Let's also define the $n\times n$ symmetric (non-Hankel) matrix $R^{(n)}(p)$ in the following way:
\begin{equation}\label{matRhankel}
  R^{(n)}_{ab}(p)=[(n+1)\,\min(a,b)-a\,b\,]\,p_{a-1}\,p_{b-1}\,,\qquad a, b=1,\ldots,n\,,
\end{equation}
in which one has to consider $p_0=0$.
For example,
$$
 R^{(5)}(p)=\left(
                             \begin{array}{ccccc}
                               0 & 0 & 0 & 0 & 0 \\
                               0 & 8p_1^2 & 6p_1 p_2 & 4p_1p_3 & 2p_1p_4 \\
                               0 & 6p_1 p_2 &9p_2^2 & 6 p_2p_3 & 3p_2p_4 \\
                               0 & 4p_1p_3 & 6 p_2p_3 & 8 p_3^2 & 4p_3p_4 \\
                               0 & 2p_1p_4 & 3p_2p_4 & 4p_3p_4 & 5p_4^2 \\
                             \end{array}
                           \right)
$$

For the groups $A_n$ one proves the following theorem.

\noindent\begin{theorem}\label{hankelAn}
If the basic invariant polynomials of $A_n$ are those defined by Eqs. (\ref{elemsymmpolSn+1})--(\ref{basicpolynAn}), the corresponding \wPm, that can be calculated using Eq. (\ref{PjkdiAn}), can be written in the following way:
\begin{equation}\label{PAnHankel}
  \widehat P(p)=R^{(n)}(p)+2\,\sum_{a=1}^n\,(a+1)\,p_a\,Y_a^{(n)}
  -(n+1)\,\sum_{a=1}^n\,p_a\,K_a^{(n)}(p)
\end{equation}
\end{theorem}

\noindent\textbf{Proof}. In Section \ref{proofhankelAn}.\scat

Regarding the groups of type $S_n$, I could not find any basis of invariant polynomials such that the corresponding \wPms\ have a simple decomposition in terms of Hankel matrices. The problem arises with the variable $p_1$, of weight 1, that cannot be written in the \wPm\ of $S_n$ in a simple ``symmetric'' way. Obviously, there exist a basis of invariant polynomials of $S_n$ in which the \wPm\ ($p_1$ excluded) has the same simple form of the \wPm\ of $A_{n-1}$, that can be written as specified by Theorem \ref{hankelAn} (but with the labels of the basic invariant polynomials of $A_{n-1}$ that have to be increased by 1, and with one more column and row in front of the $n-1$ others), but it contains the variable $p_1$ in a complicated way.

\section{Exceptional groups and non-crystallographic groups\label{exceptional}}
The main results presented in this article are the generating formulas given in Theorems \ref{lambdaSn}, \ref{lambdaAn}, \ref{lambdaBn} and \ref{lambdaDn}. These are useful (and sometimes necessary) to determine the \wPms\ and the $\lambda$-vectors in the case of the finite reflection groups in the infinite series $S_n$, $A_n$, $B_n$, and $D_n$, respectively, for which there exist a group for each value of the rank $n$. On the contrary, the exceptional groups $E_6$, $E_7$, $E_8$, $F_4$, $G_2$ and the non-crystallographic groups $H_3$, $H_4$, are defined only for specific values of the rank and a generating formula for them is not necessary. Their \wPms\ can be found in some published articles, for example in Refs. \cite{SYS1980,YaSek1981} (groups $E_6$, $F_4$, $G_2$, $H_3$, $H_4$), \cite{giv1980} (groups $E_6$, $F_4$), \cite{tal-jmp2010} (groups $E_7$, $E_8$).\\

There is another infinite series of finite reflection groups, that one of the dihedral groups $I_2(m)$, $m\geq 2$, the groups of symmetries of the regular $m$-sided polygon (that includes the crystallographic groups $A_1\times A_1\simeq I_2(2)$, $A_2\simeq I_2(3)$, $B_2\simeq I_2(4)$, and $G_2\simeq I_2(6)$). The roots, the basic invariant polynomials, the \wPms\ and the $\lambda$-vectors of the groups of type $I_2(m)$ are well known, see for instance Refs. \cite{Humphreys1990,SYS1980,YaSek1981,schw-ihes,Sar-Tal1991}. For completeness, I report briefly, for the groups of type $I_2(m)$, $m\geq 2$, one of the possible choices for the simple roots, the roots, the basis of invariant polynomials, the corresponding \wPm, the irreducible active factors and their $\lambda$-vectors.
\begin{itemize}
  \item Simple roots. The 2 simple roots can be defined as the vector $\alpha_1=(0,1)$ and the vector $\alpha_2$ obtained from $\alpha_1$ with a rotation about the origin $(0,0)$ by an angle $\varphi_m=-\left(\pi-\frac{\pi}{m}\right)$. The two simple roots obtained in this way have the same length. This is necessary if $m$ is odd, but not necessary if $m$ is even, because if $m$ is even two different root lengths are possible. To be crystallographic, the groups $I_2(4)\simeq B_2$ and $I_2(6)\simeq G_2$ require the lengths of the two simple roots in a ratio equal to $\sqrt{2}$ and $\sqrt{3}$, respectively.\\
  \item Roots. The $2m$ roots are obtained by rotating $\alpha_1$ and $\alpha_2$ by the angles $\frac{2\pi i}{m}$, $\forall\,i=0,1,\ldots,m-1$. The positive roots are the roots with a positive number as their first non-zero coordinate.\\
  \item Basic invariant polynomials.
$$p_1(x)=x_1^2+x_2^2\,,$$
$$ p_2(x)={\rm Re}\left[(x_1+i\,x_2)^m\right]=\frac{1}{2}\,\left[(x_1+i\,x_2)^m+(x_1-i\,x_2)^m\right]\,,\qquad i=\sqrt{-1}\,.$$
The polynomials $p_1(x)$ and $p_2(x)$ are invariant for the reflections corresponding to the two simple roots $\alpha_1$ and $\alpha_2$, and then for all the group $I_2(m)$ generated by them.\\
  \item \wPm. The \wPm\ corresponding to the basic invariant polynomials $p_1(x)$ and $p_2(x)$ defined above is the following one:
$$\widehat P(p)=\left(
    \begin{array}{cc}
      4\,p_1 &\ 2\,m\,p_2 \\
      2\,m\,p_2 &\ m^2\,p_1^{m-1} \\
    \end{array}
  \right).
$$
One has $\det(\widehat P(p))=4\,m^2\,(p_1^m-p_2^2)$. If $m$ is even, $\det(\widehat P(p))=a_+(p)\,a_-(p)$, with $a_\pm(p)=p_1^{m/2}\pm p_2$ two distinct irreducible active polynomials.\\
  \item $\lambda$-vectors.
$$\lambda^{(\det(\widehat P))}_a(p)=(4\,m,\ 0)\,.$$
If $m$ is even one also has:
$$\lambda^{(a_\pm)}_a(p)=(2\,m,\ \pm \,m^2\,p_1^{m/2-1})\,.$$\\
\end{itemize}
The basis of invariant polynomials for the group $I_2(m)$ just described is at the same time a flat basis, a $\det(\widehat P)$-basis and a canonical basis. If $m$ is even, the two irreducible active factors of $\det(\widehat P(p))$ define two algebraic curves $a_\pm(p)=0$ that contain the boundary of the set ${\cal S}$ representing the orbit space of $I_2(m)$. The algebraic curve $a_+(p)=0$ contains the images through the orbit map of the $m/2$ reflecting lines orthogonal to the $m$ roots obtained by rotating the simple root $\alpha_1$ by the angles $\frac{2\pi i}{m}$, $\forall\,i=0,1,\ldots,m-1$. The algebraic curve $a_-(p)=0$ has the same meaning, but is related to the simple root $\alpha_2$. This can be seen by picking a point in a reflecting line and by seeing which algebraic curve contains its image. For example the point $(x_1,\,x_2)=(x_1,\,0)$ lies in the reflecting line orthogonal to $\alpha_1$, and its image through the orbit map is the point $(p_1,\,p_2)=(x_1^2,\,x_1^m)$, that satisfies the equation $a_-(p)=0$, but not $a_+(p)=0$. Vice versa, the point $(x_1,x_2)=(x_2\,\cos\frac{\pi}{m},\,x_2\,\sin\frac{\pi}{m})$ lies in the reflecting line orthogonal to $\alpha_2$, and its image through the orbit map is the point $(p_1,p_2)=(x_2^2,\,-x_2^m)$, that satisfies the equation $a_+(p)=0$, but not $a_-(p)=0$. If $m$ is even there are two different conjugacy classes in $I_2(m)$ and two different singular orbit type strata of dimension 1 that are represented in ${\cal S}$ by two primary strata lying in the two curves, respectively. If $m$ is odd there is a single conjugacy class in $I_2(m)$ and a single singular orbit type stratum of dimension 1. The image in ${\cal S}$ of this orbit type stratum is formed by two different primary, lying in the two curves, respectively, that are separated by the origin of $\R^2$.

\section{Examples \label{examples}}
\noindent\textbf{Example 1}.\\
This example shows the construction of the rotation matrix $R_4$, with Eq. (\ref{matRn+1}), of the basic invariant polynomials of $A_3$, with Eqs. (\ref{elemsymmpolSn+1})--(\ref{basicpolynAn}), and of the \wPm\ of $A_3$, with Eq. (\ref{PjkdiAn}).\\
The matrix $R_{n+1}$, defined in Eq. (\ref{matRn+1}), and its transpose,  in the case $n+1=4$, that is $n=3$, are  the following:
$$
R_4=\left(%
\begin{array}{cccc}
  \ \;{1}/{\sqrt{2}}  &\ \;{-1}/{\sqrt{2}}   & \ \;0 & \ 0 \\
 {1}/{\sqrt{6}}  &  \ \;{1}/{\sqrt{6}}  & \ \; {-2}/{\sqrt{6}}& \ 0 \\
 {1}/{\sqrt{12}}       & {1}/{\sqrt{12}} & \ \;{1}/{\sqrt{12}} & \ {-3}/{\sqrt{12}} \\
  {1}/{2}                   &     \ {1}/{2}        & {1}/{2}  & \ {1}/{2} \\
\end{array}\right),\qquad R_4^\top=\left(%
\begin{array}{cccc}
  \ \;{1}/{\sqrt{2}}  &\ \;{1}/{\sqrt{6}}   & \ \;{1}/{\sqrt{12}} & \ {1}/{2} \\
  {-1}/{\sqrt{2}} &  \ \;{1}/{\sqrt{6}}  & \ \;{1}/{\sqrt{12}} & \ {1}/{2} \\
  0                   & {-2}/{\sqrt{6}} & \ \;{1}/{\sqrt{12}} & \ {1}/{2} \\
  0                   & 0            &  {-3}/{\sqrt{12}} & \ {1}/{2} \\
\end{array}\right)
$$
One may easily verify that $R_4 R_4^\top=R_4^\top R_4=I_4$, that $R_4$ transforms the column unit vector $u_4=(\frac{1}{2},\frac{1}{2},\frac{1}{2},\frac{1}{2})$ into the column unit vector $e_4=(0,0,0,1)$, and that $R_4^\top$ makes the opposite transformation.\\
To obtain the basic polynomial invariants $s'_a(x')$, $a=1,\ldots,4$, in Eq. (\ref{trasfAn1}), one has to make the variable substitutions ${x\to R_{4}^\top x'}$ in the basic polynomial invariants $s_a(x)$, $a=1,\ldots,4$, given by Eq. (\ref{elemsymmpolSn+1}), with $n=3$. One finds so the following
basic invariant polynomials of $S_4$:
{\footnotesize
\begin{eqnarray*}
   s'_1(x') &=& \left(\frac{1}{\sqrt{2}} x'_1+ \frac{1}{\sqrt{6}} x'_2+  \frac{1}{\sqrt{12}} x'_3+  \frac{1}{2} x'_4\right)+\left(\frac{-1}{\sqrt{2}} x'_1+  \frac{1}{\sqrt{6}}  x'_2  +\frac{1}{\sqrt{12}}
  x'_3+  \frac{1}{2} x'_4\right)+\left(\frac{-2}{\sqrt{6}} x'_2+  \frac{1}{\sqrt{12}} x'_3 + \frac{1}{2} x'_4 \right)+\left(\frac{-3}{\sqrt{12}}x'_3 + \frac{1}{2} x'_4\right)=\\
  &=&2x'_4\\
   s'_2(x')  &=&  \left(\frac{1}{\sqrt{2}} x'_1+ \frac{1}{\sqrt{6}} x'_2+  \frac{1}{\sqrt{12}} x'_3+  \frac{1}{2} x'_4\right)\left(\frac{-1}{\sqrt{2}} x'_1+  \frac{1}{\sqrt{6}}  x'_2  +\frac{1}{\sqrt{12}}
  x'_3+  \frac{1}{2} x'_4\right)+
  \ldots +
   \left(\frac{-2}{\sqrt{6}} x'_2+  \frac{1}{\sqrt{12}} x'_3 + \frac{1}{2} x'_4 \right)\left(\frac{-3}{\sqrt{12}}x'_3 + \frac{1}{2} x'_4\right)=\\
  &=&-\frac{1}{2}\,\left( {x'_1}^2 +  {x'_2}^2 +  {x'_3}^2\right) + \frac{3}{2}\, {x'_4}^2\\
  s'_3(x')  &=&\ldots =\\
  &=&\frac{1}{18}
  \left(3 \sqrt{6}\, {x'_1}^2 {x'_2} - \sqrt{6} \,{x'_2}^3 + 3 \sqrt{3} \,{x'_1}^2 {x'_3} + 3 \sqrt{3}\, {x'_2}^2 {x'_3} - 2 \sqrt{3}\, {x'_3}^3 - 9 \,{x'_1}^2 {x'_4} - 9 \,{x'_2}^2 {x'_4} - 9\, {x'_3}^2 {x'_4} +
   9\, {x'_4}^3\right)\\
 s'_4(x') &=&\ldots =\\
 &=&\frac{1}{144}\left(-36 \sqrt{2}\, {x'_1}^2 {x'_2} {x'_3} + 12 \sqrt{2} \,{x'_2}^3 {x'_3} + 18\, {x'_1}^2 {x'_3}^2 + 18\, {x'_2}^2 {x'_3}^2 - 3 \,{x'_3}^4 + 12 \sqrt{6}\, {x'_1}^2 {x'_2} {x'_4} - 4 \sqrt{6}\, {x'_2}^3 {x'_4} +
   12 \sqrt{3}\, {x'_1}^2 {x'_3} {x'_4} +\right. \\
   &+&\left. 12 \sqrt{3}\, {x'_2}^2 {x'_3} {x'_4} - 8 \sqrt{3} \,{x'_3}^3 {x'_4} - 18\, {x'_1}^2 {x'_4}^2 - 18\, {x'_2}^2 {x'_4}^2 - 18 \, {x'_3}^2 {x'_4}^2 + 9\, {x'_4}^4\right)
\end{eqnarray*}
}
The basic invariant polynomials of $A_3$ are obtained from the basic invariant polynomials $s'_a(x')$, $a=1,\ldots,4,$ of $S_4$ in the way specified in Eqs. (\ref{trasfAn2}) and (\ref{basicpolynAn}). One so finds:
\begin{eqnarray}\label{ex1inv}
\nonumber  p_1(\bar x') &=&{ {x'_1}^2 +  {x'_2}^2 +  {x'_3}^2 }, \\
  p_2(\bar x') &=&-
\frac{2\sqrt{3}}{9}\left({3\sqrt{2}\,{x'_1}^2 x'_2 -\sqrt{2}\, {{x'_2}^3} +3\,{{x'_1}^2 x'_3} +3\,{{x'_2}^2  x'_3} -
2\,{x'_3}^3}\right) , \\
\nonumber  p_3(\bar x') &=& {\frac{1}{6}\,  x'_3 \left(12 \sqrt{2}\, {x'_1}^2 x'_2 -4 \sqrt{2}\, {x'_2}^3-6 {x'_1}^2 x'_3-6 {x'_2}^2 x'_3+ {x'_3}^3 \right)}.
\end{eqnarray}
The corresponding \wPm\ and $\lambda$-vector are obtained with Eqs. (\ref{PjkdiAn}) and (\ref{lambdaAnEq}) and are the following:
\begin{equation}\label{ex1eq}
  \widehat P(p)=\left(
                  \begin{array}{ccc}
                    4\,p_1 &\quad 6\,p_2 &\quad 8\,p_3 \\
                    6\,p_2 &\quad 4\,p_1^2+8\,p_3 &\quad 2\,p_1p_2 \\
                    8\,p_3 &\quad 2\,p_1p_2 &\quad 3\,p_2^2-8\,p_1p_3 \\
                  \end{array}
                \right),\qquad\lambda^{(\det(\widehat P))}(p) = (24,\,0,\,- 8\,p_{1})\,.
\end{equation} \\

\noindent\hypertarget{ex2}{\textbf{Example 2}}.\\
This example shows the construction of a flat basis of invariant polynomials (recall the discussion with Eq. (\ref{flatcondition})), together with its \wPm\ $\widehat P(p)$ and $\lambda$-vector of $\det(\widehat P(p))$. Let's consider the case of $A_3$, so we can use the basic invariant polynomials, the \wPm\ and the $\lambda$-vector obtained in Example 1.\\
In a flat basis, the condition expressed by Eq. (\ref{flatcondition}) is true. Then one has to find the basis transformation that transforms the \wPm\ in Eq. (\ref{ex1eq}) to a \wPm\ satisfying Eq. (\ref{flatcondition}). In the present case one has to eliminate the variable $p_3$ from the element $P_{33}(p)$, only. One then writes the most general allowed basis transformation $p'={\widehat p\,}'(p)$ that transforms the variable $p_3$, only, that preserves homogeneity and such that $J_{33}(p)=1$. One then writes:
\begin{equation}\label{ex2generaltransf}
  {\widehat p\,}'_1(p)=p_1\,,\qquad {\widehat p\,}'_2(p)=p_2\,,\qquad {\widehat p\,}'_3(p)=p_3+c_1\, p_1^2\,,
\end{equation}
with $c_1$ a real constant.
This transformation implies the following jacobian matrix:
\begin{equation}\label{ex2generaljacobian}
  J(p)=\left(
         \begin{array}{ccc}
           1 & 0 & 0 \\
           0 & 1 & 0 \\
           2c_1p_1 & 0 & 1 \\
         \end{array}
       \right),
\end{equation}
that gives, using Eq. (\ref{Ptransf2}), the following expression:
$$\left.\widehat P_{33}(p')\right|_{p'={\widehat p'\,}(p)}=J_{3a}(p)\,\widehat P_{ab}(p)\,J^\top_{b3}(p)=4\,c_1^2\,p_1^2\,\widehat P_{11}(p)+4\,c_1\,p_1\,\widehat P_{13}(p)+\widehat P_{33}(p)=$$
$$=4\,c_1^2\,p_1^2\,4\,p_1+4\,c_1\,p_1\,8\,p_3+(3\,p_2^2-8\,p_1p_3)=16\,c_1^2\,p_1^3+3\,p_2^2+8\,(4\,c_1-1)\,p_1p_3
$$
In the above calculation, it is not necessary to write the \wPm\ with the new variables $p'$, using the inverse transformation $\widehat p(p')$ and Eq. (\ref{Ptransf}), because the condition (\ref{flatcondition}) can be obtained either by deriving the matrix $\widehat P(p')|_{p'={\widehat p'\,}(p)}$ with respect to $p_3$ or by deriving the matrix $\widehat P(p')$ with respect to $p'_3$.
Then condition (\ref{flatcondition}) is fulfilled if $4c_1-1=0$, that is, if $c_1=\frac{1}{4}$, and Eq. (\ref{ex2generaltransf}) gives the following transformation from the basis (\ref{ex1inv}) to a flat basis:
$${\widehat p\,}'_1(p)=p_1\,,\qquad {\widehat p\,}'_2(p)=p_2\,,\qquad {\widehat p\,}'_3(p)=p_3+\frac{1}{4}\, p_1^2\,.$$
With the transformation found, Eqs. (\ref{Ptransf}) and (\ref{lambdatransf}) give:
$$\widehat P(p')=\left(
                  \begin{array}{ccc}
                    4\,p'_1 &\quad 6\,p'_2 &\quad 8\,p'_3 \\
                    6\,p'_2 &\quad 2\,{p'_1}^2+8\,p'_3 &\quad 5\,p'_1p'_2 \\
                    8\,p'_3 &\quad 5\,p'_1p'_2 &\quad {p'_1}^3+3\,{p'_2}^2 \\
                  \end{array}
                \right),\qquad\lambda^{(\det(\widehat P))}(p') = (24,\,0,\,4\,p'_{1})\,.$$
When the rank $n$ is large it may be somewhat difficult to find the transformation to a flat basis and a systematic method to follow is necessary. First of all, in place of the \wPm\ $\widehat P(p)$ in Eqs. (\ref{Ptransf}) or (\ref{Ptransf2}), one can use the part of it (let's call it $\widehat Q(p)$) containing (linearly) the highest weight variable $p_n$, only. A quick look at the Hankel matrix decomposition in Eqs. (\ref{PBnHankel}), (\ref{PDnHankel}) or (\ref{PAnHankel}), is sufficient to see what this matrix $\widehat Q(p)$, containing $p_n$, is. For example, in the case of the groups of type $A_n$ this matrix is $\widehat Q(p)=2\,(n+1)\,p_n\,Y_n^{(n)}-(n+1)\,p_n\,K_n^{(n)}(p)$. We then use Eq. (\ref{Ptransf2}) with the matrix $\widehat Q(p)$, that is: $\left.\widehat Q(p')\right|_{p'= {\widehat p\,}'(p)}=J(p)\; \widehat Q(p)\; J^\top(p)$, and take the derivative of the resulting matrix with respect to $p_n$. Let's define $A(p)=\frac{\partial}{\partial p_n} \left(J(p)\; \widehat Q(p)\; J^\top(p)\right)=J(p)\; \frac{\partial \widehat Q(p)}{\partial p_n} \; J^\top(p)$. If $p'$ is a flat basis, $A(p)$ is a constant matrix, and this means that all its elements of positive weight are zero. By requiring this, 
one finds the basis transformation. When $n$ is large, it is convenient to proceed in small steps. Let $i$ be the smallest index such that $A_{i,i}(p)$ is of positive weight.
Write the most general allowed basis transformation $p_a'={\widehat p\,}'_a(p)$, $\forall\,a=2,\ldots,i$, with $\frac{\partial {\widehat p\,}'_a(p)}{\partial p_a}=1$, and determine all the free coefficients of this transformation formula necessary to obtain $A_{a,i}(p)=0$, $\forall\,a=1,\ldots,i$. Then, for each $b=i+1,\ldots,n$, write also the most general allowed basis transformation $p_b'={\widehat p\,}'_b(p)$, with $\frac{\partial {\widehat p\,}'_b(p)}{\partial p_b}=1$ and determine all the free coefficients (in ${\widehat p\,}'_b(p)$ as well as in ${\widehat p\,}'_a(p)$, $a=2,\ldots,b-1$, that are still undetermined) necessary to have $A_{a,b}(p)=0$, $\forall\,a=1,\ldots,b$. In this way one arrives at a flat basis. It is convenient to program a computer to do the transformation of Eqs. (\ref{Ptransf}) or  (\ref{Ptransf2}). When applying this method with the groups of type $D_n$, one has to remember that the highest weight variable is $p_{n-1}$ and not $p_n$, in particular in Eq. (\ref{PDnHankel}) and in Eq. (\ref{flatcondition}).\\

\noindent\hypertarget{ex3}{\textbf{Example 3}}.\\
This example shows the construction of a $\det(\widehat P)$-basis of invariant polynomials (recall the discussion with Eq. (\ref{abasiscondition})), together with its \wPm\ $\widehat P(p)$ and $\lambda$-vector $\lambda^{(\det(\widehat P))}(p)$.
Let's consider the most general allowed basis transformation $p'={\widehat p\,}'(p)$, with $J_{aa}(p)=1$, $\forall\,a=1,\ldots,n$. Eq. (\ref{lambdatransf}) states how a $\lambda$-vector transforms. To obtain a $\det(\widehat P))$-basis, one has to require that all components, except the first one, of the transformed $\lambda$-vector of $\lambda^{(\det(\widehat P))}(p)$ be equal to zero. Using the formula:
\begin{equation}\label{ex3lambda1}
  \left.\lambda^{(\det(\widehat P))}_a(p')\right|_{p'=\widehat{p\,}'(p)}=\sum_{b=1}^n\,J_{ab}(p)\,\lambda^{(\det(\widehat P))}_b(p)\,,\qquad \forall\,a=1,\ldots,n\,,
\end{equation}
equivalent to that one in Eq. (\ref{lambdatransf}), one has to require:
\begin{equation}\label{ex3lambda}
  \left.\lambda^{(\det(\widehat P))}_a(p')\right|_{p'=\widehat{p\,}'(p)}=0\,,\qquad \forall\,a=2,\ldots,n\,.
\end{equation}
Let's consider the case of $A_3$, so we can use the basic invariant polynomials, the \wPm\ and the $\lambda$-vector obtained in Example 1. The most general allowed basis transformation with the required properties is written in Eq. (\ref{ex2generaltransf}), and its jacobian matrix is written in Eq. (\ref{ex2generaljacobian}). The \wPm\ $\widehat P(p)$ and the $\lambda$-vector $\lambda^{(\det(\widehat P))}(p)$, corresponding to the basic invariant polynomials in Eq. (\ref{ex1inv}), are written in Eq. (\ref{ex1eq}). The transformed $\lambda$-vector, obtained with Eq. (\ref{lambdatransf}), is then the following one:
$$\left.\lambda^{(\det(\widehat P))}_a(p')\right|_{p'=\widehat{p\,}'(p)}=(24,\,0,\,8(6\,c_1-1)p_1)\,,$$
and one obtains a $\det(\widehat P)$-basis if $c_1=\frac{1}{6}$. Using $c_1=\frac{1}{6}$ in Eq. (\ref{ex2generaltransf}), one finds the following transformation from the basis (\ref{ex1inv}) to a $\det(\widehat P)$-basis:
$${\widehat p\,}'_1(p)=p_1\,,\qquad {\widehat p\,}'_2(p)=p_2\,,\qquad {\widehat p\,}'_3(p)=p_3+\frac{1}{6}\, p_1^2\,.$$
With the transformation found, Eqs. (\ref{Ptransf}) and (\ref{lambdatransf}) give:
$$\widehat P(p')=\left(
                  \begin{array}{ccc}
                    4\,p'_1 &\quad 6\,p'_2 &\quad 8\,p'_3 \\ \\
                    6\,p'_2 &\quad \frac{8}{3}\,{p'_1}^2+8\,p'_3 &\quad 4\,p'_1p'_2 \\ \\
                    8\,p'_3 &\quad 4\,p'_1p'_2 &\quad \frac{8}{9}\,{p'_1}^3-\frac{8}{3}\,p'_1p'_3+3\,{p'_2}^2 \\
                  \end{array}
                \right),\qquad\lambda^{(\det(\widehat P))}(p') = (24,\,0,\,0)\,.$$
If the rank $n$ is large it may be convenient to transform one by one the variables from that one of the greatest weight $d_n$ to that one of the lowest weight greater than 2. When writing the most general transformation, one may ignore the terms not depending on $p_1$, because they do not transform a non-$a$-basis into an $a$-basis or vice versa. When one is transforming the variable indexed with $a$, one has to find the free coefficients of this transformation in such a way for which $\lambda^{(\det(\widehat P))}_a(p)=0$, only. What is here said for a $\det(\widehat P)$-basis can be performed for a general $a$-basis, as well.\\

\noindent\hypertarget{ex4}{\textbf{Example 4}}.\\
This example shows the construction of a canonical basis of invariant polynomials, $q_1(x),\ldots,q_n(x)$, (recall the discussion with Eqs. (\ref{flattocondition1}) and (\ref{flattocondition2})), together with its \wPm\ $\widehat P(q)$ and $\lambda$-vector $\lambda^{(\det(\widehat P))}(q)$ of $\det(\widehat P(q))$. Let's write $p_1(x),\ldots,p_n(x)$ for the basic invariant polynomials of the groups of type $A_n$, $B_n$, $D_n$, given in Sections \ref{An}--\ref{Dn}, with the same labeling conventions, and $p_{i_1}(x),\ldots,p_{i_n}(x)$ for the same basic invariant polynomials in which the order is according to their degrees: $d_{i_j}\leq d_{i_{j+1}}$, $\forall\,j=1,\ldots,n-1$. (Except for the groups of type $D_n$, even $n$, one has $i_j=j$, and the inequality is always strict). One has then $q_1(x)=p_1(x)=\|x\|^2$. Suppose that the $k$ basic invariant polynomials of lowest degree of a canonical basis are already known, with $k=1,2,\ldots,n-1$, that is, suppose that $q_{i_1}(x),\ldots,q_{i_k}(x)$ are already known, and that one wishes to find the next canonical invariant $q_{i_{k+1}}(x)$. One has then to write the most general homogeneous polynomial of degree $d_{i_{k+1}}$, $r(x)$, written in terms of $p_{i_1}(x),\ldots,p_{i_{k+1}}(x)$, such that the coefficient of $p_{i_{k+1}}(x)$ is 1 (except in the case of $D_n$, even $n$, that when $d_{i_{k+1}}=n$ and $p_{i_{k+1}}(x)$ is the second basic invariant of degree $n$, one has to keep the coefficient of $p_{i_{k+1}}(x)$ arbitrary). By requiring the conditions expressed by Eq. (\ref{flattocondition1}), that is $\langle q_{i_a},r\rangle=0$, $\forall\,a=1,\ldots,k$, one finds the arbitrary coefficients that render $r(x)$ canonical. Then one defines $q_{k+1}(x)=r(x)$ and proceed in the same way with the next value of $k$. In the case of $D_n$, even $n$, one has the two more conditions expressed by Eq. (\ref{flattocondition2}).\\
Let's consider the case of $A_3$, so we can use the basic invariant polynomials, the \wPm\ and the $\lambda$-vector of $\det(\widehat P(q))$ obtained in Example 1. The most general allowed basis transformation is expressed by Eq. (\ref{ex2generaltransf}). For $k=1$ one has $r(x)=p_2(x)$, without arbitrary coefficients, so one defines $q_2(x)=p_2(x)$. For $k=2$ one has $r(x)=p_3(x)+c_1\,p_1(x)^2$, with $c_1$ an arbitrary coefficient. The condition $\langle q_1,r\rangle=0$ gives $0=\langle q_1,p_3+c_1\,p_1^2\rangle=\langle q_1,p_3\rangle+c_1\,\langle q_1,p_1^2\rangle=2(-1 + 10 c_1)\,q_1(x)$.
The condition $\langle q_2,r\rangle=0$ is identically satisfied and gives no conditions. One then has a canonical basis if $c_1=\frac{1}{10}$. Using $c_1=\frac{1}{10}$ in Eq. (\ref{ex2generaltransf}), one finds the following transformation from the basis (\ref{ex1inv}) to a canonical basis:
$${\widehat q\,}_1(p)=p_1\,,\qquad {\widehat q\,}_2(p)=p_2\,,\qquad {\widehat q\,}_3(p)=p_3+\frac{1}{10}\, p_1^2\,.$$
With the transformation found, Eqs. (\ref{Ptransf}) and (\ref{lambdatransf}) give:
$$\widehat P(q)=\left(
                  \begin{array}{ccc}
                    4q_1 &\quad 6q_2 &\quad 8q_3 \\ \\
                    6q_2 &\quad \frac{16}{5}\,{q_1}^2+8q_3 &\quad \frac{16}{5}\,q_1q_2 \\ \\
                    8q_3 &\quad \frac{16}{5}\,q_1q_2 &\quad \frac{16}{25}\,{q_1}^3-\frac{24}{5}\,q_1q_3+3{q_2}^2 \\
                  \end{array}
                \right)\,,\qquad\lambda^{(\det(\widehat P))}(q) = \left(24,\,0,\,-\frac{16}{5}\,q_1\right)\,.$$
This method can be followed also if the rank $n$ is large, but the calculations may become very long.\\

\section{Proofs\label{proofs}}

\subsection{Proof of Theorem \ref{lambdaSn} for $S_n$\label{proofSn}}

\noindent\textbf{Proof of the generating formula (\ref{PjkdiSn}).}\\
The matrix obtained from the generating formula (\ref{PjkdiSn}) is symmetric, so we can simplify the formula and the proof by supposing $a\leq b$. If we prove that Eq. (\ref{PjkdiSn}) is true for $a\leq b$ then it is true also for $b<a$, hence for all allowed values of $a,b$. A further simplification is possible if we define, in addition to $p_0(x)=1$, that $p_a(x)=0$, $\forall\,a<0$ and $\forall\,a>n$. We will then prove the validity of the following formula for $S_n$, equivalent to the generating formula (\ref{PjkdiSn}):
\begin{equation}\label{PjkdiSnbis}
   {\widehat P}_{ab}(p)={\widehat P}_{ba}(p)= (n+1-a)\;p_{a-1}\;p_{b -1}- \sum_{i=a+b-n-1}^{a} \,(a+ b - 2i)\;p_{i-1}\;p_{a+b - 1-i}\,,\qquad \forall\,a\leq b=1,\ldots,n\,,
\end{equation}
in which one has to consider $p_0=1$ and $p_a=0$, $\forall\,a<0$ and $\forall\,a>n$.\\

The proof can be obtained by induction on the rank $n$. The lowest possible value for $n$ is $n=1$, and with $n=1$ Eq. (\ref{PjkdiSnbis}) is trivially verified: the only basic invariant polynomial is $p_1(x)=x_1$, with which Eq. (\ref{matriceP(x)}) gives $P_{11}(x)=1$, and Eq.  (\ref{PjkdiSnbis}) gives ${\widehat P}_{11}(p)=1$. Let's verify Eq. (\ref{PjkdiSnbis}) also for the less trivial value $n=2$. From Eq. (\ref{elemsymmpolSn}), for $n=2$, the basic invariant polynomials are $p_1(x)=x_1+x_2$ and $p_2(x)=x_1x_2$, their gradients are then $\nabla p_1(x)=(1,1)$ and  $\nabla p_2(x)=(x_2,x_1)$, and   Eq. (\ref{matriceP(x)}) gives: $P_{11}(x)=2$,  $P_{12}(x)=P_{21}(x)=x_1+x_2=p_1(x)$,  $P_{22}(x)=x_1^2+x_2^2=p_1(x)^2-2p_2(x)$. Using Eq. (\ref{PjkdiSnbis}) one finds the same form of the corresponding \wPm: ${\widehat P}_{11}(p)=2$,  ${\widehat P}_{12}(p)={\widehat P}_{21}(p)=p_1$, ${\widehat P}_{22}(p)=p_1^2-2p_2$. We have then verified that Eq. (\ref{PjkdiSnbis}) correctly gives the matrix elements of the \wPm\ $\widehat P(p)$ for the basic invariant polynomials (\ref{elemsymmpolSn}) of $S_n$, for $n=1,2$.\\

Using the induction principle, we suppose, for a general $n\geq 2$, that Eq. (\ref{PjkdiSnbis}) is true for $S_n$, and prove in all generality that Eq. (\ref{PjkdiSnbis}) is true also for $n$ substituted by $n+1$, that is for $S_{n+1}$. This implies then that Eq. (\ref{PjkdiSnbis}) is true for $S_n$ for all $n\geq 2$.\\

We use here the symbols $x$ and $y$ for generic vectors of $\R^n$ and $\R^{n+1}$, that is: $x=(x_1,x_2,\ldots,x_n)$, and $y=(x_1,x_2,\ldots,x_n,x_{n+1})$.
$S_n$ acts in $\R^n$ and its basic invariant polynomials are given by Eq. (\ref{elemsymmpolSn}): $p_a(x)$, $a=1,\ldots,n$. $S_{n+1}$ acts in $\R^{n+1}$ and its basic invariant polynomials are also  given by Eq. (\ref{elemsymmpolSn}), but with $n$ substituted by $n+1$, and we write them as $p_a(y)$, $a=1,\ldots,n+1$.\\

The basic invariant polynomials of $S_{n+1}$ can be expressed in terms of those of $S_{n}$ in the following way:
$$
\begin{array}{ccl}
   p_{1}(y)            & = &   p_1(x)+x_{n+1}\,,  \\
   p_{2}(y)            & = &   p_2(x)+p_1(x)\,x_{n+1}\,,  \\
   p_{3}(y)            & = &   p_3(x)+p_2(x)\,x_{n+1}\,,  \\
   \vdots            &  &  \\
   p_{n}(y)            & = &   p_n(x)+p_{n-1}(x)\,x_{n+1}\,,  \\
   p_{n+1}(y)            & = &   p_n(x)\,x_{n+1}\,.  \\
   \end{array}
$$
Using the definitions $p_0(x)=1$ and $p_{n+1}(x)=0$, one has the following general formula:
\begin{equation}\label{p(y)diSn}
   p_{a}(y)             =    p_a(x)+p_{a-1}(x)\,x_{n+1}\,\qquad \forall\,a=1,2,\ldots,n+1.
\end{equation}
Let's define the following ``nabla'' operators that map a scalar function into a vector of $\R^{n+1}$:
\begin{equation}\label{nablaxy}
  \nabla_{\!x}=\left(\frac{\partial}{\partial x_1},\frac{\partial}{\partial x_2},\ldots,\frac{\partial}{\partial x_n},0\right)\,,\qquad
\nabla_{\!y}=\left(\frac{\partial}{\partial x_1},\frac{\partial}{\partial x_2},\ldots,\frac{\partial}{\partial x_n},\frac{\partial}{\partial x_{n+1}}\right)\,.
\end{equation}
One has then, for Eqs. (\ref{p(y)diSn}) and (\ref{nablaxy}), $\forall a=1,\ldots,n+1$:
\begin{equation}\label{nablaYp(y)diSn}
  \nabla_{\!y}p_a(y)=\nabla_{\!y}\left[ p_a(x)+p_{a-1}(x)\,x_{n+1}\right]=\nabla_{\!x}p_a(x)+x_{n+1}\,\nabla_{\!x}p_{a-1}(x)+p_{a-1}(x)\,e_{n+1}
  \,,
\end{equation}
where $e_{n+1}$ is the canonical unit vector of $\R^{n+1}$ (with $1$ in the last, $(n+1)$-th, coordinate and $0$ elsewhere).\\
We can now calculate a general element of the matrix $P(y)$ from Eqs. (\ref{matriceP(x)}) and (\ref{nablaYp(y)diSn}), using the basic invariant polynomials $p_a(y)$, $a=1,\ldots,n+1$, of $S_{n+1}$. For all $a,b=1,\ldots,n+1$, we have:
$$P_{ab}(y)=\nabla_{\!y}p_a(y)\cdot \nabla_{\!y}p_b(y)=$$
$$=\left[\nabla_{\!x}p_a(x)+x_{n+1}\,\nabla_{\!x}p_{a-1}(x)+p_{a-1}(x)\,e_{n+1}\right]\cdot
\left[\nabla_{\!x}p_b(x)+x_{n+1}\,\nabla_{\!x}p_{b-1}(x)+p_{b-1}(x)\,e_{n+1}\right]=
$$
$$=\nabla_{\!x}p_a(x)\cdot \nabla_{\!x}p_b(x)+x_{n+1}\,
\nabla_{\!x}p_a(x)\cdot \nabla_{\!x}p_{b-1}(x)+0+x_{n+1}\,\nabla_{\!x}p_{a-1}(x)\cdot \nabla_{\!x}p_b(x)\,+$$
$$+\,x_{n+1}^2\,\nabla_{\!x}p_{a-1}(x)\cdot \nabla_{\!x}p_{b-1}(x)+0+0+0+p_{a-1}(x)\,p_{b-1}(x)=$$
\begin{equation}\label{dimSnsimm}
  =P_{ab}(x)+x_{n+1}\,\left[P_{a,b-1}(x)+P_{a-1,b}(x)\right]+x_{n+1}^2\,P_{a-1,b-1}(x)+p_{a-1}(x)\,p_{b-1}(x)\,.
\end{equation}
Of course, in the preceding expression, $P_{cd}(x)=0$, if one of the indices $c,d$ (that in the above expression can be equal to $a,b,a-1,b-1$) is equal to $0$ or to $n+1$, because for $S_n$ the allowed indices run from 1 to $n$, and the gradients of $p_0(x)=1$, and $p_{n+1}(x)=0$ are null vectors. By the induction hypothesis, Eqs. (\ref{PjkdiSnbis}) and (\ref{PjkdiSn}) are true for $S_n$, so we can use them in the preceding equation. We obtain then, $\forall\,a\leq b=1,\ldots,n+1$:
$$P_{ab}(y)=(n+1-a)\;p_{a-1}(x)\;p_{b -1}(x)- \sum_{i=a+b-n-1}^{a} \,(a+ b - 2i)\;p_{i-1}(x)\;p_{a+b - 1-i}(x)\,+$$
$$+\,x_{n+1}\,\left\{
(n+1-\min(a,b-1))\;p_{a-1}(x)\;p_{b -2}(x)- \sum_{i=a+b-n-2}^{\min(a,b-1)} \,(a+ b-1 - 2i)\;p_{i-1}(x)\;p_{a+b - 2-i}(x)\,+\right.$$
$$\left.+\,
(n+2-a)\;p_{a-2}(x)\;p_{b -1}(x)- \sum_{i=a+b-n-2}^{a-1} \,(a+ b-1 - 2i)\;p_{i-1}(x)\;p_{a+b - 2-i}(x)
\right\}+
$$
$$ +\,x_{n+1}^2\,\left\{(n+2-a)\;p_{a-2}(x)\;p_{b -2}(x)- \sum_{i=a+b-n-3}^{a-1} \,(a+ b-2 - 2i)\;p_{i-1}(x)\;p_{a+b - 3-i}(x)\right\}\,+$$
\begin{equation}\label{Pijuno}
+\, p_{a-1}(x)\;p_{b -1 }(x)\,.
\end{equation}
The first four lines of the above formula correspond to the terms of Eq. (\ref{dimSnsimm}) containing the matrix elements $P_{ab}(x)$, $P_{a,b-1}(x)$, $P_{a-1,b}(x)$ and $P_{a-1,b-1}(x)$, respectively. In the second line instead of Eq. (\ref{PjkdiSnbis}) we used the original formula (\ref{PjkdiSn}) and could not simplify further $\min(a,b-1)$, because in the present case $a\leq b$ (our hypothesis) we have to distinguish the case $a<b$ from the case $a=b$. We shall do that later.
The limitations on the indices regarding these four lines are as follows: 1st line: $a\leq b\leq n$, 2nd line: $a\leq b$ and $a\leq n$ and $2\leq b$, 3rd line: $2\leq a\leq b\leq n$, 4th line: $2\leq a$ and $2\leq b$ and $a\leq b$, respectively. These limitations are automatically satisfied, that is, if $b=n+1$, the first and third line vanish, if $a=b=n+1$ the second line vanishes, if $a=1$ the third and fourth line vanish, and if $a=b=1$  the second, in addition to the third and fourth line vanish, using $p_{-1}=0$. We can then avoid to take into account in the following the limitations on the indices. We can then write, $\forall\, a\leq b=1,\ldots,n+1$:
$$P_{ab}(y)=(n+2-a)\;p_{a-1}(x)\;p_{b -1}(x)- \sum_{i=a+b-n-1}^{a} \,(a+ b - 2i)\;p_{i-1}(x)\;p_{a+b - 1-i}(x)\,+$$
$$+\,x_{n+1}\,\left\{
(n+1-\min(a,b-1))\;p_{a-1}(x)\;p_{b -2}(x)- \sum_{i=a+b-n-2}^{\min(a,b-1)} \,(a+ b-1 - 2i)\;p_{i-1}(x)\;p_{a+b - 2-i}(x)\,+\right.$$
$$\left.+\,
(n+2-a)\;p_{a-2}(x)\;p_{b -1}(x)- \sum_{i=a+b-n-2}^{a-1} \,(a+ b-1 - 2i)\;p_{i-1}(x)\;p_{a+b - 2-i}(x)
\right\}+
$$
$$ +\,x_{n+1}^2\,\left\{(n+2-a)\;p_{a-2}(x)\;p_{b -2}(x)- \sum_{i'=a+b-n-2}^{a} \,(a+ b-2i') \;p_{i'-2}(x)\;p_{a+b - 2-i'}(x)\right\}\,.$$
The term inside the first curly brackets, that one multiplying $x_{n+1}$, can be further simplified.\\
One notes that both sums that appear inside the curly brackets contain equal terms up to the value $i=a-1$, because $a\leq b$. Moreover the first sum contains one more term if $a-1\neq \min(a,b-1)$, that is if $a<b$ (hence not if $a=b$), that corresponds to $i=a$. One has then:
$$- \sum_{i=a+b-n-2}^{\min(a,b-1)} \,(a+ b-1 - 2i)\;p_{i-1}(x)\;p_{a+b - 2-i}(x)- \sum_{i=a+b-n-2}^{a-1} \,(a+ b-1 - 2i)\;p_{i-1}(x)\;p_{a+b - 2-i}(x)=$$
$$=- \sum_{i=a+b-n-2}^{a-1} \,2\,(a+ b-1 - 2i)\;p_{i-1}(x)\;p_{a+b - 2-i}(x)- (1-\delta_{ab})(a+ b-1 - 2a)\;p_{a-1}(x)\;p_{a+b - 2-a}(x)=$$
$$= (a-b+1-\delta_{ab})\;p_{a-1}(x)\;p_{b-2}(x)- \sum_{i=a+b-n-2}^{a-1} \,2\,(a+ b-1 - 2i)\;p_{i-1}(x)\;p_{a+b - 2-i}(x)\,,$$
where $\delta_{ab}$ is the Kronecker delta.\\
The two terms without the sum inside the curly brackets multiplying $x_{n+1}$ can be transformed in the following way, that can be easily verified by distinguishing the different cases $a<b$ and $a=b$:
$$(n+1-\min(a,b-1))\;p_{a-1}(x)\;p_{b -2}(x)+(n+2-a)\;p_{a-2}(x)\;p_{b -1}(x)=
$$
$$=(n+1-a+\delta_{ab})\;p_{a-1}(x)\;p_{b -2}(x)+(n+2-a)\;p_{a-2}(x)\;p_{b -1}(x)\,.
$$
One can then write the term inside the curly brackets multiplying $x_{n+1}$  as follows:
$$x_{n+1}\,\{\ldots\}=x_{n+1}\,\Biggl\{(n+1-a+\delta_{ab})\;p_{a-1}(x)\;p_{b -2}(x)+(n+2-a)\;p_{a-2}(x)\;p_{b -1}(x)\,+$$
$$+\,(a-b+1-\delta_{ab})\;p_{a-1}(x)\;p_{b-2}(x)- \sum_{i=a+b-n-2}^{a-1} \,2\,(a+ b-1 - 2i)\;p_{i-1}(x)\;p_{a+b - 2-i}(x)\Biggr\}=
$$
$$=x_{n+1}\,\Biggl\{(n+2-b)\;p_{a-1}(x)\;p_{b -2}(x)+(n+2-a)\;p_{a-2}(x)\;p_{b -1}(x)\,+$$
$$-\,\sum_{i=a+b-n-2}^{a-1} \,2\,(a+ b-1 - 2i)\;p_{i-1}(x)\;p_{a+b - 2-i}(x)\Biggr\}\,.
$$
We have then, finally, the following expression, $\forall\,a,b=1,\ldots,n+1$:
$$P_{ab}(y)=(n+2-a)\;p_{a-1}(x)\;p_{b -1}(x)- \sum_{i=a+b-n-1}^{a} \,(a+ b - 2i)\;p_{i-1}(x)\;p_{a+b - 1-i}(x)\,+$$
$$+\,x_{n+1}\,\Biggl\{(n+2-b)\;p_{a-1}(x)\;p_{b -2}(x)+(n+2-a)\;p_{a-2}(x)\;p_{b -1}(x)-\sum_{i=a+b-n-2}^{a-1} \,2\,(a+ b-1 - 2i)\;p_{i-1}(x)\;p_{a+b - 2-i}(x)\Biggr\}\,+
$$
\begin{equation}\label{dimSnfinale1}
  +\,x_{n+1}^2\,\Biggl\{(n+2-a)\;p_{a-2}(x)\;p_{b -2}(x)- \sum_{i=a+b-n-2}^{a} \,(a+ b-2i) \;p_{i-2}(x)\;p_{a+b - 2-i}(x)\Biggr\}\,.
\end{equation}
Using Eq. (\ref{PjkdiSnbis}) in the case of $S_{n+1}$, that is for $n$ substituted by $n+1$,  and Eq. (\ref{p(y)diSn}), one would instead find, $\forall\,a\leq b=1,\ldots,n+1$:
$$  {\widehat P}_{ab}(p(y))= (n+2-a)\;p_{a-1}(y)\;p_{b-1}(y)- \sum_{i=a+b-(n+1)-1}^{a} \,(a+ b - 2i)\;p_{i -1}(y)\;p_{a+b - 1-i}(y)=
$$
$$= (n+2-a)\;(p_{a-1}(x)+p_{a-2}(x)\,x_{n+1})\;(p_{b-1}(x)+p_{b-2}(x)\,x_{n+1})\,+$$
 $$-\,\sum_{i=a+b-n-2}^{a} \,(a+ b - 2i)\;(p_{i - 1}(x)+p_{{i - 2}}(x)\,x_{n+1})\;(p_{a+b - 1-i}(x)+p_{a+b - 2-i}(x)\,x_{n+1})=
$$
$$= (n+2-a)\,\left\{\;p_{a-1}(x)\;p_{b-1}(x)+x_{n+1}\,\left[p_{a-1}(x)\;p_{b-2}(x)+p_{a-2}(x)\;p_{b-1}(x)\right]+
x_{n+1}^2\;p_{a-2}(x)\;p_{b-2}(x)\right\}+$$
 $$-\,\sum_{i=a+b-n-2}^{a} \,(a+ b - 2i)\;p_{i - 1}(x)\;p_{a+b - 1-i}(x)-x_{n+1}\;\sum_{i=a+b-n-2}^{a} \,(a+ b - 2i)\;p_{i-1}(x)\;p_{a+b - 2-i}(x)\,+$$
$$-\,x_{n+1}\;\sum_{i=a+b-n-2}^{a} \,(a+ b - 2i)\;p_{{i - 2}}(x)\;p_{a+b - 1-i}(x)-x_{n+1}^2\;\sum_{i=a+b-n-2}^{a} \,(a+ b - 2i)\;p_{{i - 2}}(x)\;p_{a+b - 2-i}(x)=
$$
$$=(n+2-a)\,p_{a-1}(x)\;p_{b-1}(x)-\sum_{i=a+b-n-1}^{a} \,(a+ b - 2i)\;p_{i - 1}(x)\;p_{a+b - 1-i}(x)\,+$$
$$+\,x_{n+1}\,\Biggl\{(n+2-a)\,\left[p_{a-1}(x)\;p_{b-2}(x)+p_{a-2}(x)\;p_{b-1}(x)\right]-
\sum_{i=a+b-n-2}^{a} \,(a+ b - 2i)\;p_{i-1}(x)\;p_{a+b - 2-i}(x)\,+$$
$$-\,\sum_{i=a+b-n-1}^{a} \,(a+ b - 2i)\;p_{{i - 2}}(x)\;p_{a+b - 1-i}(x)\Biggr\}\,+
$$
$$
+\,x_{n+1}^2\;\Biggl\{(n+2-a)\,p_{a-2}(x)\;p_{b-2}(x)-\sum_{i=a+b-n-2}^{a} \,(a+ b - 2i)\;p_{{i - 2}}(x)\;p_{a+b - 2-i}(x)\Biggr\}\,.
$$
The lower limits $i=a+b-n-2$ of the sums in the last expression have sometimes been increases, because when there is a factor  $p_{a+b - 1-i}(x)$ inside the sums, we have the limitation $a+b - 1-i\leq n$, that is, $i\geq a+b-n- 1$, otherwise this factor is zero.
The terms in this expression for ${\widehat P}_{ab}(p(y))$ that do not factorize $x_{n+1}$ and those that factorize $x_{n+1}^2$ are the same as those in Eq. (\ref{dimSnfinale1}). The terms that factorize $x_{n+1}$ must be further transformed to see that they also coincide with those in Eq.  (\ref{dimSnfinale1}). This can be done in the following way:
$$(n+2-a)\,\left[p_{a-1}(x)\;p_{b-2}(x)+p_{a-2}(x)\;p_{b-1}(x)\right]-
\sum_{i=a+b-n-2}^{a} \,(a+ b - 2i)\;p_{i-1}(x)\;p_{a+b - 2-i}(x)\,+$$
$$-\,\sum_{i=a+b-n-1}^{a} \,(a+ b - 2i)\;p_{{i - 2}}(x)\;p_{a+b - 1-i}(x)=$$
$$=(n+2-a)\;p_{a-1}(x)\;p_{b-2}(x)+(n+2-a)\;p_{a-2}(x)\;p_{b-1}(x)-\sum_{i=a+b-n-2}^{a-1} \,(a+ b - 2i)\;p_{i-1}(x)\;p_{a+b - 2-i}(x)\,+$$
$$-\,(b-a)\;p_{a-1}(x)\;p_{b - 2}(x)-\sum_{i'=a+b-n-2}^{a-1} \,(a+ b -2- 2i')\;p_{{i' - 1}}(x)\;p_{a+b - 2-i'}(x)=$$
$$=(n+2-b)\;p_{a-1}(x)\;p_{b-2}(x)+(n+2-a)\;p_{a-2}(x)\;p_{b-1}(x)-\sum_{i=a+b-n-2}^{a-1} \,2\,(a+ b - 2i)\;p_{i-1}(x)\;p_{a+b - 2-i}(x)\,+$$
$$-\,\sum_{i=a+b-n-2}^{a-1} \,(-2)\;p_{{i - 1}}(x)\;p_{a+b - 2-i}(x)=$$
$$=(n+2-b)\;p_{a-1}(x)\;p_{b-2}(x)+(n+2-a)\;p_{a-2}(x)\;p_{b-1}(x)-\sum_{i=a+b-n-2}^{a-1} \,2\,(a+ b - 1- 2i)\;p_{i-1}(x)\;p_{a+b - 2-i}(x)\,.$$
This expression coincides with the expression in Eq. (\ref{dimSnfinale1}) inside the curly brackets multiplying $x_{n+1}$.\\
We have so verified that the expression of ${\widehat P}_{ab}(p(y))$ obtained above, using Eq. (\ref{PjkdiSnbis}) in the case of $S_{n+1}$ coincides with the expression of $P_{ab}(y)$ written in Eq. (\ref{dimSnfinale1}), that was obtained from the explicit calculation of $P_{ab}(y)$ using Eq. (\ref{matriceP(x)}) and the assumed validity of Eq. (\ref{PjkdiSnbis}) in the case of $S_{n}$, using the induction principle. This proves that Eq. (\ref{PjkdiSnbis}) is true for $S_n$ for all values of the positive integer $n$. This holds also for the generating function (\ref{PjkdiSn}) because of the equivalence of Eqs. (\ref{PjkdiSn}) and (\ref{PjkdiSnbis}), given the definitions $p_a=0$, $\forall\,a<0$ and $\forall\,a>n$, valid for Eq. (\ref{PjkdiSnbis}).\scat

\noindent\textbf{Proof of the generating formula (\ref{lambdaSnEq}).}\\
Theorem \ref{quattro} tells us how to construct the $\lambda$-vector $\lambda_a^{(\det(\widehat P))}(p)$ corresponding to the determinant $\det(\widehat P(p))$ of the \wPm\ $\widehat P(p)$.\\
The positive roots of $S_n$ are vectors orthogonal to the reflection hyperplanes, and are for example the $n\choose 2$ vectors $e_i-e_j$, $1\leq i<j\leq n$. The corresponding linear forms vanishing on the reflecting hyperplanes are then the following: $x_i- x_j$, $1\leq i<j\leq n$.
Eq. (\ref{lambdaD}) gives then, for all $a=1,\ldots,n$:
$$  \lambda_a^{(\det(\widehat P))}(p(x))=2\sum_{g\in{\cal R}}\frac{\nabla p_a(x)\cdot r_g}{l_g(x)}=
2\sum_{i<j=1}^n\,\frac{\nabla p_a(x)\cdot(e_i-e_j)}{x_i-x_j}=
2\sum_{i<j=1}^n\,\frac{1}{x_i-x_j}\,\left(\frac{\partial p_a(x)}{\partial x_i}-\frac{\partial p_a(x)}{\partial x_j}\right)
$$
Using the basic invariant polynomials in Eq. (\ref{elemsymmpolSn}), we see that for $a=1$ one has $\frac{\partial p_1(x)}{\partial x_i}=1$, $\forall\,i=1,\ldots,n$, so that $ \lambda_1^{(\det(\widehat P))}(p(x))=0$, while for $a>1$ one has:
$$\frac{\partial p_a(x)}{\partial x_i}=\left.p_{a-1}(x)\right|_{x_i=0}\,,
$$
that gives, $\forall\,a=2,\ldots,n$:
$$  \lambda_a^{(\det(\widehat P))}(p(x))=
2\sum_{i<j=1}^n\,\frac{1}{x_i-x_j}\,\left(\left.p_{a-1}(x)\right|_{x_i=0}-\left.p_{a-1}(x)\right|_{x_j=0}\right)=$$
$$=
2\sum_{i<j=1}^n\,\frac{1}{x_i-x_j}\,\left(x_j\,\left.p_{a-2}(x)\right|_{x_i=x_j=0}-x_i\,\left.p_{a-2}(x)\right|_{x_i=x_j=0}\right)=
2\sum_{i<j=1}^n\,\frac{x_j-x_i}{x_i-x_j}\,\left.p_{a-2}(x)\right|_{x_i=x_j=0}=
$$
$$=
-2\sum_{i<j=1}^n\,\left.p_{a-2}(x)\right|_{x_i=x_j=0}
$$
The last expression is a completely symmetric $w$-homogeneous polynomial of weight $a-2$ in the variables $x_1,\ldots,x_n$, and is formed by terms contained in $p_{a-2}(x)$, so it must be a multiple of the basic invariant polynomial $p_{a-2}(x)$. To find the multiplying factor it is sufficient to count the number of terms, $c_1$, in the last expression and the number of terms, $c_2$, in $p_{a-2}(x)$. The multiplying factor is then equal to the ratio $\frac{c_1}{c_2}$. One easily finds that $c_1={n\choose 2}{n-2\choose a-2}$ and $c_2={n\choose a-2}$, so that $\frac{c_1}{c_2}=\frac{1}{2}\,{(n-a+2)(n-a+1)}$.
One then has:
$$  \lambda_a^{(\det(\widehat P))}(p(x))=-2\,\frac{1}{2}\,{(n-a+2)(n-a+1)}\,p_{a-2}(x)=-(n-a+2)(n-a+1)\,p_{a-2}(x)\,,\qquad \forall\,a=2,\ldots,n\,.
$$
Eq. (\ref{lambdaSnEq}) is then proved.
\scat

\subsection{Proof of Theorem \ref{lambdaAn} for $A_n$\label{proofAn}}
\noindent\textbf{Proof of the generating formula (\ref{PjkdiAn}).}\\
We adopt here the notation introduced in Section \ref{An}, concerning the groups $A_n$ and in particular, the definition (\ref{matRn+1}) of the rotation matrix $R_{n+1}$, and the derivation of the basic invariant polynomials $p_1(\bar x'),\ldots,p_n(\bar x')$, of $A_n$ from the basic invariant polynomials $s_1(x),\ldots,s_{n+1}(x)$, of $S_{n+1}$, as specified by Eqs. (\ref{elemsymmpolSn+1})--(\ref{basicpolynAn}).\\
We know the explicit form of the  \wPm\ $\widehat P(s)$ of $S_{n+1}$, when the basic invariant polynomials of $S_{n+1}$ are the elementary symmetric polynomials (\ref{elemsymmpolSn+1}). This explicit form is given by Eq. (\ref{PjkdiSn}), where the variables $p_a$ have to be substituted by the variables $s_a$, and $n$ has to be substituted by $n+1$. This gives the following expressions of the matrix elements of the \wPm\ $ {\widehat P}(s)$ of $S_{n+1}$:
\begin{equation}\label{PjkdiSn3}
  {\widehat P}_{ab}(s)= [n+2-\min(a,b)]\;s_{a-1}\;s_{b -1}- \sum_{i=\max(1,a+b-n-2)}^{\min(a,b)} \,(a+ b - 2i)\;s_{i -1}\;s_{a+b - 1-i}\,,\qquad \forall\,a,b=1,\ldots,n+1\,,
\end{equation}
in which $s_0=1$. The matrix $\widehat P(s)$ is clearly symmetric. Eq. (\ref{PjkdiSn3}) is a good point to start with to prove Eq. (\ref{PjkdiAn}), and this is the road followed in the present proof. This proof is quite long and will be broken in a few lemmas.\\

The \wPm\ $\widehat P(s')$, corresponding to the basic invariant polynomials $s'_a(x')=s_a(R^\top_{n+1}x')$, $a=1,\ldots,n+1$, by Theorem \ref{uno}, item \ref{th1it4}., has the same form as  $\widehat P(s)$, only the variables $s_a$ have to be substituted by the variables $s'_a$, $\forall\,a=1,\ldots,n+1$. So, $\widehat P(s')$ is also given by Eq. (\ref{PjkdiSn3}), but with the variables $s_a$ substituted by the variables $s'_a$.\\

One first proves the following Lemma.\\

\noindent\textbf{Lemma 1}. The elementary symmetric polynomials $s_1(x)$ and $s_2(x)$ of $S_{n+1}$, written in Eq. (\ref{elemsymmpolSn+1}), are transformed by the rotation matrix $R_{n+1}$ into the following polynomials:
\begin{eqnarray*}
  s_1(x)\to s'_1(x') =s_1(R_{n+1}^\top x')&=& \sqrt{n+1}\;x'_{n+1} \\
  s_2(x)\to s'_2(x') =s_2(R_{n+1}^\top x') &=& {\frac{n}{2}\, {x'}^2_{n+1}}-\frac{1}{2}\,\sum_{i=1}^{n} {x'}^2_i \\
\end{eqnarray*}\\
\noindent\textbf{Proof}.
 Let's write $R$ for $R_{n+1}$ in this proof.
Let's first obtain the form of $s'_1(x')$. One has $s_1(x)=x_1+x_2+\ldots +x_{n+1}$, so that, by Theorem \ref{uno}, item \ref{th1it3}, one has:
$$s'_1(x')=s_1(R^\top x')=
\sum_{i=1}^{n+1}\left(\sum_{j=1}^{n+1}R^\top_{i,j} x'_j\right)=
\sum_{j=1}^{n+1}\left(\sum_{i=1}^{n+1}R_{j,i}\right)x'_j$$
The sum inside brackets in the last member is the sum of the elements of the $j$-th row of the matrix $R$, and by definition of $R$ this is 0, except when $j=n+1$, in which case it is equal to $(n+1)\,\frac{1}{\sqrt{n+1}}=\sqrt{n+1}$. All this gives $s'_1(x')=\sqrt{n+1}\;x'_{n+1}$.\\
Let's now obtain the form of $s'_2(x')$. One has, by elementary algebraic transformations:
$$s_2(x)= x_1 x_2 + x_1 x_3 +\ldots + x_{n} x_{n+1}=\frac{1}{2}\,\left[(x_1+x_2+\ldots +x_{n+1})^2-\sum_{i=1}^{n+1}x_i^2\right]=
\frac{1}{2}\,\left[(s_1(x))^2-\|x\|^2
\right]\,.$$
To calculate $s'_2(x')=s_2(R^\top x')$ one has to evaluate the previous expression for $x$ substituted by $R^\top x'$. The squared norm $\|x\|^2=\sum_{i=1}^{n+1}x_i^2$ of $x\in\R^{n+1}$ is invariant under orthogonal transformations in $\R^{n+1}$, that is: $\|x\|^2=\|R^\top x'\|^2=\|x'\|^2$. Using also the equality $s'_1(x')=s_1(R^\top x')= \sqrt{n+1}\,x'_{n+1}$, obtained above, one finds:
$$s'_2(x')=s_2(R^\top x')=
\frac{1}{2}\,\left[(\sqrt{n+1}\,x'_{n+1})^2-\|x'\|^2\right]=
\frac{1}{2}\,\left[({n+1})\,{x'}_{n+1}^2-\sum_{i=1}^{n+1}{x'}_i^2\right]=\frac{1}{2}\,\left[n\,{x'}_{n+1}^2-\sum_{i=1}^{n}{x'}_i^2\right]\,.
$$
The Lemma is so completely proved.\scat

The basic invariant polynomials of $A_n$ are obtained by substituting $x_{n+1}=0$ in the polynomials $s'_a(x')$, $a=2,\ldots,n+1$. We write $\bar x'$ for the projection in the hyperplane $x'_{n+1}=0$ of a vector $x'\in\R^{n+1}$ and $p(\bar x')$ for the restriction to the hyperplane $x'_{n+1}=0$ of a function $p(x')$ defined in $\R^{n+1}$ (typically a polynomial in $x'$), that is,  $p(\bar x')=\left.p(x')\right|_{x'_{n+1}=0}$.\\
From the expression of $s'_2(x')$ recovered in Lemma 1, one obtains
$$s'_2(\bar x')=-\frac{1}{2}\,\sum_{i=1}^{n}{x'}_i^2\,.$$ This expression has not the standard form (\ref{quadraticinv}) one wishes for the quadratic invariant polynomial. To get the standard form (\ref{quadraticinv}) it is then necessary a further transformation. The simplest one is to multiply the basic polynomial $s'_2(x')$ by a factor $-2$, for example, by defining  $s''_a(x')=s_a'(x')$, $\forall \, a=1,\ldots,n+1$, $a\neq 2$, and $s''_2(x')=-2s_2'(x')$, but the resulting matrix $\widehat P(s'')$ would then have some fractional coefficients. To avoid fractional coefficients in the matrix $\widehat P(s'')$ it is convenient to multiply all the basic invariant polynomials, not only $s'_2(x')$, by the factor $-2$:
 \begin{equation}\label{scaletransf2}
  s''_a(x')=-2\,s'(x')\,,\qquad \forall \, a=1,\ldots,n+1\,.
\end{equation}
The \wPm\ $\widehat P(s'')$, can be determined from $\widehat P(s')$ in the way specified by Eq. (\ref{Ptransf}) in Theorem \ref{tre}. The Jacobian matrix $J(s')$ of this transformation is $J(s')=-2\,I$, with $I$ the unit matrix of order $n+1$, so one finds $J(s')\; \widehat P(s')\; J^\top(s')=4\, \widehat P(s')$. The substitution $p=\widehat p(p')$  in Eq. (\ref{Ptransf}), that here becomes $s'_a= -\frac{1}{2}s''_a$, $\forall \, a=1,\ldots,n+1$, in addition to the substitution of $s'_a$ with $s''_a$, brings a factor $-\frac{1}{2}$ in front of all variables $s'_a$ appearing in the matrix elements of $\widehat P(s')$. In $\widehat P(s')$ there are at most quadratic terms in the variables $s'_1,\ldots,s'_{n+1}$, so, after this scale transformation the coefficients of the quadratic terms are multiplied by a factor $\left(-\frac{1}{2}\right)^2$, those of the liner terms by a factor $-\frac{1}{2}$, and the constant terms are not multiplied at all.  Then, with all these substitutions, all the elements of the \wPm\ change some coefficients, but $\widehat P(s'')$ remains without fractions, because of the global factor 4.\\
For later use it is convenient to write explicitly the elements of the matrix $\widehat P(s'')$ corresponding to the basic invariant polynomials $s''_a(x')$ of $S_{n+1}$ (that are obtained from the basic invariant polynomials (\ref{elemsymmpolSn+1}) by performing the rotation of $\R^{n+1}$ obtained through the matrix $R_{n+1}$, defined in Eq. (\ref{matRn+1}), and the scale transformation (\ref{scaletransf2})). To obtain the matrix $\widehat P(s'')$, it is necessary to distinguish the terms containing zero, one or two variables $s'_1,\ldots,s'_{n+1}$, because they are multiplied by different coefficients, so, it is also convenient to distinguish the element $\widehat P_{11}(s'')$, the other elements of the first row and column of the \wPm\ $\widehat P(s'')$, and the rest of the \wPm\ $\widehat P(s'')$. In this last case, using Eq. (\ref{PjkdiSn3}), with $a,b\geq 2$, a linear term in the variables $s'_1,\ldots,s'_{n+1}$ appears only if there is an $i$ for which $i-1=0$, because then one finds the variable $s'_0=1$. When $i=1$, the variable $s'_{a+b-1-i}=s'_{a+b-2}$ is not equal to 0 only if $a+b-2\leq n+1$, that is when $a+b\leq n+3$.  (The only other possibility is when $a+b-1-i=1$, that is when $i=a+b-2$, and the limitations imply that that is possible only if $i=a=b=2$, but in that case its coefficient vanish: $a+b-2i=0$). We can then easily find, from Eq. (\ref{PjkdiSn3}) (in which the variable $s_a$ is substituted by the variable $s'_a$, $\forall a\,=1,\ldots,n+1$), using Theorem \ref{tre} (that implies the multiplication of all the terms by factors $4\cdot 1$, $4\cdot\left(-\frac{1}{2}\right)$, $4\cdot \frac{1}{4}$, as specified above), the following expressions for the elements of the \wPm\ $\widehat P(s'')$:
\begin{eqnarray}\label{PjkdiSntrasf-2}
\nonumber  {\widehat P}_{11}(s'')&=& 4\, (n+1)\,,\\
\nonumber  {\widehat P}_{1a}(s'')={\widehat P}_{a1}(s'')&=& -2 (n+2-a)\;s''_{a-1}\,,\qquad\qquad\qquad\qquad\qquad\qquad\qquad\qquad {\forall\,a=2,\ldots,n+1\,,}\\
\nonumber  {\widehat P}_{ab}(s'')&=& [n+2-\min(a,b)]\;s''_{a-1}\;s''_{b-1}+2 \,(a+ b -2)\;s''_{a+b-2}+ \\
&-&  \sum_{i=\max(2,a+b-n-2)}^{\min(a,b)} \,(a+ b - 2i)\;s''_{i -1}\;s''_{a+b - 1-i}\,,\qquad\qquad\quad\ \forall\,a,b=2,\ldots,n+1\,,\qquad
\end{eqnarray}
The term with $s''_{a+b-2}$ is present only if $a+b\leq n+3$, that is if $a+b-2\leq n+1$. To  take into account this condition it is convenient to define, in addition to $s''_0=1$, that was required by Eq. (\ref{PjkdiSn3}), also $s''_a=0$, $\forall\,a>n+1$.
We can observe that in Eq. (\ref{PjkdiSntrasf-2}) the variable $s''_0$ cannot be present, because of the limitations, so it is simpler to consider for Eq. (\ref{PjkdiSntrasf-2}) only the constraints $s''_a=0$, $\forall\,a>n+1$.\\

The two transformations performed up to now commute. It is possible first to rotate the system of coordinates and then to multiply by $-2$ the basic invariant polynomials or vice versa, first to multiply by $-2$ the basic invariant polynomials and then to rotate the system of coordinates. In both cases one ends with the same expressions both for the basic invariants $s''_a(x')$, and for the matrix $\widehat P(s'')$. Actually, only the multiplication by $-2$ is sufficient to obtain the form (\ref{PjkdiSntrasf-2}) of the \wPm.\\

The basic invariant polynomials of $A_n$ are easily found from the basic invariant polynomials $s''_a(x')$, $a=1,\ldots,n+1$, by taking $x'_{n+1}=0$. This leaves the $n$ basic invariant polynomials $s''_a(\bar x')$, $a=2,\ldots,n+1$, of degrees $2,3,\ldots,n$, only, because $s''_1(\bar x')=0$. Moreover, the quadratic invariant $s''_2(\bar x')$ has the desired standard form (\ref{quadraticinv}), equal to the squared norm: $s''_2(\bar x')=\|\,\bar x'\|^2$. A renaming the basic invariant polynomials is then convenient, like in the following way:
\begin{equation}\label{rinomina}
  p_a(\bar x')= s''_{a+1}(\bar x')\,,\qquad \forall \,a=1,\ldots,n\,.
\end{equation}

The $n\times n$ matrix $\widehat P(p)$ can be obtained through Eqs. (\ref{matriceP(x)}) and (\ref{invmatP}) from the basic invariant polynomials $p_a(\bar x')$, $a=1,\ldots,n$, written in Eq. (\ref{rinomina}), but this is not what we want to do here. The same matrix $\widehat P(p)$ can also be determined from the $(n+1)\times(n+1)$  matrix $\widehat P(s'')$, and this is what we are going to do, but with  caution. The first idea is to substitute $s''_1=0$ in the matrix $\widehat P(s'')$ (because $s''_1(x')\propto x'_{n+1}$, and when $x'_{n+1}=0$, one has $s''_1(x')=0$), to remove its first row and column (because there is no more an invariant of degree 1, and the first row and column of the matrix $\left.\widehat P(s'')\right|_{s''_1=0}$ have no more meaning), and to rename the variables as suggested by Eq. (\ref{rinomina}): $s''_{a+1}=p_a$, $\forall \,a=1,\ldots,n$. However, doing that way one generally does not find the right \wPm\ $\widehat P(p)$, corresponding to the basic invariant polynomials $p_a(\bar x')$, $a=1,\ldots,n$, that can also be calculated from its definition, that is from Eqs. (\ref{matriceP(x)}) and (\ref{invmatP}). We are now going to see when and how the knowledge of the matrix $\widehat P(s'')$ can determine the matrix $\widehat P(p)$.\\

Let's define a matrix operator $[\:\cdot\;]^p$ that applied to a $(n+1)\times(n+1)$ matrix $\widehat P(s'')$, dependent on $n+1$ variables $s''_1,\ldots,s''_{n+1}$, gives an $n\times n$ matrix $[\widehat P(s'')]^p$, dependent on $n$ variables $p_1,\ldots,p_{n}$, obtained from the matrix $\widehat P(s'')$ by applying the following rules:
\begin{enumerate}
  \item make the substitution ${s''_1=0}$ in the matrix $\widehat P(s'')$;
  \item delete the first row and column of the resulting matrix;
  \item rename the variables like in Eq. (\ref{rinomina}): $s''_{a+1}=p_a$, $\forall \,a=1,\ldots,n$.
\end{enumerate}

One would desire that $[\widehat P(s'')]^p=\widehat P(p)$, with $\widehat P(p)$ the \wPm\ corresponding (through Eqs. (\ref{matriceP(x)}) and (\ref{invmatP})) to the basic invariant polynomials of $A_n$ in Eq. (\ref{rinomina}). However, this is in general not true, because the elements of the matrix $[\widehat P(s'')]^p$ can be obtained by making the scalar products of gradients in $\R^{n+1}$, $\nabla s''_{a+1}(x')\cdot\nabla s''_{b+1}(x')$, $\forall\,a,b=1,\ldots,n$, by expressing the results in terms of the basic invariant polynomials $s''_{a}(x')$, $a=1,\ldots,n+1$, and then by doing the substitutions $s''_1(x')=0$, $s''_{a+1}(x')= p_a$, $\forall\,a=1,\ldots,n$, while the elements of the matrix $\widehat P(p)$ can be obtained by making the scalar products of gradients in $\R^{n}$, $\nabla p_a(\bar x')\cdot\nabla p_b(\bar x')$,  $\forall\,a,b=1,\ldots,n$, by expressing the results in terms of the basic invariant polynomials $p_{a}(\bar x')$, $a=1,\ldots,n$, and then by doing the substitutions $p_{a}(\bar x')= p_a$, $\forall\,a=1,\ldots,n$, and these two procedures give generally different results. There are, actually, infinitely many choices of a set of $n+1$ basic invariant polynomials $z_a(x')$ of $S_{n+1}$, $a=1,\ldots,n+1$, with $z_1(x')\propto x'_{n+1}$, that give rise, by taking $x'_{n+1}=0$, to the same set of $n$ basic invariant polynomials $p_a(\bar x')$ of $A_n$, but that give rise to different matrices $[\widehat P(z)]^p$. So, only clever choices of the basic invariant polynomials $z_a(x')$ are such that $\widehat P(p)=[\widehat P(z)]^p$, as one desires.\\

Let's prove the following Lemma.\\

\noindent\textbf{Lemma 2}. A set of basic invariant polynomials of $S_{n+1}$, $z_a(x')$, $a=1,\ldots,n+1$, with $z_1(x')\propto x'_{n+1}$, by taking $x'_{n+1}=0$, can determine a set of basic invariant polynomials of $A_n$, $p_a(\bar x')=z_{a+1}(\bar x')$, $a=1,\ldots,n$, with the property that $\widehat P(p)=[\widehat P(z)]^p$, where $\widehat P(p)$ and $\widehat P(z)$ are determined, through Eqs. (\ref{matriceP(x)}) and (\ref{invmatP}), by the sets of basic invariant polynomials $p_a(\bar x')$, $a=1,\ldots,n$, of $A_n$, and $z_a(x')$, $a=1,\ldots, n+1$, of $S_{n+1}$, respectively, if and only if $\left.\widehat P_{1a}(z)\right|_{z_{1}=0}=0$, $\forall\,a=2,\ldots, n+1$.\\

\noindent\textbf{Proof}. Let's use the symbols $\nabla_{(n)}$ and $\nabla_{(n+1)}$ for the gradients in $\R^n$ and $\R^{n+1}$, respectively. Adopting the usual notation $x'\in\R^{n+1}$, and $\bar x'=\left.x'\right|_{x'_{n+1}=0}\in\R^n$, one then has:
$$\nabla_{(n)}=\left(\frac{\partial}{\partial \bar x'_1},\ldots,\frac{\partial}{\partial \bar x'_{n}}\right)\,,\qquad \nabla_{(n+1)}=\left(\frac{\partial}{\partial x'_1},\ldots,\frac{\partial}{\partial x'_{n+1}}\right)\,.$$
Consider the matrix element $\widehat P_{ab}(p)$, $\forall\,a,b=1,\ldots,n$. One has:
$$\widehat P_{ab}(p(\bar x'))=\nabla_{(n)}p_a(\bar x')\cdot \nabla_{(n)}p_b(\bar x')\,.$$
If the equality $\widehat P(p)=[\widehat P(z)]^p$ is true $\forall\,a,b=1,\ldots,n$, the matrix element $\widehat P_{ab}(p(\bar x'))$, after making in it the substitution $p_a(\bar x')=z_{a+1}(\bar x')$, $\forall\,a=1,\ldots,n$, must be equal to the following matrix element:
$$\left.\widehat P_{a+1,b+1}(z(x'))\right|_{x'_{n+1}=0}=\left.\left(\nabla_{(n+1)}z_{a+1}(x')\cdot \nabla_{(n+1)}z_{b+1}(x')\right)\right|_{x'_{n+1}=0}=$$
$$=\nabla_{(n)}z_{a+1}(\bar x')\cdot \nabla_{(n)}z_{b+1}(\bar x')+\left.\left(\frac{\partial z_{a+1}(x')}{\partial x'_{n+1}}\cdot \frac{\partial z_{b+1}(x')}{\partial x'_{n+1}}\right)\right|_{x'_{n+1}=0}
$$
The first term in the last member, with $z_{a+1}(\bar x')$ substituted by $p_a(\bar x')$, $\forall\,a=1,\ldots,n$, is equal to $\nabla_{(n)}p_{a}(\bar x')\cdot \nabla_{(n)}p_{b}(\bar x')=\widehat P_{ab}(p(\bar x'))$, and, for the hypothesis $\widehat P(p)=[\widehat P(z)]^p$, is also equal to $\left.\widehat P_{a+1,b+1}(z(x'))\right|_{x'_{n+1}=0}$. So, if the equality $\widehat P(p)=[\widehat P(z)]^p$ is true, one has:
$$\left.\left(\frac{\partial z_{a+1}(x')}{\partial x'_{n+1}}\cdot \frac{\partial z_{b+1}(x')}{\partial x'_{n+1}}\right)\right|_{x'_{n+1}=0}=0\,,\qquad \forall\, a,b=1,\ldots, n\,.$$
In particular, by taking $a=b$ one finds:
\begin{equation}\label{lemma3uno}
   \left.\left.\frac{\partial z_{a+1}(x')}{\partial x'_{n+1}}\right.\right|_{x'_{n+1}=0}=0\,,\qquad \forall\, a=1,\ldots, n\,.
\end{equation}
From the hypothesis $z_1(x')\propto x'_{n+1}$, one has:
\begin{equation}\label{lemma3due}
 \nabla_{(n+1)}z_{1}(x')
 =\frac{\partial z_1(x')}{\partial x'_{n+1}}\, e_{n+1}=c\;e_{n+1}\,,
\end{equation}
where $e_{n+1}$ is the canonical unit vector of $\R^{n+1}$ orthogonal to the hyperplane $x'_{n+1}=0$, and $c$ is a non-zero real number. One then obtains, using Eqs. (\ref{lemma3due}) and (\ref{lemma3uno}), $\forall\,a=1,\ldots,n$:
$$\left.\widehat P_{1,a+1}(z(x'))\right|_{x'_{n+1}=0}=\left.\left(\nabla_{(n+1)}z_{1}(x')\cdot \nabla_{(n+1)}z_{a+1}(x')\right)\right|_{x'_{n+1}=0}=\left.\left(c\;\frac{\partial z_{a+1}(x')}{\partial x'_{n+1}} \right)\right|_{x'_{n+1}=0}=0\,.$$
This condition is equivalent to  $\left.\widehat P_{1a}(z)\right|_{z_{1}=0}=0$, $\forall\,a=2,\ldots, n+1$, because $z_1(x')\propto x'_{n+1}$ implies that the condition $x'_{n+1}=0$ is equivalent to the condition $z_1(x')=0$.
The necessity (only if part) of the statement of the Lemma is proved.\\
Vice versa, supposing  $\left.\widehat P_{1a}(z)\right|_{z_{1}=0}=0$, $\forall\,a=2,\ldots, n+1$, that, because of $z_1(x')\propto x'_{n+1}$, is equivalent to:
$$\left.\widehat P_{1,a+1}(z(x'))\right|_{x'_{n+1}=0}=0\,,\qquad \forall\,a=1,\ldots,n\,,$$
one obtains the following conditions: $$\left.\left(\nabla_{(n+1)}z_{1}(x')\cdot \nabla_{(n+1)}z_{a+1}(x')\right)\right|_{x'_{n+1}=0}=0\,,\quad \forall\,a=1,\ldots,n\,.$$
For the hypothesis $z_1(x')\propto x'_{n+1}$, that implies Eq. (\ref{lemma3due}), the previous equation implies Eq. (\ref{lemma3uno}), but then, $\forall\,a,b=1,\ldots,n$, one has:
$$\left.\widehat P_{a+1,b+1}(z(x'))\right|_{x'_{n+1}=0}=\left.\left(\nabla_{(n+1)}z_{a+1}(x')\cdot \nabla_{(n+1)}z_{b+1}(x')\right)\right|_{x'_{n+1}=0}=\left.\left(\nabla_{(n)}z_{a+1}(x')\cdot \nabla_{(n)}z_{b+1}(x')\right)\right|_{x'_{n+1}=0}=
$$
$$=\nabla_{(n)}z_{a+1}(\bar x')\cdot \nabla_{(n)}z_{b+1}(\bar x')=\nabla_{(n)}p_{a}(\bar x')\cdot \nabla_{(n)}p_{b}(\bar x')=\widehat P_{ab}(p(\bar x'))\,,$$ with $p_a(\bar x')$ substituted by $z_{a+1}(\bar x')$, $\forall\,a=1,\ldots,n$,
and this implies $\widehat P(p)=[\widehat P(z)]^p$. The sufficiency (if part) of the statement of the Lemma is proved.\scat

Now we have to find the transformation sending the basis $s''_a(x')$, $a=1,\ldots,n+1$, of invariant polynomials of $S_{n+1}$, to a basis $z_a(x')$, $a=1,\ldots,n+1$, of invariant polynomials of $S_{n+1}$ that satisfies the hypothesis  of Lemma 2, that is: $\left.\widehat P_{1a}(z)\right|_{z_{1}=0}=0$, $\forall\,a=2,\ldots, n+1$. To this end, one proves the following Lemma.\\

\noindent\textbf{Lemma 3}. The basic invariant polynomials $z_a(x')$, $a=1,\ldots,n+1$, of $S_{n+1}$, that are obtained from the basic invariant polynomials  $s''_a(x')$, $a=1,\ldots,n+1$, of $S_{n+1}$ (that are defined by Eqs. (\ref{elemsymmpolSn+1}), (\ref{trasfAn1}), and (\ref{scaletransf2})), in the following way:
\begin{eqnarray}\label{lintrasfinlemma3}
\nonumber  z_1(x')&=&s''_1(x'),\\
  z_a(x')&=&s''_a(x')+ c_a\,s''_1(x')\, s''_{a-1}(x')\,,\qquad c_a=\frac{n+2-a}{2\,(n+1)}\,,\qquad\qquad \forall\,a=2,\ldots,n+1\,,
\end{eqnarray}
have the property that $[\widehat P(z)]^p=\widehat P(p)$, where $p_a(\bar x')=z_{a+1}(\bar x')$, $\forall\,a=1,\ldots,n$, is a basis of invariant polynomials of $A_n$, and the matrix $\widehat P(p)$ coincides with their \wPm, as can be calculated from Eqs. (\ref{matriceP(x)}) and (\ref{invmatP}). The only other possibilities to find basic invariant polynomials $z_a(x')$, $a=1,\ldots,{n+1}$, with the property that $\widehat P(p)=[\widehat P(z)]^p$, is obtained by adding terms like ${s''_1}^2(x')\,q_{a-2}(s''(x'))$ to the polynomials $z_a(x')$, defined in Eq. (\ref{lintrasfinlemma3}), in which $a\geq 2$ and $q_{a-2}(s''(x'))$ is an arbitrary invariant homogeneous polynomial of degree $a-2$.\\

\noindent\textbf{Proof}.
Because of Lemma 2, to prove that the basic invariant polynomials $z_a(x')$, $a=1,\ldots,n+1$, satisfy the equation $[\widehat P(z)]^p=\widehat P(p)$, it is sufficient to prove that they satisfy the conditions $\left.\widehat P_{1a}(z)\right|_{z_{1}=0}=0$, $\forall\,a=2,\ldots, n+1$.\\
As $z_1(x')=s''_1(x')\propto x'_{n+1}$, one has $\left.z_1(x')\right|_{x'_{n+1}=0}=\left.s''_1(x')\right|_{x'_{n+1}=0}=z_1(\bar x')=s''_1(\bar x')=0$, and $\left.\widehat P_{1a}(z)\right|_{z_{1}=0}=\left.\widehat P_{1a}(z(x'))\right|_{x'_{n+1}=0}$, $\forall\,a=2,\ldots, n+1$.
Using the transformation (\ref{lintrasfinlemma3}), and the matrix $\widehat P(s'')$ written in Eq. (\ref{PjkdiSntrasf-2}), one finds, $\forall a=2,\ldots,n+1$:
$$\left. \widehat P_{1a}(z(x'))\right|_{x'_{n+1}=0}=\left.\left(\nabla z_1(x')\cdot \nabla z_a(x')\right)\right|_{x'_{n+1}=0}=\left.\left(\nabla s''_1(x')\cdot \nabla \left[s''_a(x')+ c_{a}\,s''_1(x')\, s''_{a-1}(x')\right]\right)\right|_{x'_{n+1}=0}=$$
$$=\left.\left(\nabla s''_1(x')\cdot \nabla s''_a(x')\right)\right|_{x'_{n+1}=0}+ c_{a}\,\left.\left( s''_{a-1}(x')\,\nabla s''_1(x')\cdot \nabla s''_1(x')\right)\right|_{x'_{n+1}=0}+ c_{a}\,\left.\left( s''_1(x')\,\nabla s''_1(x')\cdot \nabla s''_{a-1}(x')\right)\right|_{x'_{n+1}=0}=$$
$$=\left.\widehat P_{1a}(s''(\bar x'))\right.+ c_{a}\, \left.s''_{a-1}(\bar x')\,\widehat P_{11}(s''(\bar x'))\right.+0=
-2 (n+2-a)\;s''_{a-1}(\bar x')+ c_{a}\, s''_{a-1}(\bar x')\,4\, (n+1)=
$$
$$=\left(-2 (n+2-a)+ \frac{n+2-a}{2\,(n+1)}\, 4\, (n+1)\right)\;s''_{a-1}(\bar x')=0\,.
$$
The conditions just proved imply  $\left.\widehat P_{1a}(z)\right|_{z_{1}=0}=0$, $\forall\,a=2,\ldots, n+1$, and Lemma 2 assures that the equation $[\widehat P(z)]^p=\widehat P(p)$ is true. An additional term like ${s''_1}^2(x')\,q_{a-2}(s''(x'))$, added to a polynomial $z_a(x')$ in Eq. (\ref{lintrasfinlemma3}), $a=2,\ldots,n+1$, implies the following additional term to the gradient $\nabla z_a(x')$: $2s\,''_1(x')\,q_{a-2}(s''(x'))\,\nabla s\,''_1(x')+ {s''_1}^2(x')\,\nabla q_{a-2}(s''(x'))$, that, however, vanishes when $s''_1(x')=0$, and hence originates the same matrix $[\widehat P(z)]^p$. The same additional term ${s''_1}^2(x')\,q_{a-2}(s''(x'))$ vanishes when $x'_{n+1}=0$ and has no influence on the definition of the polynomials $p_a(\bar x')$, $\forall\,a=1,\ldots,n$, too.\scat

It is interesting to note that the basis transformation (\ref{lintrasfinlemma3}) was used also in Section (2.4.8) of Ref. \cite{SYS1980}, inside a proof concerning the \wPms\ of the groups of type $A_n$, although the proof in Ref. \cite{SYS1980} was done in view to prove the existence of a flat basis (recall the discussion with Eq. (\ref{flatcondition})) for the groups of type $A_n$, and not to find a generating function for the complete structure of the \wPms\ for the groups of type $A_n$, as the present one.\\

We should now determine the matrix $\widehat P(z)$, corresponding to the basic invariant polynomials of $S_{n+1}$ defined in Eq. (\ref{lintrasfinlemma3}), because from the knowledge of $\widehat P(z)$ it is possible to determine the matrix $\widehat P(p)$ from the equation $\widehat P(p)=[\widehat P(z)]^p$. The matrix elements $\widehat P_{ab}(p)$ would then be equal to the matrix elements $\left.\widehat P_{a+1,b+1}(z)\right|_{z_1=0}$, with the variables $z_{a+1}$ substituted by the variables $p_a$, $\forall\,a=1,\ldots,n$. The determination of the matrix  $\widehat P(z)$ is however somewhat difficult, and we can simplify the calculation if we determine in one step the matrix $\widehat P(p)$ from the matrix $\widehat P(s'')$, without finding first the matrix $\widehat P(z)$. This is done in the following Lemma.\\

\noindent\textbf{Lemma 4}. The \wPm\ $\widehat P(p)$ corresponding to the basic invariant polynomials of $A_{n}$ $p_a(\bar x')$, $a=1,\ldots,n$, restrictions to the hyperplane $x'_{n+1}=0$ of the basic invariant polynomials $z_a(x')$, $a=2,\ldots,n+1$, of $S_{n+1}$, specified in Eq. (\ref{lintrasfinlemma3}), and of the basic invariant polynomials $s''_a(x')$, $a=2,\ldots,n+1$, of $S_{n+1}$, specified in Eq. (\ref{scaletransf2}), that is  $p_a(\bar x')=z_{a+1}(\bar x')=s''_{a+1}(\bar x')$,  $\forall\,a=1,\ldots,n$, is the following one:
\begin{eqnarray}\label{matriceP(p)4}
\nonumber  \widehat P_{ab}(p)&=&\left[\min(a,b)-\frac{a\,b}{n+1}\right]\;p_{a-1}\;p_{b-1}+2 \,(a+ b)\;p_{a+b-1}- \sum_{i=\max(2,a+b-n-1)}^{\min(a,b)-1} \,(a+ b - 2i)\;p_{i-1}\;p_{a+b-1-i}\,,\\ && \qquad\qquad\qquad\qquad\qquad\qquad\qquad\qquad\qquad\qquad\qquad\qquad\qquad\qquad\qquad\qquad\qquad\forall\,a,b=1,\ldots,n\,,
\end{eqnarray}
in which one has to consider $p_0=0$ and $p_a=0$, $\forall\,a>n$.\\

\noindent\textbf{Proof}.
For Lemma 3, the matrix elements $\widehat P_{ab}(p)$, $\forall\,a,b=1,\ldots,n$, are equal to the matrix elements $\left.\widehat P_{a+1,b+1}(z)\right|_{z_1=0}$, $a,b=1,\ldots,n$, with the variables $z_{a+1}$ substituted by the variables $p_a$, $\forall\,a=1,\ldots,n$.\\
It is easier to determine the matrix $\widehat P(p(\bar x'))$ through Eqs. (\ref{matriceP(x)}) and (\ref{invmatP}), instead of using Theorem \ref{tre}. The explicit form of the matrix $\widehat P(s'')$, written in Eq. (\ref{PjkdiSntrasf-2}), will also be used.\\
The gradients of the basic invariant polynomials $z_{a+1}(x')$, $\forall\,a=1,\ldots,n$, are calculated, from Eq. (\ref{lintrasfinlemma3}), in terms of those of $s''_{a+1}(x')$, $\forall\,a=0,\ldots,n$, and are as follows, $\forall\,a=1,\ldots,n$:
$$\nabla z_{a+1}(x')=\nabla \left[s''_{a+1}(x')+ c_{{a+1}}\,s''_1(x')\, s''_{a}(x')\right]=
\nabla s''_{a+1}(x')+ c_{{a+1}}\, s''_{a}(x')\,\nabla s''_1(x')+ c_{{a+1}}\,s''_1(x')\,\nabla s''_{a}(x')\,,
$$
and when $x'_{n+1}=0$, that imply $s''_1(\bar x')=z_1(\bar x')=0$, they become:
$$\left.\nabla z_{a+1}(x')\right|_{x'_{n+1}=0}=
\left.\nabla s''_{a+1}(x')\right|_{x'_{n+1}=0}+ c_{a+1}\, s''_{a}(\bar x')\,\left.\nabla s''_1(x')\right|_{x'_{n+1}=0}\,,\qquad \forall\,a=1,\ldots,n\,.
$$
We know that $[\widehat P(z)]^p=\widehat P(p)$, so we know that:
$$\left.\Bigl(\nabla  z_{a+1}(x')\cdot \nabla  z_{b+1}(x')\Bigr)\right|_{x'_{n+1}=0}=\left(\left.\nabla  z_{a+1}(x')\right|_{x'_{n+1}=0}\right)\cdot \left(\left.\nabla  z_{b+1}(x')\right|_{x'_{n+1}=0}\right)\,.
$$
Then, $\forall\,a,b=1,\ldots,n$, one has:
$$\left.\widehat P_{a+1,b+1}(z(x'))\right|_{x'_{n+1}=0}=\left.\Bigl(\nabla  z_{a+1}(x')\cdot \nabla  z_{b+1}(x')\Bigr)\right|_{x'_{n+1}=0}=$$
$$=\left( \left.\nabla s''_{a+1}(x')\right|_{x'_{n+1}=0}+ c_{a+1}\, s''_{a}(\bar x')\,\left.\nabla s''_1(x')\right|_{x'_{n+1}=0}\right)\cdot\left( \left.\nabla s''_{b+1}(x')\right|_{x'_{n+1}=0}+ c_{b+1}\, s''_{b}(\bar x')\,\left.\nabla s''_1(x')\right|_{x'_{n+1}=0}\right)=
$$
$$=\widehat P_{a+1,b+1}(s''(\bar x'))+c_{b+1}\, s''_{b}(\bar x')\,\widehat P_{1,a+1}(s''(\bar x'))+
c_{a+1}\, s''_{a}(\bar x')\,\widehat P_{1,b+1}(s''(\bar x'))+c_{a+1}\,c_{b+1}\, s''_{a}(\bar x')\, s''_{b}(\bar x')\,\widehat P_{1,1}(s''(\bar x'))=
$$
\begin{equation}\label{dimAn1}
  =\left.\widehat P_{a+1,b+1}(s'')\right|_{s''_1=0}+\left.\left[c_{b+1}\, s''_{b}\,\widehat P_{1,a+1}(s'')+
c_{a+1}\, s''_{a}\,\widehat P_{1,b+1}(s'')+c_{a+1}\,c_{b+1}\, s''_{a}\, s''_{b}\,\widehat P_{1,1}(s'')\right]\right|_{s''_1=0},
\end{equation}
where in the last member the dependence on $\bar x'$ has been understood, and the condition $x'_{n+1}=0$ has been replaced by the equivalent condition $s''_1=0$.\\
The terms inside the square brackets can be simplified by substituting the definitions of the coefficients $c_a$, given in Eq. (\ref{lintrasfinlemma3}), and the explicit form of the matrix elements of the first row and column of the matrix $\widehat P(s'')$, given in Eq. (\ref{PjkdiSntrasf-2}). In this way the content of the square brackets becomes as follows:
$$\left.\left[\ldots\right]\right|_{s''_1=0}=\left.\left[\frac{n+1-b}{2\,(n+1)}\, s''_{b}\,\widehat P_{1,a+1}(s'')+
 \frac{n+1-a}{2\,(n+1)}\, s''_{a}\,\widehat P_{1,b+1}(s'')+\frac{(n+1-a)(n+1-b)}{4\,(n+1)^2}\, s''_{a}\, s''_{b}\,\widehat P_{1,1}(s'')\right]\right|_{s''_1=0}=
$$
$$=\left.\left[\frac{n+1-b}{2\,(n+1)}\, s''_{b}\,\left(-2(n+1-a)\,s''_{a}\right)+
 \frac{n+1-a}{2\,(n+1)}\, s''_{a}\,\left(-2(n+1-b)\,s''_{b}\right)+\frac{(n+1-a)(n+1-b)}{4\,(n+1)^2}\, s''_{a}\, s''_{b}\,4\,(n+1)\right]\right|_{s''_1=0}=
$$
$$=-\frac{(n+1-a)(n+1-b)}{n+1}\;\left.\left( s''_{a}\, s''_{b}\right)\right|_{s''_1=0}\,,
$$
expression that is equal to 0 if $a=1$ or $b=1$, because $s''_1=0$.\\
To simplify the explicit expression of the matrix element $\left.\widehat P_{a+1,b+1}(s'')\right|_{s''_1=0}$, that appears in the first term of the last member of Eq. (\ref{dimAn1}), one has to take into account the third line of Eq. (\ref{PjkdiSntrasf-2}) and the conditions valid for Eq. (\ref{PjkdiSntrasf-2}), that is: $s''_a=0$, $\forall\,a>n+1$. One so finds, $\forall\,a,b=1,\ldots,n$:
$$\left.\widehat P_{a+1,b+1}(s'')\right|_{s''_1=0}=\left.\left( [n+1-\min(a,b)]\;s''_{a}\;s''_{b}+2 \,(a+ b)\;s''_{a+b}
- \sum_{i=\max(2,a+b-n)}^{\min(a,b)+1} \,(a+ b+2 - 2i)\;s''_{i-1}\;s''_{a+b+1-i}\right)\right|_{s''_1=0}=
$$
$$
= [n+1-\min(a,b)]\;\left.\left(s''_{a}\;s''_{b}\right)\right|_{s''_1=0}+2 \,(a+ b)\;s''_{a+b}- \sum_{i=\max(3,a+b-n)}^{\min(a,b)+1} \,(a+ b+2 - 2i)\;\left.\left(s''_{i-1}\;s''_{a+b+1-i}\right)\right|_{s''_1=0}=
$$
$$
= [n+1-\min(a,b)]\;\left.\left(s''_{a}\;s''_{b}\right)\right|_{s''_1=0}+2 \,(a+ b)\;s''_{a+b}-[a+b-2\,\min(a,b)]\;\left.\left(s''_{\min(a,b)}\;s''_{\max(a,b)}\right)\right|_{s''_1=0}\,+$$
$$-\, \sum_{i=\max(3,a+b-n)}^{\min(a,b)} \,(a+ b+2 - 2i)\;s''_{i-1}\;s''_{a+b+1-i}=
$$
$$
= [n+1-\max(a,b)]\;\left.\left(s''_{a}\;s''_{b}\right)\right|_{s''_1=0}+2 \,(a+ b)\;s''_{a+b}- \sum_{i'=\max(2,a+b-n-1)}^{\min(a,b)-1} \,(a+ b - 2i')\;s''_{i'}\;s''_{a+b-i'}\,,
$$
where in the third expression the lower limit of the sum has been increased from 2 to 3 because the condition $s''_1=0$ implies $s''_{i-1}=0$, if $i=2$, and where in the last expression one one could consider $\,s''_{\min(a,b)}\;s''_{\max(a,b)}=s''_{a}\;s''_{b}$. We can then write for Eq. (\ref{dimAn1}) the following expression, where, as before, the dependence on $\bar x'$ is understood:
$$\left.\widehat P_{a+1,b+1}(z)\right|_{z_1=0}=\left(n+1-\max(a,b)-\frac{(n+1-a)(n+1-b)}{n+1}\right)\;\left.\left(s''_{a}\;s''_{b}\right)\right|_{s''_1=0}+2 \,(a+ b)\;s''_{a+b}\,+$$
$$
-\, \sum_{i=\max(2,a+b-n-1)}^{\min(a,b)-1} \,(a+ b - 2i)\;s''_{i}\;s''_{a+b-i}=
$$
$$=\left[\min(a,b)-\frac{a\,b}{n+1}\right]\;\left.\left(s''_{a}\;s''_{b}\right)\right|_{s''_1=0}+2 \,(a+ b)\;s''_{a+b}
- \sum_{i=\max(2,a+b-n-1)}^{\min(a,b)-1} \,(a+ b - 2i)\;s''_{i}\;s''_{a+b-i}\,,\qquad \forall\,a,b=1,\ldots,n\,.
$$
In the last member, one has to substitute the polynomials $s''_a(\bar x')$, $\forall\,a=1,\ldots,n+1$, with their expressions in terms of the polynomials $z_a(\bar x')$, $\forall\,a=1,\ldots,n+1$, that can be obtained from Eq. (\ref{lintrasfinlemma3}), but for $s''_1(\bar x')=0$ this is easy, because Eq. (\ref{lintrasfinlemma3}) gives $z''_1(\bar x')=0$ and $s''_a(\bar x')=z_a(\bar x')$, $\forall \,a=2,\ldots,n+1$. To find the matrix $\widehat P(p)$ one has, finally, to substitute $p_a$ in place of $z_{a+1}$ (that is of $s''_{a+1}$),  $\forall \,a=1,\ldots,n$. One so finds:
$$\widehat P_{ab}(p)=\left[\min(a,b)-\frac{a\,b}{n+1}\right]\;p_{a-1}\;p_{b-1}+2 \,(a+ b)\;p_{a+b-1}
-\!\! \sum_{i=\max(2,a+b-n-1)}^{\min(a,b)-1} \,(a+ b - 2i)\;p_{i-1}\;p_{a+b-1-i}\,,\quad \forall\,a,b=1,\ldots,n\,,
$$
where the conditions $z_1=0$ (corresponding to $x'_{n+1}=0$) and $z_a=0$, $\forall\, a>n+1$, become $p_0=0$ and $p_a=0$, $\forall\, a>n$. The proof of the Lemma is complete.\scat

The matrix $\widehat P(p)$ in Eq. (\ref{matriceP(p)4}) has fractional coefficients, because of the denominator $n+1$.
If one wishes to have a \wPm\ with only integer coefficients one can make a scale transformation of the invariant polynomials $p_a(\bar x')$, $a=2,\ldots,n$, avoiding to modify the quadratic invariant $p_1(\bar x')$ that has its standard form given in Eq. (\ref{quadraticinv}). This is done in the following Lemma.\\

\noindent\textbf{Lemma 5}. The basic invariant polynomials of $A_{n}$ $p'_a(\bar x')$, $a=1,\ldots,n$, defined by the scale transformation
\begin{equation}\label{scalap}
  p'_a(\bar x')=(n+1)^{\frac{a-1}{2}}\,p_a(\bar x')\,,\qquad \forall\,a=1,\ldots,n\,,
\end{equation}
in terms of the basic invariant polynomials of $A_{n}$ $p_a(\bar x')$, $a=1,\ldots,n$, given in Lemma 4,
are such that $p'_1(\bar x')$, has the standard form of Eq. (\ref{quadraticinv}): $p'_1(\bar x')=\|\bar x'\|^2$, and the corresponding  \wPm\ $\widehat P(p')$ can be obtained with the following formula:
\begin{eqnarray}\label{matriceP(p)5}
\nonumber  \widehat P_{ab}(p')&=&[(n+1)\,\min(a,b)-a\,b\,]\;p'_{a-1}\;p'_{b-1}+2 \,(a+ b)\;p'_{a+b-1}
+ \\
&-&(n+1)\,\sum_{i=\max(2,a+b-n-1)}^{\min(a,b)-1}\,(a+ b - 2i)\;p'_{ i-1}\;p'_{a+b-1-i}\,,\qquad\qquad\qquad \forall\,a,b=1,\ldots,n\,,
\end{eqnarray}
where one has to consider $p'_0=0$ and $p'_a=0$, $\forall\,a>n$. The matrix $\widehat P(p')$ has then only integer coefficients.\\

\noindent\textbf{Proof}. We first note that for $a=1$, Eq. (\ref{scalap}) gives $p'_1(\bar x')=p_1(\bar x')$, hence $p'_1(\bar x')=\|\bar x'\|^2$.
To obtain the matrix $\widehat P(p')$ we use Theorem \ref{tre}. The Jacobian matrix of the transformation (\ref{scalap}) is diagonal:
$$J_{ab}(p)=(n+1)^{\frac{a-1}{2}}\,\delta_{ab}\,,\qquad \forall\, a,b=1,\ldots,n\,,$$
that implies $J^\top(p)=J(p)$. The substitution $p= \widehat p(p')$ in Eq. (\ref{Ptransf}), using Eq. (\ref{scalap}), becomes the following one:
$$  p_a=(n+1)^{\frac{1-a}{2}}\,p'_a\,,\qquad \forall\,a=1,\ldots,n\,.
$$
Using Eq. (\ref{matriceP(p)4}), $\forall\,a,b=1,\ldots,n$, Eq. (\ref{Ptransf}) gives:
$$  \widehat P_{ab}(p')=\left.J_{ac}(p)\; \widehat
P_{cd}(p)\; J^\top_{db}(p)\right|_{p= \widehat p(p')}=\left.\left(
(n+1)^{\frac{a-1}{2}}\,\delta_{ac}\; \widehat
P_{cd}(p)\;(n+1)^{\frac{b-1}{2}}\,\delta_{db}\right)\right|_{p= \widehat p(p')}=(n+1)^{\frac{a+b-2}{2}}\,\left.
 \widehat
P_{ab}(p)\right|_{p= \widehat p(p')}=
$$
$$=(n+1)^{\frac{a+b-2}{2}}\,\Biggl\{\left[\min(a,b)-\frac{a\,b}{n+1}\right]\;(n+1)^{\frac{1-(a-1)}{2}}\,p'_{a-1}\;(n+1)^{\frac{1-(b-1)}{2}}\,p'_{b-1}+2 \,(a+ b)\;(n+1)^{\frac{1-(a+b-1)}{2}}\,p'_{a+b-1}\,+$$
$$
-\, \sum_{i=\max(2,a+b-n-1)}^{\min(a,b)-1} \,(a+ b - 2i)\;(n+1)^{\frac{1-(i-1)}{2}}\,p'_{i-1}\;(n+1)^{\frac{1-(a+b-1-i)}{2}}\,p'_{a+b-1-i}\Biggr\}=
$$
$$=(n+1)^{\frac{a+b-2}{2}}\,\Biggl\{\left[(n+1)\,\min(a,b)-a\,b\right]\;(n+1)^{\frac{2-a-b}{2}}\,p'_{a-1}\;p'_{b-1}+2 \,(a+ b)\;(n+1)^{\frac{2-a-b}{2}}\,p'_{a+b-1}\,+$$
$$
-\, (n+1)^{\frac{4-a -b}{2}}\; \sum_{i=\max(2,a+b-n-1)}^{\min(a,b)-1} \,(a+ b - 2i)\;p'_{i-1}\;p'_{a+b-1-i}\Biggr\}=
$$
$$=\left[(n+1)\,\min(a,b)-a\,b\right]\;p'_{a-1}\;p'_{b-1}+2 \,(a+ b)\;p'_{a+b-1}- (n+1)\, \sum_{i=\max(2,a+b-n-1)}^{\min(a,b)-1} \,(a+ b - 2i)\;p'_{i-1}\;p'_{a+b-1-i}\,.
$$
This is the \wPm\ in Eq. (\ref{matriceP(p)5}) and is clear that all the coefficients of the monomials in the variables $p'_a$, $a=1,\ldots,n$, that appear in it, are integer numbers.\scat

The \wPm\ in Eq. (\ref{matriceP(p)5}) is the \wPm\ corresponding to the basic invariant polynomials $p'_a(\bar x')$, $a=1,\ldots,n$, of $A_n$, and is the one also reported in Eq. (\ref{PjkdiAn}). Let's review how, in this proof, the polynomials $p'_a(\bar x')$, $a=1,\ldots,n$, are obtained from the basic invariant polynomials $s_a(x)$, $a=1,\ldots,n+1$, of $S_{n+1}$, given in Eq. (\ref{elemsymmpolSn+1}).
\begin{enumerate}
  \item We performed a rotation of the space $\R^{n+1}$ by the matrix $R_{n+1}$ given in Eq. (\ref{matRn+1}): $x'=Rx$. The basic polynomials of $S_{n+1}$ were transformed into $s'_a(x')=s_a(R^\top_{n+1}x)$, $\forall\,a=1,\ldots,n+1$.
  \item We performed the scale transformation by the factor $-2$ as written in Eq. (\ref{scaletransf2}). The basic polynomials of $S_{n+1}$ were transformed into $s''_a(x')=-2\,s'_a(x')$, $\forall\,a=1,\ldots,n+1$.
  \item  We performed the basis transformation written in Eq. (\ref{lintrasfinlemma3}). The basic polynomials of $S_{n+1}$ were transformed into $z_1(x')=s''_1(x')$, $z_a(x')=s''_a(x')+ \frac{n+2-a}{2\,(n+1)}\,s''_1(x')\, s''_{a-1}(x')$, $\forall\,a=2,\ldots,n+1$.
  \item We took the restrictions to the hyperplane $x'_{n+1}=0$ of the basic invariants $z_a(x')$, $\forall a=1,\ldots,n+1$, of $S_{n+1}$. One would obtain the same polynomials by taking the restrictions to the hyperplane $x'_{n+1}=0$ of the basic invariant polynomials $s''_a(x')$, $\forall a=1,\ldots,n+1$, of $S_{n+1}$ (although $[\widehat P(z)]^p\neq [\widehat P(s'')]^p$). One so finds $s''_{1}(\bar x')=\left.s''_{1}(x')\right|_{x'_{n+1}=0}=0$, and the restrictions to the hyperplane $x'_{n+1}=0$ of the non-linear polynomials $s''_a(x')$, $a=2,\ldots,n+1$, form a basis of invariant polynomials of $A_n$. These basic invariant polynomials of $A_n$, after renaming and renumbering, are specified in Eq. (\ref{rinomina}): $p_a(\bar x')=s''_{a+1}(\bar x')=\left.s''_{a+1}(x')\right|_{x'_{n+1}=0}$, $\forall\,a=1,\ldots,n$.
  \item We performed the scale transformation reported in Eq. (\ref{scalap}), in which the basic invariant polynomials of $A_{n}$ were transformed into $p'_a(\bar x')=(n+1)^{\frac{a-1}{2}}\,p_a(\bar x')$, $\forall\,a=1,\ldots,n$.
\end{enumerate}
Regarding the construction of the basic invariant polynomials of $A_n$ $p'_a(\bar x')$, $a=1,\ldots,n$, the relevant transformations are those in items 1., 2., 4., and 5. The order is not important, except for the restriction to the hyperplane $x'_{n+1}=0$ (item 4.) that must be performed after the rotation (item 1.). For example, one obtains the same basic invariant polynomials following the operations in the following order: item 2., item 5, item 1., item 4., or in the following one: item 1., item 2., item 5., item 4., that is the one reported in Section \ref{An}.  The combination of the scale transformations in items 2. and 5., if done before the renaming of item 4, results in the following scale transformation:
$$  t_a(x')=-2\,(n+1)^{\frac{a-2}{2}}\,s'_a(x')\,,\qquad a=1,\ldots,n+1\,,
$$
where the index $a$ in Eq. (\ref{scalap}) has been changed into $a-1$ because the scale transformation is done before the renaming of item 4.. The scale transformation has been extended also to $s'_1(x')$ without any problem, because the restriction to the hyperplane $x'_{n+1}=0$ implies in any case that $t_1(\bar x')=0$. This scale transformation is that one reported in Eq. (\ref{trasfAn2}).\\

Without the restriction to the hyperplane $x'_{n+1}=0$ specified in item 4. (and with or without the rotation specified in item 1.), one obtains a set of basic invariant polynomials of $S_{n+1}$, but both the invariant polynomials and the corresponding \wPms\ depend on the order these transformations are performed. As we have seen the operator $[\;\cdot\;]^p$, when applied to the \wPm\ corresponding to the transformed invariant polynomials of $S_{n+1}$, gives a \wPm\ for $A_n$, only if the hypothesis of Lemma 2 are satisfied.\\

We started this proof from the \wPm\ of $S_{n+1}$ in Eq. (\ref{PjkdiSn3}). The final matrix reported in Eq. (\ref{matriceP(p)5}), is the \wPm\ corresponding to the basic invariants $p'_a(\bar x')$, $a=1,\ldots,n$, of $A_n$, obtained in this proof and coincides with the matrix reported in Eq. (\ref{PjkdiAn}), that is the \wPm\ corresponding to the basic invariants $p_a(\bar x')$, $a=1,\ldots,n$ of $A_n$ obtained in Eqs. (\ref{elemsymmpolSn+1})--(\ref{basicpolynAn}). This completes the proof of the first part of Theorem \ref{lambdaAn}.\scat

\noindent\textbf{Proof of the generating formula (\ref{lambdaAnEq}).}\\
The basis of invariant polynomials $p_a(\bar x')$, $a=1,\ldots,n$, of $A_n$, to which it corresponds the matrix $\widehat P(p)$ of $A_n$ that can be generated by Eq. (\ref{PjkdiAn}), was derived from the basis of invariant polynomials $s_a(x)$, $a=1,\ldots,n+1$, of $S_{n+1}$, that is written in Eq. (\ref{elemsymmpolSn+1}), through the transformations defined in Eqs. (\ref{trasfAn1}), (\ref{trasfAn2}) and (\ref{basicpolynAn}). We shall now determine the $\lambda$-vector corresponding to the determinant $\det(\widehat P(p))$ of the \wPm\ of $A_n$, using Theorem \ref{trasflambda}, starting from the $\lambda$-vector corresponding to the determinant $\det(\widehat P(s))$ of the \wPm\ of $S_{n+1}$, that is written in Eq. (\ref{lambdaSnEq}), but in which $n$ must be substituted by $n+1$, and the variables $p_a$ must be substituted with the variables $s_a$, $\forall\,a=1,\ldots,n+1$, that is from the following $\lambda$-vector of $\det(\widehat P(s))$:
\begin{equation}\label{lambdaSn+1}
  \lambda^{(\det(\widehat P))}(s) = (\lambda^{(\det(\widehat P))}_1(s),\ldots,\lambda^{(\det(\widehat P))}_{n+1}(s)),\qquad \lambda^{(\det(\widehat P))}_{a}(s)=-(n - a+3)(n - a+2)\,s_{a - 2}\,,\quad a=1,\ldots,n+1\,,
\end{equation}
in which one has to consider $s_{-1}=0$ and $s_0=1$.\\

With the basis transformation written in Eq. (\ref{trasfAn1}), consequent to the rotation produced by the matrix $R_{n+1}$,  one has, by Theorem \ref{uno}, item \ref{th1it4}., $\widehat P(s')=\left.\widehat P(s)\right|_{s=s'}$, and Theorem \ref{trasflambda}, item \ref{thlit1}, asserts that
$$\lambda^{(\det(\widehat P))}(s')=\left.\lambda^{(\det(\widehat P))}(s)\right|_{s=s'}\,,$$
so that $\lambda^{(\det(\widehat P))}(s')$ is given by Eq. (\ref{lambdaSn+1}), with $s$ substituted by $s'$.\\

To the basis transformation of the basic invariant polynomials of $S_{n+1}$ written in Eq. (\ref{trasfAn2}), it corresponds the $(n+1)\times(n+1)$ diagonal jacobian matrix $J(s')$, whose elements are the following:
$$J_{ab}(s')=\frac{\partial t_a}{\partial s'_b}=-2\, \sqrt{(n+1)^{a-2}}\,\delta_{ab}\,,\qquad a,b=1,\ldots,n+1\,,
$$
with $\delta_{ab}$ the Kronecker delta.\\
The determinant of the \wPm\ of $S_{n+1}$ $\widehat P(t)$, corresponding to the basic invariant polynomials $t_1(x'),\ldots,t_{n+1}(x')$, satisfies the boundary equation (\ref{boundaryeqndet}) with the $\lambda$-vector $\lambda_a^{(\det(\widehat P))}(t)$ that can be determined from $\lambda^{(\det(\widehat P))}(s')$ using Theorem  \ref{trasflambda}, item \ref{thlit2}. This calculation gives, $\forall\,  a=1,\ldots,n+1$:
$$\lambda_a^{(\det(\widehat P))}(t)=\left.J_{ab}(s')\;\lambda_b^{(\det(\widehat P))}(s')\right|_{s'=\widehat{s}'(t)}=\left.-2\, \sqrt{(n+1)^{a-2}}\;\delta_{ab}\;\lambda_b^{(\det(\widehat P))}(s')\right|_{s'=\widehat{s}'(t)}=
$$
$$=\left.-2\, \sqrt{(n+1)^{a-2}}\;\lambda_a^{(\det(\widehat P))}(s')\right|_{s'=\widehat{s'}(t)}=
\left.2\, \sqrt{(n+1)^{a-2}}\,(n - a+3)\,(n - a+2)\,s'_{a - 2}\right|_{s'=\widehat{s'}(t)}$$
The inverse transformation of the transformation in Eq. (\ref{trasfAn2}) is the following one:
 $$\widehat{s'}_a(t)=-\frac{1}{2}\,\sqrt{(n+1)^{2-a}}\;t_{a}\,,\qquad a=1,\ldots,n+1\,,$$
that gives $\left.s'_{a - 2}\right|_{s'=\widehat{s'}(t)}=\widehat{s'}_{a - 2}(t)=-\frac{1}{2}\sqrt{(n+1)^{4-a}}\,t_{a-2}$. In the formula for $\lambda_a^{(\det(\widehat P))}(t)$, we have however to consider separately the cases $a=1$ and $a=2$ from those with $a>2$, because when $a=1$ in the formula one finds $s'_{a - 2}=s'_{-1}=0$, and when $a=2$ one finds $s'_{a - 2}=s'_{0}=1$, so in both these cases no substitution has to be performed because no variables $s'_a$, $a=1,\ldots,n+1$ are present in the expression for $\lambda_a^{(\det(\widehat P))}(t)$. One finds then:
 $$\lambda_1^{(\det(\widehat P))}(t)=0\,,\qquad\qquad\lambda_2^{(\det(\widehat P))}(t)=2\,n\, (n + 1)\,,$$
 and $\forall\,a=3,\ldots,n+1$, one has:
 $$\lambda_a^{(\det(\widehat P))}(t)=
2\, \sqrt{(n+1)^{a-2}}\,(n - a+3)(n - a+2)\,\left(-\frac{1}{2}\,\sqrt{(n+1)^{4-a}}\,t_{a-2}\right)=
-(n+1)(n - a+3)(n - a+2)\,t_{a-2}\,.
$$

The last transformation to do is that one in Eq. (\ref{basicpolynAn}), that can be performed in two steps. In the first one one has to take $x'_{n+1}=0$, that means $t_{1}(\bar x')=0$, and implies $\lambda_3^{(\det(\widehat P))}(t)=0$. In the second and last step one has to rename the basic polynomials and the variables in the following way:
  $t_{a+1}(\bar x')=p_a(\bar x')\,,$ $a=1,\ldots,n\,,$ because $t_{1}(\bar x')$ is no longer an invariant polynomial, so that it does no longer interest, $\lambda_{a+1}^{(\det(\widehat P))}(t)=\lambda_{a}^{(\det(\widehat P))}(p)$, $\forall\,a=1,\ldots,n$, and this gives the following final form of the $\lambda$-vector of $\det(\widehat P(p))$, where its first component has been dropped:
$$\lambda^{(\det(\widehat P))}(p) = (2\,n\, (n + 1),\,\lambda^{(\det(\widehat P))}_2(p),\,\ldots,\,\lambda^{(\det(\widehat P))}_{n}(p))\,,$$
$$ \lambda^{(\det(\widehat P))}_{a}(p)=- (n + 1)(n - a+2)(n - a+1)\,p_{a - 2}\,,\qquad \forall\,a=2,\ldots,n\,,$$
in which one has to consider $p_0=0$.
This is Eq.  (\ref{lambdaAnEq}), so the proof is complete. Using $p_0=0$ one sees that $\lambda^{(\det(\widehat P))}_2(p)=0$.
\scat

\subsection{Proof of Theorem \ref{lambdaBn} for $B_n$\label{proofBn}}
\noindent\textbf{Proof of the generating formula (\ref{PjkdiBn}).}\\

The proof can be obtained by induction on the rank $n$. The lowest possible value for $n$ is $n=2$ (because of the isomorphism $B_1\simeq A_1$), so we first prove that Eq. (\ref{PjkdiBn}) is true for $n=2$. \\
The basic invariant polynomials of $B_2$, from Eq. (\ref{basicpolynBn}), are the following:
$\;  p_1(x)=x_1^2+x_2^2\,$, $\;  p_2(x)=x_1^2 x_2^2\,$, their gradients are the following:
$\nabla p_1(x)=(2x_1,2x_2)$, $\nabla p_2(x) =(2x_1x_2^2,2x_1^2,x_2)$,
and the elements of the matrix $P(x)$, calculated with Eqs. (\ref{matriceP(x)}) and (\ref{invmatP}), are the following:
$   P_{11}(x)             =   4(x_1^2+x_2^2)             =   4p_1(x)$, $  P_{12}(x)=P_{21}(x)   =  4(x_1^2x_2^2+x_1^2x_2^2)   =   8p_2(x)$,
$   P_{22}(x)            =  4(x_1^2x_2^4+x_1^4x_2^2)   =   4p_1(x)p_2(x)$.
Using  Eq. (\ref{PjkdiBn}) one finds the following expression for the \wPm\ elements of $B_n$:
$\widehat P_{11}(p)=4p_1$,  $\widehat P_{12}(p)=\widehat P_{21}(p)=8p_2$,
$\widehat P_{22}(p)= 4p_1p_2$.
We see then that for $n=2$ Eq. (\ref{PjkdiBn}) is true.\\

Using the induction principle, we suppose that Eq. (\ref{PjkdiBn}) is true for a general $n\geq 2$, and prove in all generality that it is true also for $n$ substituted by $n+1$. This implies then that Eq. (\ref{PjkdiBn}) is true for all $n\geq 2$.\\

We use the symbols $x$ and $y$ for general vectors of $\R^n$ and $\R^{n+1}$, that is: $x=(x_1,x_2,\ldots,x_n)$, and $y=(x_1,x_2,\ldots,x_n,x_{n+1})$.
$B_n$ acts in $\R^n$ and its basic invariant polynomials are $p_a(x)$, $a=1,\ldots,n$, defined in Eq. (\ref{basicpolynBn}), and $B_{n+1}$ acts in $\R^{n+1}$ and its basic invariant polynomials are $p_a(y)$, $a=1,\ldots,n+1$, also defined in Eq. (\ref{basicpolynBn}), but with $n$ substituted by $n+1$.\\

The basic invariant polynomials of $B_{n+1}$ can be expressed in terms of those of $B_{n}$ in the following way:
$$
\begin{array}{ccl}
   p_{1}(y)            & = &   p_1(x)+x_{n+1}^2\,,  \\
   p_{2}(y)            & = &   p_2(x)+p_1(x)\,x_{n+1}^2\,,  \\
   p_{3}(y)            & = &   p_3(x)+p_2(x)\,x_{n+1}^2\,,  \\
   \vdots            &  &  \\
   p_{n}(y)            & = &   p_n(x)+p_{n-1}(x)\,x_{n+1}^2\,,  \\
   p_{n+1}(y)            & = &   p_n(x)\,x_{n+1}^2\,,  \\
   \end{array}
$$
By defining $p_0(x)=1$ and $p_{n+1}(x)=0$, one has the following general formula:
\begin{equation}\label{p(y)diBn}
   p_{a}(y)             =    p_a(x)+p_{a-1}(x)\,x_{n+1}^2\,\qquad \forall\,a=1,2,\ldots,n+1.
\end{equation}
Let's use the ``nabla'' operators defined in Eq. (\ref{nablaxy}):
$$\nabla_{\!x}=\left(\frac{\partial}{\partial x_1},\frac{\partial}{\partial x_2},\ldots,\frac{\partial}{\partial x_n},0\right)\,,\qquad
\nabla_{\!y}=\left(\frac{\partial}{\partial x_1},\frac{\partial}{\partial x_2},\ldots,\frac{\partial}{\partial x_n},\frac{\partial}{\partial x_{n+1}}\right)\,.$$
One has then, $\forall a=1,\ldots,n+1$, for Eqs. (\ref{p(y)diBn}) and (\ref{nablaxy}):
\begin{equation}\label{nablaYp(y)}
  \nabla_{\!y}p_a(y)=\nabla_{\!y}\left[ p_a(x)+p_{a-1}(x)\,x_{n+1}^2\right]=\nabla_{\!x}p_a(x)+x_{n+1}^2\,\nabla_{\!x}p_{a-1}(x)+p_{a-1}(x)\,2\,x_{n+1}\,e_{n+1}\,,
\end{equation}
where $e_{n+1}$ is the canonical unit vector of $\R^{n+1}$.\\
We can now calculate a general element of the matrix $P(y)$ of $B_{n+1}$, using the basic invariant polynomials $p_a(y)$, $a=1,\ldots,n+1$, of $B_{n+1}$, and Eqs. (\ref{matriceP(x)}) and (\ref{nablaYp(y)}). For all $a,b=1,\ldots,n+1$ we have:
$$P_{ab}(y)=\nabla_{\!y}p_a(y)\cdot \nabla_{\!y}p_b(y)=$$
$$=\left[\nabla_{\!x}p_a(x)+x_{n+1}^2\,\nabla_{\!x}p_{a-1}(x)+p_{a-1}(x)\,2\,x_{n+1}\,e_{n+1}\right]\cdot
\left[\nabla_{\!x}p_b(x)+x_{n+1}^2\,\nabla_{\!x}p_{b-1}(x)+p_{b-1}(x)\,2\,x_{n+1}\,e_{n+1}\right]=
$$
$$=\nabla_{\!x}p_a(x)\cdot \nabla_{\!x}p_b(x)+x_{n+1}^2\,\nabla_{\!x}p_{a-1}(x)\cdot \nabla_{\!x}p_b(x)+0+x_{n+1}^2\,
\nabla_{\!x}p_a(x)\cdot \nabla_{\!x}p_{b-1}(x)+
x_{n+1}^4\,\nabla_{\!x}p_{a-1}(x)\cdot \nabla_{\!x}p_{b-1}(x)\,+
$$
$$+\,0+0+0+4\,x_{n+1}^2\,p_{a-1}(x)\,p_{b-1}(x)=$$
$$=P_{ab}(x)+x_{n+1}^2\,\left[P_{a-1,b}(x)+P_{a,b-1}(x)+4\,p_{a-1}(x)\,p_{b-1}(x)\right]+x_{n+1}^4\,P_{a-1,b-1}(x)\,.$$
Of course, in the preceding expression, $P_{cd}(x)=0$, if one of the indices $c,d$ (that in the above expression can be equal to $a,b,a-1,b-1$) is equal to $0$ or to $n+1$, because for $B_n$ the allowed indices run from 1 to $n$, and the gradients of $p_0(x)=1$, and $p_{n+1}(x)=0$ are null vectors. By the induction hypothesis, Eq. (\ref{PjkdiBn}) is true for $B_n$, so we can use it in the preceding equation. We obtain then, $\forall\,a,b=1,\ldots,n+1$:
$$P_{ab}(y)=\sum_{i=\max(0,a+b-n-1)}^{\min(a,b)-1}\;4\,(a+b-1-2i)\;p_i(x)\;p_{a+b-1-i}(x)\,+$$
$$+\,4\,x_{n+1}^2\,\left\{\sum_{i=\max(0,a+b-n-2)}^{\min(a-1,b)-1}\,(a+b-2-2i)\;p_i(x)\;p_{a+b-2-i}(x)\,+\right.$$
$$\left.+\,\sum_{i=\max(0,a+b-n-2)}^{\min(a,b-1)-1}\,(a+b-2-2i)\;p_i(x)\;p_{a+b-2-i}(x)+p_{a-1}(x)\,p_{b-1}(x)\right\}+
$$
$$+\,4\,x_{n+1}^4\,\sum_{i=\max(0,a+b-n-3)}^{\min(a-1,b-1)-1}\,(a+b-3-2i)\;p_i(x)\;p_{a+b-3-i}(x)\,.
$$
The term inside the curly brackets multiplying $x_{n+1}^2$ can be simplified. One notes that both sums contain equal terms up to the value $i=\min(a-1,b-1)-1=\min(a,b)-2$. Moreover the first sum contains one more term if $\min(a-1,b-1)\neq \min(a-1,b)$, that is if $a>b$, that corresponds to $i=b-1$ and the second sum  contains one more term if $\min(a-1,b-1)\neq \min(a,b-1)$, that is if $a<b$, that corresponds to $i=a-1$. These two terms never appear simultaneously and one can write that the additional term appears for $i=\min(a,b)-1$. One can then write the term inside the curly brackets multiplying $x_{n+1}^2$  as follows:
$$4\,x_{n+1}^2\,\{\ldots\}=4\,x_{n+1}^2\,\left\{\sum_{i=\max(0,a+b-n-2)}^{\min(a,b)-2}\,2\,(a+b-2-2i)\;p_i(x)\;p_{a+b-2-i}(x)\,+\right.$$
$$\left.+\,\left(a+b-2-2(\min(a,b)-1)\right)\;p_{\min(a,b)-1}(x)\;p_{a+b-2-(\min(a,b)-1)}(x)+p_{a-1}(x)\,p_{b-1}(x)\right\}
$$
One can see from this expression that if $a=b$ there is no additional term originating from the two sums, as it should be.
The terms in the second line of the above expression, if $a< b$ can be written in the following way:
$$\left.\left(a+b-2-2(a-1)\right)\;p_{a-1}(x)\;p_{a+b-2-(a-1)}(x)+p_{a-1}(x)\,p_{b-1}(x)\right.=
$$
\begin{equation}\label{modulo}
  =\left(-a+b\right)\;p_{a-1}(x)\;p_{b-1}(x)+p_{a-1}(x)\,p_{b-1}(x)=\left(b-a+1\right)\;p_{a-1}(x)\;p_{b-1}(x)\,,
\end{equation}
and if $a>b$ can be written in the following way, obtained by interchanging $a$ and $b\,$: $\left(a-b+1\right)\;p_{a-1}(x)\;p_{b-1}(x)$. In all cases, one can then write the term inside the curly brackets multiplying $x_{n+1}^2$  as follows:
$$4\,x_{n+1}^2\,\{\ldots\}=4\,x_{n+1}^2\,\left\{\sum_{i=\max(0,a+b-n-2)}^{\min(a,b)-2}\,2\,(a+b-2-2i)\;p_i(x)\;p_{a+b-2-i}(x)+\left(|b-a|+1\right)\;p_{a-1}(x)\;p_{b-1}(x)\right\}
$$
In the sum multiplying $4\,x_{n+1}^4$, it is convenient to change the summation variable in $i'=i+1$. One so finds:
$$\sum_{i=\max(0,a+b-n-3)}^{\min(a-1,b-1)-1}\,(a+b-3-2i)\;p_i(x)\;p_{a+b-3-i}(x)=
\sum_{i'=\max(1,a+b-n-2)}^{\min(a,b)-1}\,(a+b-1-2i')\;p_{i'-1}(x)\;p_{a+b-2-i'}(x)\,.$$
All this gives:
$$P_{ab}(y)=\sum_{i=\max(0,a+b-n-1)}^{\min(a,b)-1}\;4\,(a+b-1-2i)\;p_i(x)\;p_{a+b-1-i}(x)\,+$$
$$+\,4\,x_{n+1}^2\,\left\{\sum_{i=\max(0,a+b-n-2)}^{\min(a,b)-2}\,2\,(a+b-2-2i)\;p_i(x)\;p_{a+b-2-i}(x)+\left(|a-b|+1\right)\;p_{a-1}(x)\;p_{b-1}(x)\right\}+
$$
\begin{equation}\label{mp2}
  +\,4\,x_{n+1}^4\,\sum_{i=\max(1,a+b-n-2)}^{\min(a,b)-1}\,(a+b-1-2i)\;p_{i-1}(x)\;p_{a+b-2-i}(x)\,.
\end{equation}
This expression coincides exactly with that one one would find from Eq. (\ref{PjkdiBn}) in the case of $B_{n+1}$. In fact, Eq. (\ref{PjkdiBn}) gives, $\forall\,a\leq b=1,\ldots,n+1$:
$$  {\widehat P}_{ab}(p(y))= \sum_{i=\max(0,a+b-(n+1)-1)}^{\min(a,b)-1}\;4\,(a+b-1-2i)\;p_i(y)\;p_{a+b-1-i}(y)=
$$
$$=\sum_{i=\max(0,a+b-n-2)}^{\min(a,b)-1}\;4\,(a+b-1-2i)\;\left( p_{ i}(x)+p_{i - 1}(x)\,x_{n+1}^2\right)\;\left( p_{a+b-1-i}(x)+p_{a+b-2-i}(x)\,x_{n+1}^2\right)=$$
$$= \sum_{i=\max(0,a+b-n-1)}^{\min(a,b)-1}\;4\,(a+b-1-2i)\; p_{ i}(x)\;p_{a+b-1-i}(x)\,+$$
$$+\,4\,x_{n+1}^2\,\left\{\sum_{i=\max(0,a+b-n-2)}^{\min(a,b)-1}\,(a+b-1-2i)\; \,p_{i}(x)\;p_{a+b-2-i}(x)\,+\right.$$
$$+\left.\,\sum_{i=\max(1,a+b-n-1)}^{\min(a,b)-1}\;(a+b-1-2i)\;p_{i-1}(x)\;p_{a+b-1-i}(x)\right\}+$$
$$+\,4\,x_{n+1}^4\;\sum_{i=\max(1,a+b-n-2)}^{\min(a,b)-1}\;(a+b-1-2i)\; p_{i-1}(x)\;p_{a+b-2-i}(x)\,.
$$
In the first and third sums the lower limit of the sum has been changed from $\;\max(0,a+b-n-2)\;$ to $\;\max(0,a+b-n-1)$, because the factor $p_{a+b-1-i}(x)$ vanish when $i=a+b-n-2$, and in the two last sums the lower limit of the sum has been changed from $\;\max(0,a+b-n-2)\;$ (or from $\;\max(0,a+b-n-1)\;$) to $\;\max(1,a+b-n-2)$ (or to $\;\max(1,a+b-n-1)\;$), because the factor $p_{i-1}(x)$ vanish when $i=0$.\\
The first and the last sums coincide exactly with the first and the last sums in Eq. (\ref{mp2}). The expression inside the curly brackets needs to be transformed somewhat to see that it is identical to the expression inside the curly brackets in Eq. (\ref{mp2}). The road to follow is indicated by the fact one wants to obtain the sum that appears inside the curly brackets in Eq. (\ref{mp2}). One has:
$$\sum_{i=\max(0,a+b-n-2)}^{\min(a,b)-1}\,(a+b-1-2i)\; \,p_{i}(x)\;p_{a+b-2-i}(x)+\sum_{i=\max(1,a+b-n-1)}^{\min(a,b)-1}\;(a+b-1-2i)\;p_{i-1}(x)\;p_{a+b-1-i}(x)=$$
$$=\sum_{i=\max(0,a+b-n-2)}^{\min(a,b)-1}\,(a+b-1-2i)\; \,p_{i}(x)\;p_{a+b-2-i}(x)+\sum_{i'=\max(0,a+b-n-2)}^{\min(a,b)-2}\;(a+b-1-2(i'+1))\;p_{i'}(x)\;p_{a+b-1-(i'+1)}(x)=$$
$$=\sum_{i=\max(0,a+b-n-2)}^{\min(a,b)-1}\,(a+b-1-2i)\; \,p_{i}(x)\;p_{a+b-2-i}(x)+\sum_{i=\max(0,a+b-n-2)}^{\min(a,b)-2}\;(a+b-3-2i)\;p_{i}(x)\;p_{a+b-2-i}(x)=$$
$$=\sum_{i=\max(0,a+b-n-2)}^{\min(a,b)-1}\,(a+b-2-2i)\; \,p_{i}(x)\;p_{a+b-2-i}(x)+\sum_{i=\max(0,a+b-n-2)}^{\min(a,b)-1}\;p_{i}(x)\;p_{a+b-2-i}(x)\,+$$
$$+\,\sum_{i=\max(0,a+b-n-2)}^{\min(a,b)-2}\;(a+b-2-2i)\;p_{i}(x)\;p_{a+b-2-i}(x)-\sum_{i=\max(0,a+b-n-2)}^{\min(a,b)-2}\;p_{i}(x)\;p_{a+b-2-i}(x)=$$
$$=\sum_{i=\max(0,a+b-n-2)}^{\min(a,b)-2}\,2\,(a+b-2-2i)\; \,p_{i}(x)\;p_{a+b-2-i}(x)\,+$$
$$+\,\biggl\{\,\Bigl[\,a+b-2-2(\min(a,b)-1)\,\Bigr]+1\,\biggr\}\; \,p_{\min(a,b)-1}(x)\;p_{a+b-2-(\min(a,b)-1)}(x)=$$
$$=\sum_{i=\max(0,a+b-n-2)}^{\min(a,b)-2}\,2\,(a+b-2-2i)\; \,p_{i}(x)\;p_{a+b-2-i}(x)+(|a-b|+1)\; \,p_{a-1}(x)\;p_{b-1}(x)\,.$$
The simplification in the last line is obtained by distinguishing the cases $a>b$, $a=b$ and $a<b$.
This expression is identical to the expression inside the curly brackets in Eq. (\ref{mp2}). We have so proved that the whole expression for the matrix element $P_{ab}(y)$ obtained from the application of Eq. (\ref{PjkdiBn}) for the group $B_{n+1}$ coincides with Eq. (\ref{mp2}), that is obtained from the induction principle and the assumed validity of Eq. (\ref{PjkdiBn}) for the group $B_{n}$. This proves that Eq. (\ref{PjkdiBn}) is valid for $B_n$ for all values of $n$.\scat

\noindent\textbf{Proof of the generating formulas (\ref{lambdaBnEq1}), (\ref{lambdaBnEq2}) and (\ref{lambdaBnEq3}).}\\
The groups of type $B_n$ have two different root systems (and as crystallographic groups have two different root lengths in the ratio $\sqrt{2}$). Theorem \ref{cinque} then asserts that there are two different active factors $s(p)$ and $l(p)$ of $\det(\widehat P(p))$ that are the $w$-homogeneous polynomials in $p_1,\ldots,p_n$, that represent the invariant polynomials product of the squares of the linear forms $ l_r(x)=r\cdot x$, corresponding to the $n$ positive short roots and to the $2 {n\choose 2}$ positive long roots of $B_n$, respectively.
 Theorem  \ref{cinque} also tells us how to construct the corresponding $\lambda$-vectors.\\
Let's start with the short roots. As recalled in the beginning of Section \ref{Bn}, a set of $n$ positive short roots 
are $e_i$, $i=1,\ldots,n$
. The linear forms vanishing in the reflection hyperplanes of the short roots are the variables themselves: $x_i$, $i=1,\ldots,n$, and the polynomial $s(p(x))$, obtained as specified in Theorem \ref{cinque}, coincides with the highest degree basic invariant polynomial $p_n(x)$:
$$s(p(x))=\prod_{i=1}^n x_i^2=p_n(x)\,.
$$
Its $\lambda$-vector can be obtained from the first of Eqs. (\ref{lambdasl}). For all $a=1,\ldots,n$, one has:
$$  \lambda_a^{(s)}(p(x))=2\,\sum_{r\in{\cal R}_{+,s}}\frac{\nabla p_a(x)\cdot r}{l_r(x)}=
2\,\sum_{i=1}^n\frac{\nabla p_a(x)\cdot e_i}{x_i}=2\,\sum_{i=1}^n\frac{1}{x_i}\,\frac{\partial p_a(x)}{\partial x_i}\,.
$$
The basic invariant polynomial $p_a(x)$, written in Eq. (\ref{basicpolynBn}), has ${n\choose a}$ terms, and the number of those containing $x_i$ (actually $x_i^2$) is ${n-1\choose a-1}$. One then finds:
$$\frac{1}{x_i}\,\frac{\partial p_a(x)}{\partial x_i}=\frac{1}{x_i}\,2\,x_i\left(\sum_{(i_1,i_2,\ldots,i_{a-1})\subset (1,2,\ldots,\hat i,\ldots,n)} x_{i_1}^2x_{i_2}^2\cdots x_{i_{a-1}}^2 \right)
=2\left(\sum_{(i_1,i_2,\ldots,i_{a-1})\subset (1,2,\ldots,\hat i,\ldots,n)} x_{i_1}^2x_{i_2}^2\cdots x_{i_{a-1}}^2 \right),$$
where the sum is over all the combinations of $a-1$ integers $(i_1,i_2,\ldots,i_{a-1})$ among the first $n$ integers, excluding the integer $i$ (the hat in $\hat i$ means exclusion, and writing the symbol $(i_1,i_2,\ldots,i_{a-1})$ we automatically suppose $i_1<i_2<\ldots<i_{a-1}$). There are $n-1 \choose a-1$ such combinations. One then has:
$$  \lambda_a^{(s)}(p(x))=4\,\sum_{i=1}^n\,\left(\sum_{(i_1,i_2,\ldots,i_{a-1})\subset (1,2,\ldots,\hat i,\ldots,n)} x_{i_1}^2x_{i_2}^2\cdots x_{i_{a-1}}^2 \right).
$$
The second member is a completely symmetric polynomial with terms that are products of $a-1$ different variables squared, among $x_1,\ldots,x_n$. It must then be a multiple of the basic invariant polynomial $p_{a-1}(x)$. To find its coefficient one can count the number of terms in the expression above and in the expression of $p_{a-1}(x)$. The ratio of these two numbers gives the coefficient of $p_{a-1}(x)$ one is searching in. The expression at the second member above has
$c_1=n{n-1\choose a-1}$ terms, while the explicit expression of $p_{a-1}(x)$, given in Eq. (\ref{basicpolynBn}), has
$c_2={n\choose a-1}$ terms. The coefficient  of $p_{a-1}(x)$ one is searching in is then the following one:
$$\frac{c_1}{c_2}=\frac{n\,(n-1)!}{(a-1)!\,(n-a)!}\cdot\frac{(a-1)!\,(n-a+1)!}{n!}=(n-a+1)\,,
$$
and one finally obtains:
$$  \lambda_a^{(s)}(p(x))=4\,(n-a+1)\,p_{a-1}(x)\,,
$$
that concludes the proof of Eq. (\ref{lambdaBnEq1}).\\
There is a much simpler proof of Eq. (\ref{lambdaBnEq1}) if one makes use of the explicit form of the \wPm\ given in Eq. (\ref{PjkdiBn}). One starts by verifying that $p_n$ is an active polynomial. To do this one has to verify if $p_n$ satisfies the boundary equation (\ref{boundaryeqn}) with a proper $\lambda$-vector $\lambda^{(p_n)}(p)$. Let's take $a(p)=p_n$ in Eq. (\ref{boundaryeqn}). The first member of Eq. (\ref{boundaryeqn}) then reduces to the following one:
$$
 \sum_{b=1}^n{\widehat P}_{ab}(p)\,\frac{\partial
p_n}{\partial p_b}=\sum_{b=1}^n{\widehat P}_{ab}(p)\,\delta_{bn}=
{\widehat P}_{an}(p)=4\,(n-a+1)\,p_{a-1}\,p_n,\qquad a=1,\ldots,n\,,
$$
If one defines the $\lambda$-vector $\lambda^{(p_n)}(p)=(\lambda^{(p_n)}_1(p),\ldots,\lambda^{(p_n)}_n(p))$ such that
$$\lambda^{(p_n)}_a(p)=4\,(n-a+1)\,p_{a-1}\,,\qquad a=1,\ldots,n\,,$$
one sees that $p_n$ is active because it satisfies the boundary equation (\ref{boundaryeqn}). Theorem 4.1  of Ref. \cite{Sar-Tal1991}, item iv), establishes then that $p_n$ is a factor of $\det(\widehat P(p))$.\\

Let's consider now the long roots. As recalled in the beginning of Section \ref{Bn}, a set of $2{n \choose 2}$ positive long roots 
are $e_i\pm e_j$, $1\leq i<j\leq n$. The linear forms vanishing in the reflection hyperplanes of the long roots are then: $x_i\pm x_j$, $1\leq i<j\leq n$, and the polynomial $l(p(x))$, obtained as specified by Theorem \ref{cinque}, is the following one:
$$l(p(x))=\prod_{i<j=1}^n (x_i+x_j)(x_i-x_j)=\prod_{i<j=1}^n x_i^2-x_j^2\,.
$$
Its $\lambda$-vector can be obtained from the second of Eqs. (\ref{lambdasl}). For all $a=1,\ldots,n$, one has:
$$  \lambda_a^{(l)}(p(x))=2\,\sum_{r\in{\cal R}_{+,l}}\frac{\nabla p_a(x)\cdot r}{l_r(x)}=
2\,\nabla p_a(x)\cdot\sum_{i<j=1}^n\frac{e_i+e_j}{x_i+x_j}+\frac{e_i-e_j}{x_i-x_j}=
2\,\nabla p_a(x)\cdot\sum_{i<j=1}^n\frac{2\,x_i\,e_i-2\,x_j\,e_j}{(x_i+x_j)(x_i-x_j)}=
$$
\begin{equation}\label{ultimalambdaal}
  =
4\,\sum_{i<j=1}^n\,\frac{1}{x_i^2-x_j^2}\,\left(x_i\,\frac{\partial p_a(x)}{\partial x_i}-x_j\,\frac{\partial p_a(x)}{\partial x_j}\right).
\end{equation}
The basic polynomial $p_a(x)$, written in Eq. (\ref{basicpolynBn}), has ${n\choose a}$ terms, and the number of those containing $x_i$ (actually $x_i^2$) is ${n-1\choose a-1}$. One then finds:
$${x_i}\,\frac{\partial p_a(x)}{\partial x_i}=2\,x_i^2\left(\sum_{(i_1,i_2,\ldots,i_{a-1})\subset (1,2,\ldots,\hat i,\ldots,n)} x_{i_1}^2x_{i_2}^2\cdots x_{i_{a-1}}^2 \right),$$
where the sum is over all the combinations of $a-1$ integers $(i_1,i_2,\ldots,i_{a-1})$ among the first $n$ integers, excluding the integer $i$. There are $n-1 \choose a-1$ such combinations.
It is convenient to distinguish the terms containing the factor $x_j^2$ from those that do not. We can then write:
$${x_i}\,\frac{\partial p_a(x)}{\partial x_i}=2\left(\sum_{(i_1,i_2,\ldots,i_{a-2})\subset (1,2,\ldots,\hat i,\ldots,\hat j,\ldots,n)} x_i^2x_j^2 \cdot x_{i_1}^2x_{i_2}^2\cdots x_{i_{a-2}}^2+\sum_{(i_1,i_2,\ldots,i_{a-1})\subset (1,2,\ldots,\hat i,\ldots,\hat j,\ldots,n)} x_i^2\cdot x_{i_1}^2x_{i_2}^2\cdots x_{i_{a-1}}^2 \right).$$
The expression for ${x_j}\,\frac{\partial p_a(x)}{\partial x_j}$is similar, with $i$ and $j$ interchanged. When making the difference of the two expressions so found, all the terms of the first sums cancel, because of the symmetry in $i$ and $j$, and one finds:
$${x_i}\,\frac{\partial p_a(x)}{\partial x_i}-{x_j}\,\frac{\partial p_a(x)}{\partial x_j}
=2(x_i^2-x_j^2)\,\left(\sum_{(i_1,i_2,\ldots,i_{a-1})\subset (1,2,\ldots,\hat i,\ldots,\hat j,\ldots,n)} x_{i_1}^2x_{i_2}^2\cdots x_{i_{a-1}}^2 \right).
$$
Inserting this result in Eq. (\ref{ultimalambdaal}), one obtains:
$$  \lambda_a^{(l)}(p(x))=8\,\sum_{i<j=1}^n\,\left(\sum_{(i_1,i_2,\ldots,i_{a-1})\subset (1,2,\ldots,\hat i,\ldots,\hat j,\ldots,n)} x_{i_1}^2x_{i_2}^2\cdots x_{i_{a-1}}^2 \right).
$$
The second member is a completely symmetric polynomial with terms that are products of $a-1$ different variables squared, among $x_1,\ldots,x_n$. It must then be a multiple of the basic invariant polynomial $p_{a-1}(x)$. To find its coefficient one can count the number of terms in the expression above and in the expression of $p_{a-1}(x)$. The ratio of these two numbers gives the coefficient of $p_{a-1}(x)$ one is searching in. The expression at the second member above has
$c_1={n\choose 2}{n-2\choose a-1}$ terms, while the expression of $p_{a-1}(x)$, given in Eq. (\ref{basicpolynBn}), has
$c_2={n\choose a-1}$ terms. The coefficient  of $p_{a-1}(x)$ one is searching in is then the following one:
$$\frac{c_1}{c_2}=\frac{n\,(n-1)}{2}\,\frac{(n-2)!}{(a-1)!\,(n-a-1)!}\,\frac{(a-1)!\,(n-a+1)!}{n!}=\frac{(n-a+1)\,(n-a)}{2}\,,
$$
and one obtains:
$$  \lambda_a^{(l)}(p(x))=8\,\frac{(n-a+1)\,(n-a)}{2}\,p_{a-1}(x)=4\,{(n-a+1)\,(n-a)}\,p_{a-1}(x)\,.
$$
Eq. (\ref{lambdaBnEq2}) is so proved.\\

From Eq. (\ref{lambdas+lambdal=lambdaD}) one then has:
$$  \lambda_a^{(\det(\widehat P))}(p)=\lambda_a^{(s)}(p)+\lambda_a^{(l)}(p)=4\,{(n-a+1)}\,p_{a-1}+4\,{(n-a+1)(n-a)}\,p_{a-1}=4\,{(n-a+1)^2}\,p_{a-1}\,.
$$
This concludes the proof of Eq. (\ref{lambdaBnEq3}).\scat

\subsection{Proof of Theorem \ref{lambdaDn} for $D_n$\label{proofDn}}
\noindent\textbf{Proof of the generating formula (\ref{PjkdiDn_divisa}).}\\
Comparing Eq. (\ref{basicpolynDn}), giving the basic invariant polynomials of $D_n$, with Eq. (\ref{basicpolynBn}), giving the basic invariant polynomials of $B_n$, one sees that,
for $a<n$ the basic invariant polynomials of $D_n$ are identical with the basic invariant polynomials of $B_n$, and the last basic invariant polynomial $p_n(x)$ of $B_n$ is just the square of the last basic invariant polynomial $p_n(x)$ of $D_n$. It is convenient in this proof to write $p_n^B(x)$ and $p_n^D(x)$ to distinguish the basic invariant polynomial $p_n(x)$ of $B_n$ and $D_n$, respectively. One has thus $(p_n^D(x))^2=p_n^B(x)$. No confusion arises for the other basic invariant polynomials, so $p_a(x)$, when $a<n$, needs no specification. This implies that for $a,b<n$, the matrix elements of the matrix $P(x)$, obtained through Eq. (\ref{matriceP(x)}), when calculated with the basic invariants of $D_n$ or of $B_n$, is the same, and when one has to use Eq. (\ref{invmatP}) to express these invariant polynomials in terms of the basic invariant polynomials, the expressions obtained for $D_n$ differ from those of $B_n$ only for the use of $(p_n^D)^2$ in place of $p_n^B$. Then the matrix elements $\widehat P_{ab}^D(p)$ of the \wPm\ of $D_n$ are the same of the matrix elements $\widehat P_{ab}^B(p)$ of the \wPm\ of $B_n$, except for the substitution $p_n^B\to (p_n^D)^2$, that is, the variable $p_n$ of $B_n$ must be substituted with the square of the variable $p_n$ of $D_n$. Using Eq. (\ref{PjkdiBn}) and the substitution $p_n\to p_n^2$, the first line of Eq. (\ref{PjkdiDn_divisa}) follows immediately.\\
To prove the second and third lines  of Eq. (\ref{PjkdiDn_divisa}) it is possible to calculate the matrix elements of the matrix $\widehat P^D(p)$ of $D_n$ using the explicit form of the matrix elements of the matrix $\widehat P^B(p)$ of $B_n$, written in Eq. (\ref{PjkdiBn}). As $p_n^B(x)=(p_n^D(x))^2$, one has $\nabla p_n^B(x)=\nabla (p_n^D(x))^2=2 p_n^D(x)\,\nabla p_n^D(x)$. Then, $\forall\,a=1,\ldots,n-1$:
$$\widehat P_{an}^D(x)=\nabla p_a(x)\cdot \nabla p_n^D(x)=\frac{1}{2 \,p_n^D(x)}\,\nabla p_a(x)\cdot \nabla p_n^B(x)=\frac{1}{2 \,p_n^D(x)}\,\widehat P_{an}^B(p(x))=\frac{1}{2\, p_n^D(x)}\,4\,(n-a+1)\,p_{a-1}(x)\,p_n^B(x)=$$
$$=\frac{1}{ p_n^D(x)}\,2\,(n-a+1)\,p_{a-1}(x)\,(p_n^D(x))^2=2\,(n-a+1)\,p_{a-1}(x)\,p_n^D(x)$$
The second line of Eq. (\ref{PjkdiDn_divisa}) is thus proved. To prove the third line, we have:
$$\widehat P_{nn}^D(x)=\nabla p_n^D(x)\cdot \nabla p_n^D(x)=\frac{1}{4 \,(p_n^D(x))^2}\,\nabla p_n^B(x)\cdot \nabla p_n^B(x)=\frac{1}{4\, p_n^B(x)}\,\widehat P_{nn}^B(p(x))=\frac{1}{4\, p_n^B(x)}\,4\,p_{n-1}(x)\,p_n^B(x)=p_{n-1}(x)$$
The third line of Eq. (\ref{PjkdiDn_divisa}) is thus proved and  (\ref{PjkdiDn_divisa}) is then completely proved.\scat

\noindent\textbf{Proof of the generating formula (\ref{lambdaDnEq}).}\\
Theorem \ref{quattro} tells us how to construct the $\lambda$-vector $\lambda_a^{(D)}(p)$ corresponding to the determinant $\det(\widehat P(p))=D(p)$ of the \wPm\ $\widehat P(p)$.\\
As recalled in the beginning of Section \ref{Dn}, a set of $2{n\choose 2}$ positive roots of $D_n$ is for example formed by the vectors $e_i\pm e_j$, $1\leq i<j\leq n$, and the linear forms vanishing in the corresponding reflection hyperplanes are $x_i\pm x_j$, $1\leq i<j\leq n$.
Eq. (\ref{lambdaD}) gives then, for all $a=1,\ldots,n$:
$$  \lambda_a^{(D)}(p(x))=2\,\sum_{r\in{\cal R}_+}\frac{\nabla p_a(x)\cdot r}{l_r(x)}=
2\,\nabla p_a(x)\cdot\sum_{i<j=1}^n\frac{e_i+e_j}{x_i+x_j}+\frac{e_i-e_j}{x_i-x_j}=
2\,\nabla p_a(x)\cdot\sum_{i<j=1}^n\frac{2\,x_i\,e_i-2\,x_j\,e_j}{(x_i+x_j)(x_i-x_j)}=
$$
$$=
4\,\sum_{i<j=1}^n\,\frac{1}{x_i^2-x_j^2}\,\left(x_i\,\frac{\partial p_a(x)}{\partial x_i}-x_j\,\frac{\partial p_a(x)}{\partial x_j}\right).
$$
For $a=1,\ldots,n-1$, the basic invariant polynomials of $D_n$ are the same as those of $B_n$, so the previous calculation has already been done in the proof of Eq. (\ref{lambdaBnEq2}). Using the result there obtained for $\lambda_a^{(l)}(p(x))$, corresponding to the long roots of $B_n$, we can write:
$$  \lambda_a^{(D)}(p)=4(n-a+1)(n-a)\,p_{a-1}\,,\qquad \forall\,a=1,\ldots,n-1\,.$$
This is equal to Eq. (\ref{lambdaDnEq}), but only for $a=1,\ldots,n-1$.
It remains to calculate the case $a=n$. One has $p_n(x)=x_1x_2,\cdots,x_n$, that implies: $\nabla p_n(x)=\left(\frac{1}{x_1},\frac{1}{x_2},\ldots,\frac{1}{x_n}\right)p_n(x)$. Then:
$$  \lambda_n^{(D)}(p(x))=
4\,\sum_{i<j=1}^n\,\frac{1}{x_i^2-x_j^2}\,\left(x_i\,\frac{\partial p_n(x)}{\partial x_i}-x_j\,\frac{\partial p_n(x)}{\partial x_j}\right)=
4\,\sum_{i<j=1}^n\,\frac{1}{x_i^2-x_j^2}\,\left(x_i\,\frac{1}{x_i}\,p_n(x)-x_j\,\frac{1}{x_j}\,p_n(x)\right)=0\,.
$$
Eq. (\ref{lambdaDnEq}) is so completely proved.\scat

\subsection{Proof of Theorem \ref{hankelBn}\label{proofhankelBn}}

Let's consider the generating Eq. (\ref{PjkdiBn}), relative to the group $B_n$, in which, for the symmetry, we can suppose $a\leq b$. The variable $p_n$, representing the basic invariant polynomial of the highest degree $2n$, can be found in the element $\widehat P_{ab}(p)$ only in $p_{a+b-1-i}$ when $i=a+b-n-1$. From the lower limit of the sum in  Eq. (\ref{PjkdiBn}) we know that $i=a+b-n-1$ is possible only if $a+b-n-1\geq 0$, that is if $a+b\geq n+1$. This tells us that $p_n$ can be found not above the auxiliary diagonal of the \wPm, only. (This is obvious also from the calculation of the weights of the matrix elements of the \wPm: $w(\widehat P_{ab}(p))=d_a+d_b-2=2a+2b-2=2(a+b)-2$, and above the auxiliary diagonal one has $a+b<n+1$ that gives $w(\widehat P_{ab}(p))<2n=d_n$). The coefficient of $p_n$ in $\widehat P_{ab}(p)\,$, supposing $a+b\geq n+1$,  is obtained from  Eq. (\ref{PjkdiBn}) with $i=a+b-n-1$, and is the following one: $4\, (a+b-1-2(a+b-n-1))\,p_{a+b-n-1}=4\, (2n-a-b+1)\,p_{a+b-n-1}$. This result depends on $a$ and $b$ only through the sum $a+b$, and this means that all the elements of the \wPm\ that are in a same up going diagonal (that is parallel to the auxiliary diagonal), contain $p_n$ with the same coefficient, the same of the element in the last row, that one corresponding to $b=n$. Eq.  (\ref{PjkdiBn}) gives $\widehat P_{na}(p)=4\, (n-a+1)\,p_{a-1}\,p_n$, so the coefficient of $p_n$ in all elements of the up going diagonal starting at $\widehat P_{na}$ is $4\, (n-a+1)\,p_{a-1}$, $\forall\,a=1,\ldots,n$. Using the definition of the Hankel matrices $H_a^{(n)}(p)$, given just after Eq. (\ref{hankelmatrixhn}), we can say that the terms in the matrix $\widehat P(p)$ containing $p_n$ form the Hankel matrix  $4\,p_n\,H_n^{(n)}(p)$.\\
Consider now the matrix $\widehat P(p)-4\,p_n\,H_n^{(n)}(p)$. It does not contain $p_n$ and its greatest weight variable is $p_{n-1}$. The variable $p_{n-1}$, when not multiplied by $p_n$, is contained in the element $\widehat P_{ab}(p)$ only in $p_{a+b-1-i}$ when $i=a+b-n$. From the lower limit of the sum in  Eq. (\ref{PjkdiBn}) we know that $i=a+b-n$ is possible only if $0\leq a+b-n\leq \min(a,b)-1=a-1$, and we obtain the inequalities $b\leq n-1$ and $a+b\geq n$. This tells us that $p_{n-1}$, when not multiplied by $p_n$, can be found only not above the diagonal of the \wPm\ connecting $\widehat P_{n-1,1}(p)$ with $\widehat P_{1,n-1}(p)$, and not in the last row (and column) of the \wPm. The coefficient of $p_{n-1}$ in $\widehat P_{ab}(p)\,$, supposing $a+b\geq n$ and $b\leq n-1$, is obtained from  Eq. (\ref{PjkdiBn}) with $i=a+b-n$, and is the following one:
$4\, (a+b-1-2(a+b-n))\,p_{a+b-n}=4\, (2n-a-b-1)\,p_{a+b-n}$. This result depends on $a$ and $b$ only through the sum $a+b$, and this means that all the elements of the matrix $\widehat P(p)-4\,p_n\,H_n^{(n)}(p)$ that are in a same up going diagonal, contain $p_{n-1}$ with the same coefficient, the same of the element in the second-last row, that one corresponding to $b=n-1$ in  Eq. (\ref{PjkdiBn}). From Eq. (\ref{PjkdiBn}) one sees that $\widehat P_{n-1,a}(p)$, $\forall\,a=1,\ldots,n-1$, contains the term $4\, (n-a)\,p_{a-1}\,p_{n-1}$, so the coefficient of $p_{n-1}$ in all elements of the up going diagonal starting at $\widehat P_{n-1,a}(p)$ is $4\, (n-a)\,p_{a-1}$. Using the definition of the Hankel matrix $H_{n-1}^{(n)}(p)$, we can say that the terms in the matrix  $\widehat P(p)-4\,p_n\,H_n^{(n)}(p)$ containing $p_{n-1}$, but not $p_n$, form the Hankel matrix  $4\,p_{n-1}\,H_{n-1}^{(n)}(p)$.\\
We can continue this way finding where $p_{n-2}$ is contained in the matrix $\widehat P(p)-4\,p_n\,H_n^{(n)}(p)-4\,p_{n-1}\,H_{n-1}^{(n)}(p)$ and again for $p_{n-3}$,  $p_{n-4}$, and so on, and the formula (\ref{PBnHankel}) easily follows.\scat

\subsection{Proof of Theorem \ref{hankelDn}\label{proofhankelDn}}
Theorem \ref{hankelBn} states that Eq. (\ref{PBnHankel}) is equivalent to Eq. (\ref{PjkdiBn}). A comparison of Eqs. (\ref{PjkdiBn}) and (\ref{PjkdiDn_divisa}) is sufficient to understand that the \wPm\ of $D_n$ in the upper left $(n-1)\times(n-1)$ block is equal to the  \wPm\ of $B_n$ except for the substitution of $p_n$ with $p_n^2$. Using Eq. (\ref{PBnHankel}), one finds the first two terms at the second member of Eq. (\ref{PDnHankel}).
A comparison of Eq. (\ref{PjkdiBn}), for $b=n$, that implies $i=a-1$, and the second line of Eq. (\ref{PjkdiDn_divisa}), is sufficient to understand that the elements in the last row and column of the \wPm\ of $D_n$, except the $(n,n)$ element, are one half of the corresponding elements of the \wPm\ of $B_n$. Using Eq. (\ref{PBnHankel}) for $a=n$, this gives the third term at the second member of Eq. (\ref{PDnHankel}). The third line of Eq. (\ref{PjkdiDn_divisa}) is represented by the fourth term at the second member of Eq. (\ref{PDnHankel}). Eq. (\ref{PDnHankel}) and Theorem \ref{hankelDn} are so proved.\scat

\subsection{Proof of Theorem \ref{hankelAn}\label{proofhankelAn}}
The \wPm\ corresponding to the basic invariant polynomials of $A_n$ defined in Eqs. (\ref{elemsymmpolSn+1})--(\ref{basicpolynAn}) can be constructed using Eq. (\ref{PjkdiAn}). Let's consider the three terms in the expression of $\widehat P_{ab}(p)$ given by Eq. (\ref{PjkdiAn}). By considering all possible values of  $a,b=1,\ldots,n$, they define three $n\times n$ matrices. The matrix corresponding to the first term in Eq. (\ref{PjkdiAn}) corresponds to the symmetric matrix $R^{(n)}(p)$ defined in Eq. (\ref{matRhankel}), and corresponds to the first term in the second member of Eq. (\ref{PAnHankel}). The matrix corresponding to the second term in Eq. (\ref{PjkdiAn}) is a Hankel matrix because it depends only on the sum $a+b$, and corresponds to the matrix $2\,\sum_{a=1}^n (a+1)\,p_a \,Y_a^{(n)}$ in Eq. (\ref{PAnHankel}). We have then just to prove that the matrix $T(p)$ corresponding to the third term in Eq. (\ref{PjkdiAn}), that one with the summation symbol, corresponds to the matrix $-(n+1)\,\sum_{a=1}^n p_a \,K_a^{(n)}(p)$ in Eq. (\ref{PAnHankel}). From Eq. (\ref{PjkdiAn}), we have:
$$T_{ab}(p)=-(n+1)\,\sum_{i=\max(2,a+b-n-1)}^{\min(a,b)-1} \,(a+b - 2i)\;p_{i- 1}\;p_{a+b-1-i}\,,\qquad \forall\,a,b=1,\ldots,n\,,$$
in which one has to consider $p_0=0$. We can simplify the formula defining the matrix $T(p)$. For the symmetry of the matrix $T(p)$ we can suppose $a\leq b$. Moreover, we can define $p_a=0$, $\forall\,a<0$ and $\forall\,a>n$, and can then write:
$$T_{ab}(p)=-(n+1)\,\sum_{i=2}^{a-1} \,(a+b - 2i)\;p_{i- 1}\;p_{a+b-1-i}\,,\qquad \forall\,a,b=1,\ldots,n\,.$$
We can now change the lower limit of the sum in the following way:
$$T_{ab}(p)=-(n+1)\,\sum_{i=1}^{a-1} \,(a+b - 2i)\;p_{i- 1}\;p_{a+b-1-i}\,,\qquad \forall\,a,b=1,\ldots,n\,,$$
because for $i=1$ there is a factor $p_0$ that is equal to 0.
If $a=1$, $T_{ab}(p)=0$, $\forall\,b=1,\ldots,n$, so $T(p)$ has its first row (and column) with elements equal to 0. (Actually, also the second row (and column) of $T(p)$ is zero, but this does not matter here).
We can define a matrix $T'(p)$ that has for its first $n-1$ rows the last $n-1$ rows of the matrix $T(p)$ (that are so shifted one position towards the top). We can write the formula for the matrix elements of $T'(p)$ just by replacing the row label $a$ of the formula for $T(p)$ with $a+1$. One then has:
$$T'_{ab}(p)=T_{a+1,b}(p)=-(n+1)\,\sum_{i=1}^{a} \,(a+b+1 - 2i)\;p_{i- 1}\;p_{a+b-i}\,,\qquad \forall\,a=1,\ldots,n-1\,,\quad \forall\,b=1,\ldots,n\,.$$
Forgetting $T(p)$ we can allow the index $a$ run up to $n$. Then:
$$T'_{ab}(p)=-(n+1)\,\sum_{i=1}^{a} \,(a+b+1 - 2i)\;p_{i- 1}\;p_{a+b-i}\,,\qquad \forall\,a,b=1,\ldots,n\,,$$
and changing the summation index from $i$ to $i'=i-1$, one can write:
$$T'_{ab}(p)=-(n+1)\,\sum_{i'=0}^{a-1} \,(a+b-1 - 2i')\;p_{i'}\;p_{a+b-1-i'}\,,\qquad \forall\,a,b=1,\ldots,n\,.$$
With the same definitions for the variables $p_a$, that is $p_0=0$ and $p_a=0$, $\forall\,a<0$ and $\forall\,a>n$, the \wPm\ of $B_n$, in Eq. (\ref{PjkdiBn}), can be written in the following way:
$$\widehat P_{ab}(p)= \sum_{i=0}^{a-1} \,4\,(a+b -1- 2i)\;p_{i}\;p_{a+b-1-i}\,,\qquad \forall\,a,b=1,\ldots,n\,.$$
Comparing with the expression of $T'(p)$ we see that the only difference, in addition to $p_0=0$ instead of $p_0=1$, is the coefficient: $-(n+1)$ in the case of $T'(p)$ and $4$ in the case of the \wPm\ of $B_n$. We can then use Theorem \ref{hankelBn} and Eq. (\ref{PBnHankel}) and write:
$$T'(p)=-(n+1)\,\sum_{a=1}^n\,p_a\,H_a^{(n)}(p)\,,$$
but we have to consider $p_0=0$ instead of $p_0=1$.
We can now obtain $T(p)$ just by shifting one position to the bottom the first $n-1$ rows of $T'(p)$, and setting all the elements in the first row of $T(p)$ equal to zero. For this it is then convenient to define the matrices $K_a^{(n)}(p)$ that are obtained from the matrices $H_a^{(n)}(p)$ just by shifting one position towards the bottom their first $n-1$ rows, and by substituting $p_0=0$ in place of $p_0=1$. We have then:
$$T(p)=-(n+1)\,\sum_{a=1}^n\,p_a\,K_a^{(n)}(p)\,,$$
and this completes the proof of Theorem \ref{hankelAn}.\scat

\section{appendix}

In this Appendix I list alternative expressions of the generating formulas  (\ref{PjkdiSn}),  (\ref{PjkdiAn}),  (\ref{PjkdiBn}) and  (\ref{PjkdiDn_divisa}) in which the symmetry of the resulting matrices is hidden, but that can, however, be of some interest.\\

Formula equivalent to Eq. (\ref{PjkdiSn}), for the groups of type $S_n$:
$$  {\widehat P}_{ab}(p)= (n+1-b)\;p_{a-1}\;p_{b -1}+ \sum_{i=1}^{a-1} \,(a- b - 2i)\;p_{a -1- i}\;p_{b - 1+i}\,,\qquad \forall\,a,b=1,\ldots,n\,,
$$
in which one has to consider $p_0=1$ and $p_a=0$, $\forall\,a>n$.\\

Formula equivalent to Eq. (\ref{PjkdiAn}), for the groups of type $A_n$:
$${\widehat P}_{ab}(p)=a(n+1-b)\,p_{a-1}\;p_{b-1}+2 \,(a+ b)\;p_{a+b-1}
 +(n+1)\,\sum_{i=1}^{a-2} \,(a- b - 2i)\;p_{a-1- i}\;p_{b-1+i}\,,\qquad \forall\,a,b=1,\ldots,n\,,
$$
in which one has to consider $p_0=0$ and $p_a=0$, $\forall\,a>n$.\\

Formula equivalent to Eq. (\ref{PjkdiBn}), for the groups of type $B_n$:
 $${\widehat P}_{ab}(p)= 4\,\sum_{i=1}^a \,(b - a - 1 + 2i)\;p_{a - i}\;p_{b -1+ i}\,,\qquad \forall\,a,b=1,\ldots,n\,,$$
in which one has to consider $p_0=1$ and $p_a=0$, $\forall\,a>n$.\\

Formula equivalent to Eq. (\ref{PjkdiDn_divisa}), for the groups of type $D_n$:
$$\left\{\begin{array}{rcll}
\nonumber  {\widehat P}_{ab}(p)=&=& 4\, \left. \sum_{i=1}^a\,
        (b -a - 1 + 2i)\,p_{a - i}\, p_{b - 1 + i}\,\right|_{p_n\to p_n^2} & \qquad \forall\,a,b=1,\ldots,n-1\,,\\
   {\widehat P}_{an}(p)={\widehat P}_{na}(p) &=& 2\,(n-a+1)\,p_{a - 1}\,p_n & \qquad \forall\,a=1,\ldots,n-1\,,\\
\nonumber  {\widehat P}_{nn}(p) &=& p_{n-1}\,, &
\end{array}\right.
$$
where one has to consider $p_0=1$ and $p_a=0$, $\forall\,a>n$.\\

To obtain the formulas written in this appendix from those in Eqs. (\ref{PjkdiSn}),  (\ref{PjkdiAn}),  (\ref{PjkdiBn}) and  (\ref{PjkdiDn_divisa}), one starts to consider $a\leq b$, for the symmetry of the \wPms\ obtained by Eqs. (\ref{PjkdiSn}),  (\ref{PjkdiAn}), (\ref{PjkdiBn}) and  (\ref{PjkdiDn_divisa}), define $p_a=0$ $\forall\,a<0$ and $\forall\, a>n$, in such a way to drop all $\min$ and $\max$ in Eqs. (\ref{PjkdiSn}),  (\ref{PjkdiAn}),  (\ref{PjkdiBn}) and  (\ref{PjkdiDn_divisa}), and then change the summation index from $i$ to $i'=a-i$. The formulas written above generate symmetric matrices (a formal proof is not short so I do not reproduce here), so it is not necessary to consider the limitation $a\leq b$, but it is anyway convenient because if one considers $a\leq b$ one finds a smaller number of terms originating from the sums. If $a\geq b$ many of these terms cancel each others.

\end{document}